\documentclass[npg,manuscript]{copernicus}

\usepackage{float}

\begin{document}

\nolinenumbers

\title{Nonlinear Ocean Waves Amplification in Straits}

\Author[1,3,4]{Andrei}{Pushkarev}
\Author[1,2,3]{Vladimir}{Zakharov}

\affil[1]{Skoltech, Bolshoy Boulevard 30, bld. 1, Moscow, 121205, Russia}
\affil[2]{Department of Mathematics, University of Arizona, Tucson, AZ 85721, USA}
\affil[3]{Waves and Solitons LLC, 1719 W. Marlette Ave., Phoenix, AZ 85015, USA}
\affil[4]{Lebedev Physical Institute RAS, Leninsky 53, Moscow 119991, Russia}

\runningtitle{RUNNINGTITLE}

\runningauthor{RUNNINGAUTHOR}

\correspondence{Andrei Pushkarev (dr.push@gmail.com)}


\firstpage{1}

\maketitle

\begin{abstract}

We study deep water ocean wind-driven waves in strait, with wind directed orthogonally to the shore, through exact Hasselmann equation.
Despite of "dissipative" shores - we do not include any reflection from the coast lines -  we show that the wave turbulence evolution can be split in time into two different regimes. During the first wave propagate along the wind, and the wind-driven sea can be described by the self-similar solution of the Hasselmann equation like in the open sea. The second regime starts later in time, after significant enough wave energy accumulation at the down-wind boundary. Since this moment the ensemble of waves propagating against the wind starts its formation.
Also, the waves, propagation along the strait start to appear. The wave system eventually reaches asymptotic stationary state in time, consisting of two co-existing states: the first, self-similar wave ensemble, propagating with the wind, and the second, quasi-monochromatic waves, propagating almost orthogonal to the wind direction and tending to slant against the wind at the angle of $15^{\circ}$ closer to the wave turbulence origination shore line. These "secondary waves" appear only due to intensive nonlinear wave interaction. The total wave energy exceeds its "expected value" approximately by the factor of two, with respect to estimated in the absence of the shores. It is expected that in the reflective shores presence this amplification will grow essentially. We propose to call this laser-like {\bf N}onlinear {\bf O}cean {\bf W}aves {\bf A}mplification mechanism by the acronym \textit{\bf NOWA}. \end{abstract}

\introduction \label{Int}

The geophysical phenomena are known to be strongly affecting the cource of humankind history. One of the relatively recent example is allied invasion of Normandy in Operation Overlord during World War II in 1944, the largest seaborne invasion in history. The conditions of the landing assumed requirements for the weather, sea  waves, the phase of the moon, the tides, and the time of the day that meant only a few days each month were deemed suitable. In the cource of the invasion, partially due to poor forecasting, rough seas significantly contributed into allied casualties reaching at least 10,000, with 4,414 confirmed dead.

In the current reasearch we are trying to shed the light on the specificity of the wind sea development in the ocean straits. While it is the widespread oceanographic community opinion that there are all necessary research tools, including operational wave forecasting models and parallel computers, we show that there are still new aspects of the related underlying physics, which understanding is highly desireble for proper construction of the wave forecasting models.

The modern ocean waves forecasting has begun with the wind driven water surface gravity waves statistical theory, described by \cite{R1,R2} kinetic equation (hearafter HE):

\begin{equation}
\label{HE}
\frac{\partial \varepsilon }{\partial t} +\frac{\partial \omega _{k} }{\partial \vec{k}} \frac{\partial \varepsilon }{\partial \vec{r}} =S_{nl} +S_{in} +S_{diss}
\end{equation}
where $\varepsilon =\varepsilon (\omega_k,\theta,\vec{r},t)$ is the wave energy spectrum, as a function of wave dispersion $\omega_k =\omega (k)$, angle $\theta$, two-dimensional real space coordinate $\vec{r}=(x,y)$ and time $t$. $S_{nl}$, $S_{in}$ and $S_{diss}$ are the nonlinear, wind input and wave-breaking dissipation source terms, respectively. Hereafter, only the deep water case, $\omega=\sqrt{g k }$ is considered, where $g$ is the gravity acceleration and $k=|\vec{k}|$ is the absolute value of the vector wavenumber $\vec{k}=(k_x,k_y)$. 

It is widely distributed opinion in the oceanographic community that Eq.({\ref{HE}}) is already well-studied object, especially in the considered deep-water case, and the further efforts have to be concentrated on the improvement of the simulation with extremely high hurricane winds, the bottom friction influence and the corrections related to the shallow water effects. In the presented research we show that Eq.({\ref{HE}}), however, in its "simple" form is still the object, deserving further detailed study, which can reveal unknown physical effects.

Such implications are related to the presence of the boundary conditions in real space, such as shore lines in straits as well as necessity of using correct nonlinear interaction source term $S_{nl}$ instead of routinely used more computationally effective DIA surrogate. As we show below, those effect can significanty modify already seems to be well understood physical picture of the wind-driven surface waves turbulence in unlimited domains.

\subsection{Recent achievements in the development of physically justified HE source terms in unlimited domains}

Numerical comparison of different historically developed $S_{in}$ source terms shows the scatter up to the factor of five, see  \cite{R6,R7}. The situation is not better with $S_{diss}$ wave energy dissipation source terms. As the result, modern operational wave forecasting models have more than two dozen of tuning parameters. The pertinent detailed analysis can be found in \cite{R7}.

The step toward creation of tuning-free wave forecasting models has been made in \cite{R8,R88}. The Eq.(\ref{HE}) in stationary limited fetch approximation

\begin{align}
\label{LFeq}
\frac{1}{2} \frac{g \cos \theta}{\omega } \frac{\partial \epsilon}{\partial x} = S_{nl} + S_{wind}
\end{align}
\noindent
has been analyzed for self-similar solutions

\begin{equation}
\label{SelfSimFetch}
\epsilon = x^{p+q} F(\omega x^q,\theta)
\end{equation}

\noindent
which gave in the assumption of power dependence on the frequency of the wind input term $S_{in} \sim \omega^{s+1} f(\theta)$:

\begin{eqnarray}
\label{qsRelation}
q &=& \frac{1}{2+s} \\
\label{psRelation}
p &=&\frac{8-s}{2(2+s)}
\end{eqnarray}

\noindent
and the  "magic relation"

\begin{equation}
\label{MagicRelation}
10q-2p = 1
\end{equation}

Eqs.(\ref{qsRelation})-(\ref{psRelation}) are incomplete and need extra relation. It is provided by the experimental regression line \citep{R20}, yielding specific indices values, defining ZRP model \cite{R8,R88}:

\begin{eqnarray}
\label{Indices}
&& p=1,\,\,q=0.3 \\
&& s=4/3 \\
&& S_{in} \sim \omega^{7/3}
\end{eqnarray}

Eq.(\ref{LFeq}) gives the dependencies of total energy and mean frequency on the fetch coordinate:

\begin{eqnarray}
\label{EnFetchSelfSim}
&& E(x) = E_0 x^p                      \\
\label{FrFetchSelfSim}
&&<\omega(x)> = \omega_0 x^{-q}
\end{eqnarray} 

ZRP model reproduces a dozen of limited fetch experimental field observations, analyzed in \cite{R18}, as well as self-similar dependencies Eqs.(\ref{Indices}), (\ref{EnFetchSelfSim}), (\ref{FrFetchSelfSim}), and does not require tuning with respect to the wind speed change from 5 m/sec to 10 m/sec \citep{R88}. While ZRP approach is not completely tuning-free, as it contains two tuning parameters in the wind input term $S_{in}$, it might be considered as the step forward toward physically justified models of HE Eq.(\ref{HE}). The facts, presented in the current subsection, consistute the argumentation for the ZRP source terms set choice as the basis for HE numerical simulation in the presented research.

\section{HE simulation for bounded domains}

The above described self-similar theory should be applied  with care for the surface wave turbulence description in limited domains, such as ocean straits. It is intuitively conceivable that mutual interplay of the wave energy advection velocity sign change in conjunction with four-wave nonlinear interaction, due to wave energy concentation at the specific locations, can modify self-similar picture of the wave turbulence in unbounded domains. The numerical simulation tools come in the situations like this in the first place for proper understanding of the underlying physics due to considered problem complexity.

Below, we consider first the geometry of the problem and formulate simple, but physical boundary conditions for the corresponding Caughy problem. Attraction of the Fourier space "pipelines" concept for interpretation of the energy advection in different directions with respect to the blowing wind, appears to be productive for the 
formulation and understanding of the problem statement. 

Next, we describe the model used for the numerical simulation. We carry out the numerical tests, demostrating the proper reproduction of the advection analytical properties of Eq.(\ref{HE}), "stripped out" of the nonlinear interaction term.

And finally, we describe and discuss the numerical simulation results in the form of integral and spectral characteristics of the studied surface wave turbulence as well as the similarities and the differences between already known self-similar and newly observed wave turbulence regimes.

\subsection{Cauchy problem statement } \label{CauchyProblem}

\begin{figure}[!]
\noindent\center\includegraphics[scale=0.6]{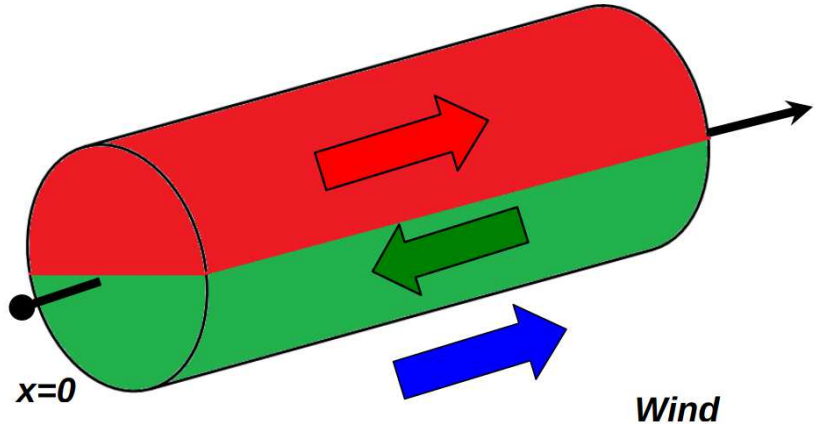} 
\caption{Schematic description of the wave energy fluxes propagation along the fetch in Real and Fourier space.}
\label{Fluxes}
\end{figure}

Let us consider Cauchy problem for the one-dimensional, time-dependent version of Eq.(\ref{HE}):

\begin{eqnarray}
\frac{\partial \varepsilon}{\partial t} + \frac{g}{2\omega} \cos{\theta} \frac{\partial \varepsilon}{\partial x} = S_{nl}+S_{in}+S_{diss}
\label{HE2}
\end{eqnarray}

\begin{figure}[ht]
\noindent\center\includegraphics[scale=0.3]{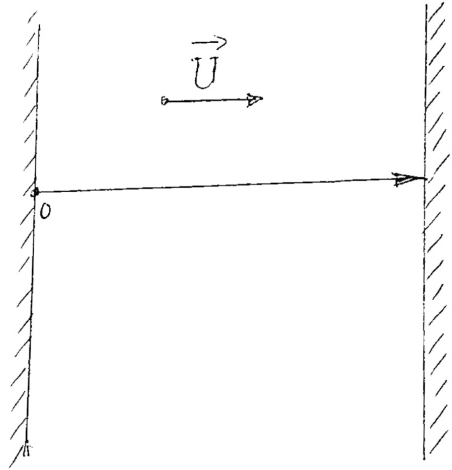} 
\caption{ The schematic presentation of the simulation domain in real space.  $\vec{U}$ is the constant wind velocity in the direction of the fetch axis $\vec{x}$.} \label{Geometry}
\end{figure}

It is assumed that $\varepsilon=\varepsilon(\omega_k,\theta,x,t)$, where $x$ is the coordinate of the axis $\vec{x}$, othogonal to the shoreline (referenced in physical ocenography as the "fetch"), the wave vector $\vec{k}=(k,\theta)$ is two-dimensional and is defined by its modulus $k=|\vec{k}|$ and the angle $\theta$ between $\vec{k}$ and $\vec{x}$. The problem is homogeneous in the direction orthogonal to the axis $\vec{x}$.  For the convenience, we call the left shoreline the "west coast", and the right shoreline the "east coast". The corresponding $R$-space domain of the length $L$ between the "west" and "east" coastlines is schematically presented on Fig.\ref{Geometry}. The wind speed $\vec{U}$ is assumed to be constant in the direction of $x$.
 
The mixed "Real-Fourier" space of the system is schematically presented on Fig.\ref{Fluxes}. The advection group velocity in front of the spatial derivative $\frac{g}{2\omega}\cos{\theta}$ in Eq.(\ref{HE2})   has different signs for waves propagating along and agains the wind. The corresponding domain is drawn as red-green cylinder with the red part corresponding to the positive advection velocity (red arrow is directed from the west to the east shoreline), and the green part corresponding to the negative advection velocity. This schematic picture suggests that the limited fetch growth consists of three processes happening in the Real and Fourier spaces:

\begin{itemize}
\item{} wave energy advection in the fetch direction (red part of the tube)
\item{} wave energy advection against the fetch direction (green part of the tube)
\item{} nonlinear interaction of the waves between the read and green part of the tube in Fourier space in any given fetch point and time
\end{itemize}
Such splitting by physical processes will further serve as the idea for HE numerical algorithm design.

Let us discuss the boundary conditions of the presented problem. It is assumed that the waves having the wave-vector component $\vec{k}$ in the wind direction $\vec{U}$, and therefore propagating in the upper red half-cylinder, have zero amplitude in the beginning of the fetch $x=0$ (i.e. 'west coast'), or left upper half-circle of the cylinder on Fig.\ref{Fluxes}):

\begin{eqnarray}
\varepsilon(\omega,\theta,x,t)|_{x=0} &=& 0, \,\,\,  -\pi/2 < \theta < \pi/2
\end{eqnarray}

We call these boundary conditions "dissipative".

The waves having the wave-vector component $\vec{k}$ in the direction, opposite to the wind $\vec{U}$, and therefore propagating in the lower green half-cylinder, are supposed to have zero amplitude at the end of the fetch $x=L$ (i.e. 'east coast'), or right lower half-circle of cylinder on Fig.\ref{Fluxes}):

\begin{eqnarray}
\varepsilon(\omega,\theta,x,t)|_{x=L} &=& 0,  \,\,\,  \pi/2 < \theta < 3\pi/2
\end{eqnarray}

There was no need to  define any boundary conditions on other half-circles of the cylinder, since the incoming waves were supposed to be free advecting through the corresponding encountered boundary. 

Such boundary conditions have the physical interpretation of perfect incoming waves absorption at the shorelines without any reflection from them, i.e. perfect wave energy absorbing beaches without any wave generation, or reflection on them. Such "dissipative "type of boundary conditions can be observed in nature as well as in experimental laboratory wave tanks in the form of gently sloping pebble beaches.

As far as concerns the initial conditions,  the low-level white noise wave energy distribution along the fetch for the waves running in the wind direction was assumed, making the nonlinear interaction negligible at the beginning of the simulation for the "red" cylinder

\begin{eqnarray}
\varepsilon(\omega,\theta,x,t)|_{t=0} = 10^{-6} x/L,  \,\,\,\,\,\,  -\pi/2 < \theta < \pi/2
\label{InCond}
\end{eqnarray}

and zero level wave energy for the waves, running against the wind in the "green" cylinder:

\begin{eqnarray}
\varepsilon(\omega,\theta,x,t)|_{t=0} &=& 0,  \,\,\,\,\,\,  \pi/2 < \theta < 3\pi/2
\end{eqnarray}




\section{Numerical simulation} \label{NumSim}

\subsection{Discretization algorithm}

Eq.(\ref{HE2}) has been numerically solved via splitting by physical processes technique \citep{R104}: the advection, $S_{nl}$, $S_{in}$ and  $S_{diss}$ terms have been considered as the different physical processes, constituting the right-hand side of the discretized equation, while temporal derivative was referred to the left-hand side.

The advection part has been approximated by uncoditionally stable second order in space and time "rectangle" numerical scheme \citep{R103}, $S_{nl}$ term has been solved by WRT (Webb-Resio-Tracy) method  \citep{R69,R68}, the wind input term was integrated analytically, and wave-breaking high-frequency energy dissipation has been taken into account in the "implicit" form as Phillips spectral continuation tale $ \sim \omega^{-5}$ to the dynamical part of the wave spectrum, adjusted on every time step.

The time advance has been cycled through via explicit first order numerical integration scheme.

\subsection{Wind wave energy input and wave-breaking dissipation terms}

The wind energy input source function $S_{in}$ was chosen in ZRP form \citep{R8,R7, R88}: 

\begin{eqnarray}
\label{ZRP}
&&S_{in}(\omega,\theta) = \gamma(\omega,\theta)\cdot \varepsilon(\omega,\theta) \\
&&\gamma(\omega,\theta ) = \left\{\begin{array}{l} {0.05 \frac{\rho _{air} }{\rho _{water} } \omega \left(\frac{\omega }{\omega _{0} } \right)^{4/3} q(\theta )  {\rm \; \; for\; } f_{min} \leq f  \leq f_{d}, \; \; \omega =2 \pi f} \\ {0{\rm \; \; otherwise}} \end{array}\right. \label{ZRP1} \\
&&q(\theta ) = \left\{\begin{array}{l} {\cos^2 \theta {\rm \; \; for\; -}\pi {\rm /2}\le \theta \le \pi {\rm /2}} \\ {0{\rm \; \; otherwise}} \end{array}\right. \label{ZRP2} \\
&&\omega _{0} = \frac{g}{U }, \,\,\,  \frac{\rho _{air} }{\rho _{water} } =1.3\cdot 10^{-3} \label{ZRP3}
\end{eqnarray} 

\noindent
where $U$ is the wind speed at $10$ meters reference level, $\rho_{air}$ and $\rho_{water}$ are the air and water density correspondingly. The frequency $f_{d} = 1.1 \;\; \rm Hz$ was found empirically from \cite{R20}, $f_{min}=0.1 \;\; \rm Hz$. 

All simulations have been performed in the frequency domain $0.1 \,\,\, Hz < f < 2.0 \,\,\, Hz$ on the logarithmic grid of 71 points in frequency and 36 equidistant points in the angle directions. The exact expression for $S_{nl}$ term has been used; the initial conditions were chosen in the form of seeding waves with initially negligible nonlinearity level, see section \ref{CauchyProblem}.

To account for wave-breaking energy absorption, the "implicit" wave energy dissipation function $S_{diss}$ \citep{R8,R7,R88} in the form of Phillips spectral tale continuation $ \sim \omega^{-5}$, starting from the frequency $f=1.1$ Hz has been introduced.

\subsection{Numerical tests}

As it was mentioned above, the presence of the advection term in Eq.(\ref{HE2}) can bring new effects due to nonlinear interaction between wave fields, contained inside upper red and lower green pipelines. It is important in this relation to test the correctness of numerical implementation of different terms in Eq.(\ref{HE2}).

As far as concerns the nonlinear part of the interaction, it was calculated by WRT approach \citep{R69,R68}, which has been proved to reproduce theoretical analysis and field experimental observation in series of works \citep{R47,R46,R86,R48,R50,R72,R63,R7,R88,R101}. 

To check the correctness of the advection term discretization, the numerical tests of the Eq.(\ref{HE2}), "stripped out" of nonlinear interaction term

\begin{eqnarray}
\frac{\partial \varepsilon}{\partial t} + \frac{g}{2\omega} \cos{\theta} \frac{\partial \varepsilon}{\partial x} = \gamma \varepsilon
\label{Adv}
\end{eqnarray}

\noindent
have been performed. 

Eq.(\ref{Adv}) yields the analytical solution

\begin{equation}
\varepsilon(x,t) = \left\{\begin{array}{l} {e^{\gamma t}, \;\;\;\;\;\;\;\;\; t< \beta x } \\
                                           {e^{\gamma \beta x}, \;\;\;\;\;\;\;\;\; t  \ge \beta x} \end{array} \right. \\
\label{AdvectionTest}
\end{equation} 
\noindent
where $\beta = \frac{2\omega}{g \cos{\theta}} $, which can be tested numerically.



\begin{figure}[!]
\noindent\center\includegraphics[scale=0.7]{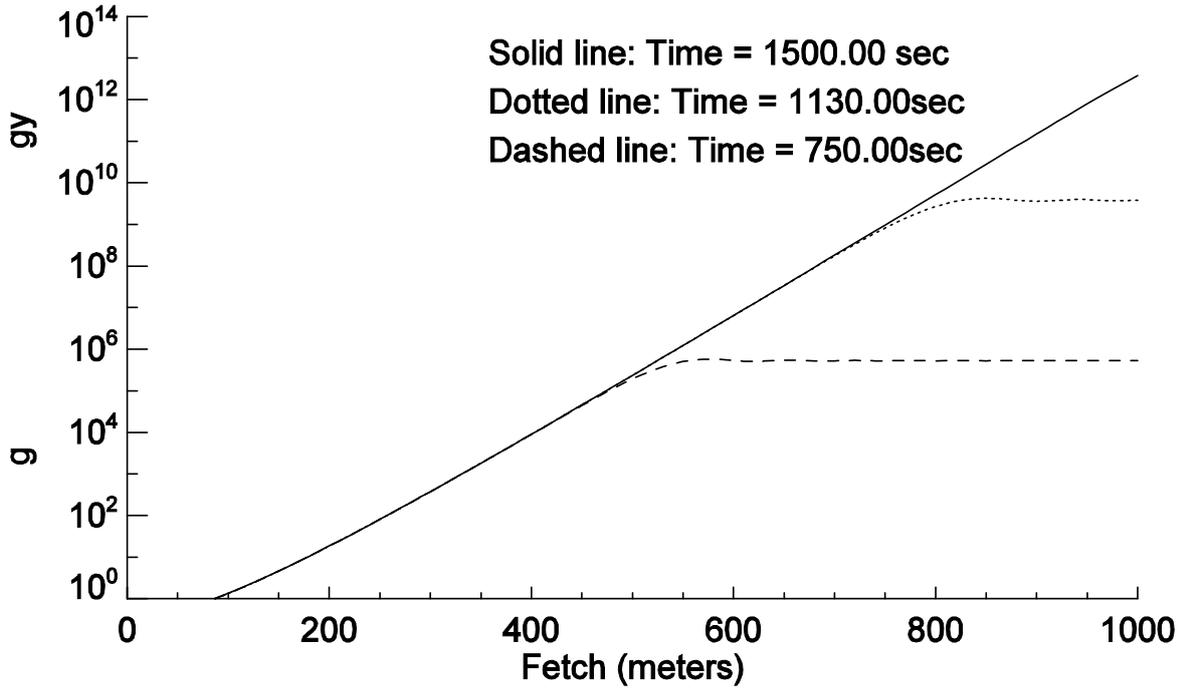} 
\caption{ Decimal logarithm of the total wave energy as the function of the fetch for three distinct  moments of time.} \label{FetchTest2}
\end{figure}

Fig.\ref{FetchTest2} presents decimal logarithm of the total wave energy as the function of the fetch for three distinct moments of time. In accordance with the analytical solution Eqs.(\ref{AdvectionTest}), it demonstrates exponential growth of the total energy along the fetch until it reaches the corresponding stable state.

\subsection{Numerical results} \label{NumRes}

The numerical simulation of Eq.(\ref{HE2}) was performed for the following set of parameters:

\begin{itemize}

\item{} Fetch size 40 km, similar to English Straight
\item{} 40 nodes in real space
\item{} 72 logithmically spaced frequencies in Fourier space
\item{} 36 equidistant points covering $2 \pi$ angle domain
\item{} wind speed at 10 m height $U=5$ m/sec

\end{itemize}

As it was mentioned in the section \ref{CauchyProblem}, the initial conditions were chosen in the form of Eq.(\ref{InCond}) distributed over frequencies and angles as the seeding waves with negligible nonlinearity, running with the wind in the red part of the cylinder on Fig.\ref{Fluxes} and zeros for the waves running agains the wind in the green part of the cylinder. 

\begin{figure}[ht]
\noindent\center\includegraphics[scale=0.8, angle=90]{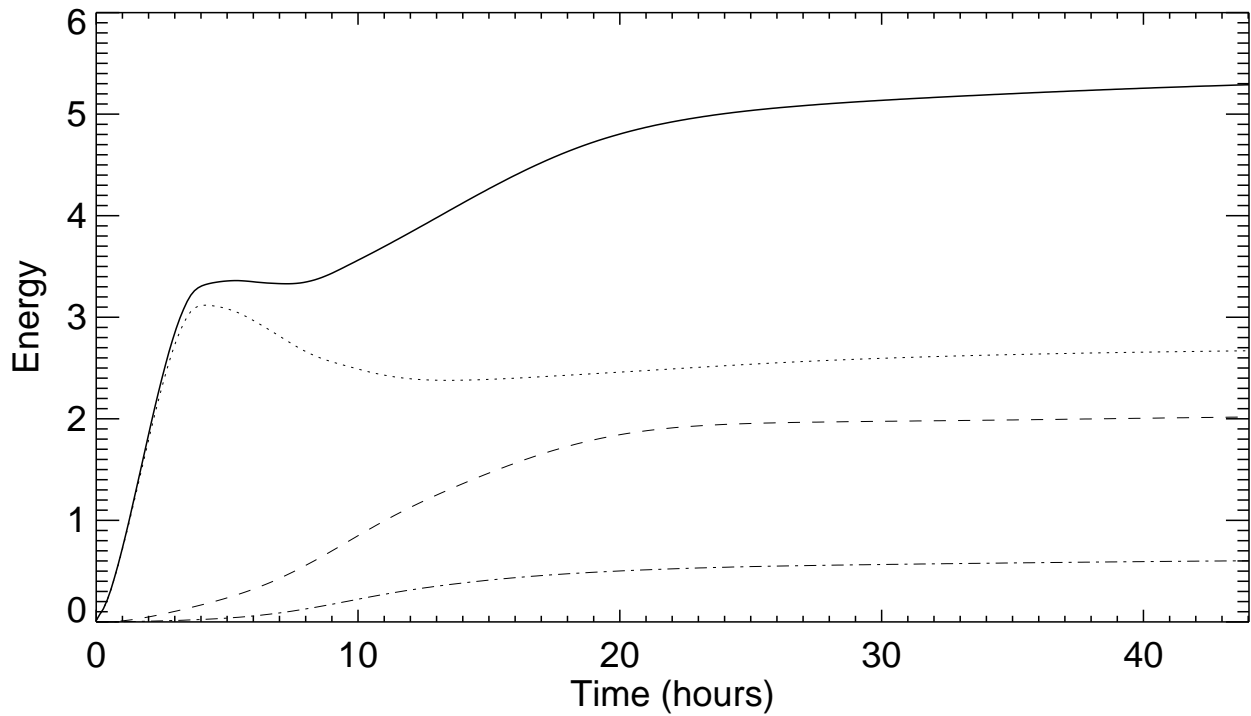} 
\caption{Total wave energy of the fetch as the function of time: solid line. Wave energy against the wind direction (green pipe) - dashed line; in the wind direction (red pipe) - dotted line; orthogonal to the wind direction: dash-dotted line.}
\label{TotalEnergyOnTime}
\end{figure}

Fig.\ref{TotalEnergyOnTime} shows the total wave energy of the fetch as the function of time. It can be split into two parts: relatively fast linear growth for time less then 5 hr and subsequent relatively  slow relaxation to the constant asymptotic value. It is quite surprizing that the system comes into the equilibrium state, which means that the wave energy input channel from the wind is balanced by two wave energy absorption channels: one due to wave-breaking and another due to incoming waves dissipation at the shore lines.

According to Fig.\ref{TotalEnergyOnTime}, the total wave energy is asymptotically directionally distributed in the following proportions:

\begin{itemize}

\item{} 54 $\%$ - in the wind direction
\item{} 35 $\%$ - against the wind direction
\item{} 11 $\%$ - orthogonal to the wind

\end{itemize}

which means that almost half of the energy in the asymptotic equlibrium state propagates against or orthogonal to the wind. Since the wind input is localized in the angle range $-\pi/2 < \theta < \pi/2$ in accordance with Eq.(\ref{ZRP2}), the wave energy appearance outside of this spread can be explained only by nonlinear wave interaction. It must be stressed out that this effect appears only if the "exact" $S_{nl}$ is chosen.

\subsection{The dynamics of the spectrum in time}

For earlier time $t=2$ hr, see Fig.\ref{Spectrum3D2h}, \ref{Polar2h}, the shape of the spectrum has the form of the single hump, growing with the distance from the west shore, with the spectral maximum having threshold shape as the function of the fetch. Similar threshold propagation effect is seen on Fig.\ref{FetchTest2}. While this similarity is qualitative, it evidences the fact that the wave system  evolves in the framework of classic limited fetch formulation, being 'unaware' yet of the east cost bound.

\begin{table}[htbp]
\centering
	\begin{center} 
		\begin{tabular}{c c}
			\includegraphics[width=0.4\linewidth]{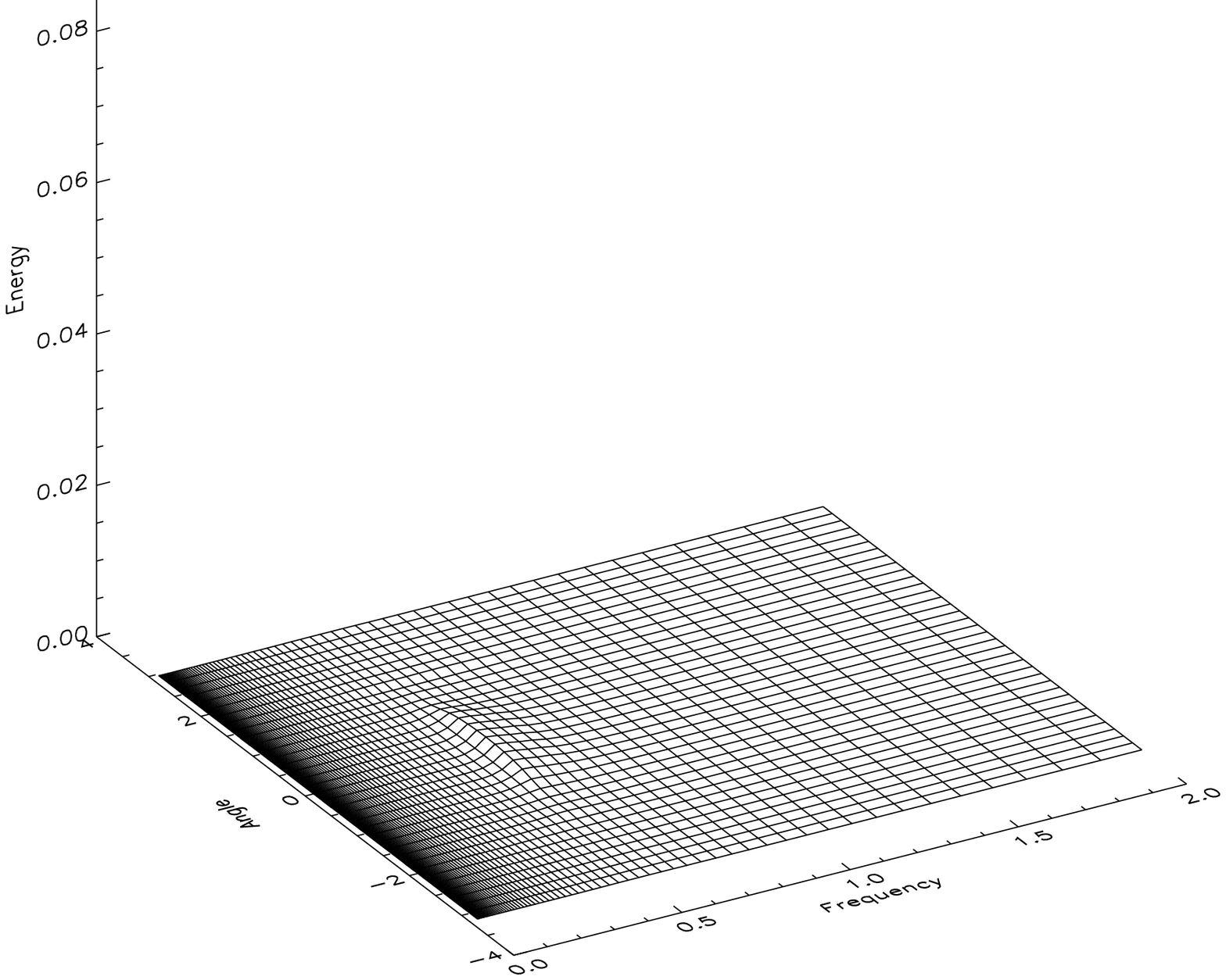} & \includegraphics[width=0.4\linewidth]{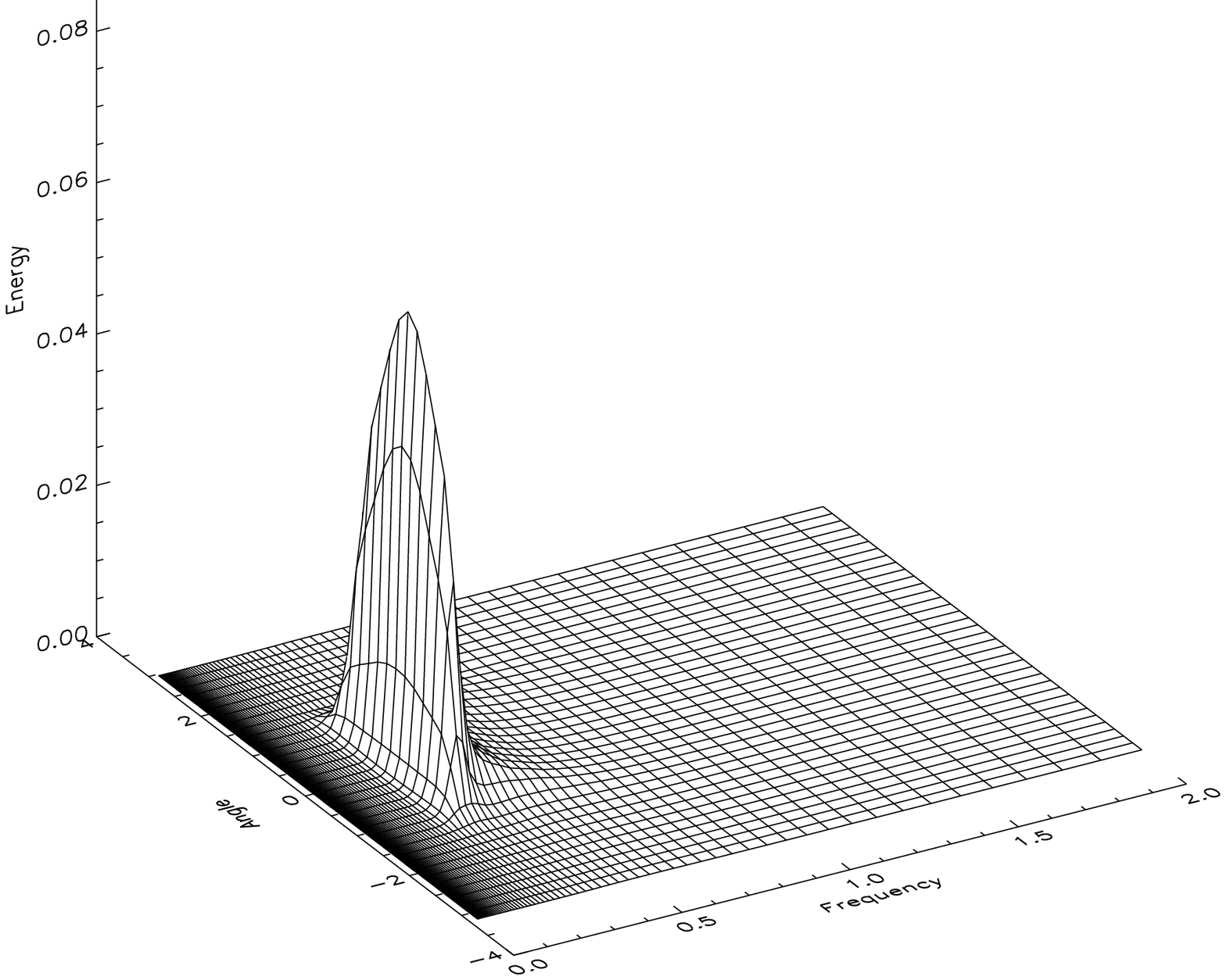}   \\ [1.8 cm]
			\includegraphics[width=0.4\linewidth]{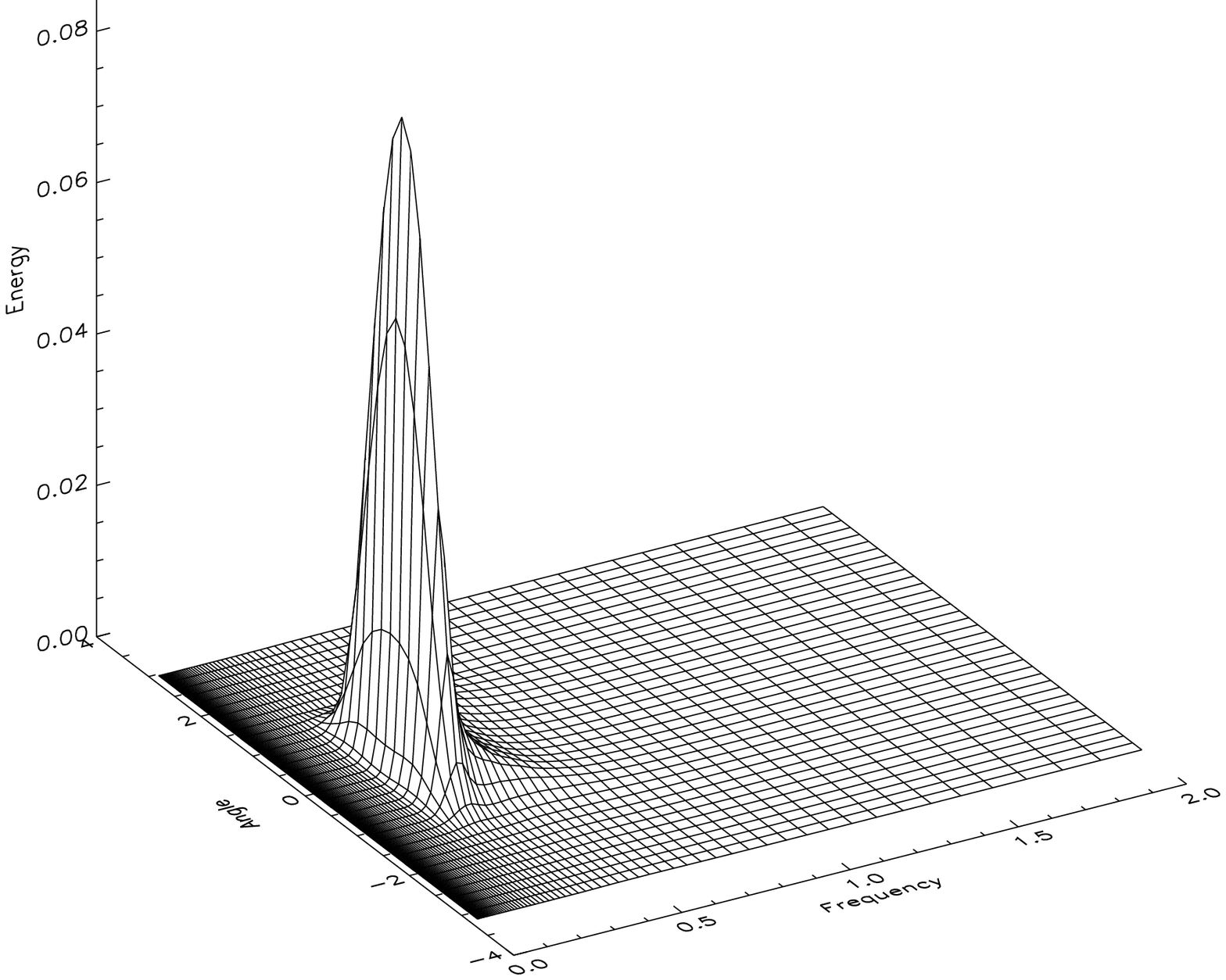} & \includegraphics[width=0.4\linewidth]{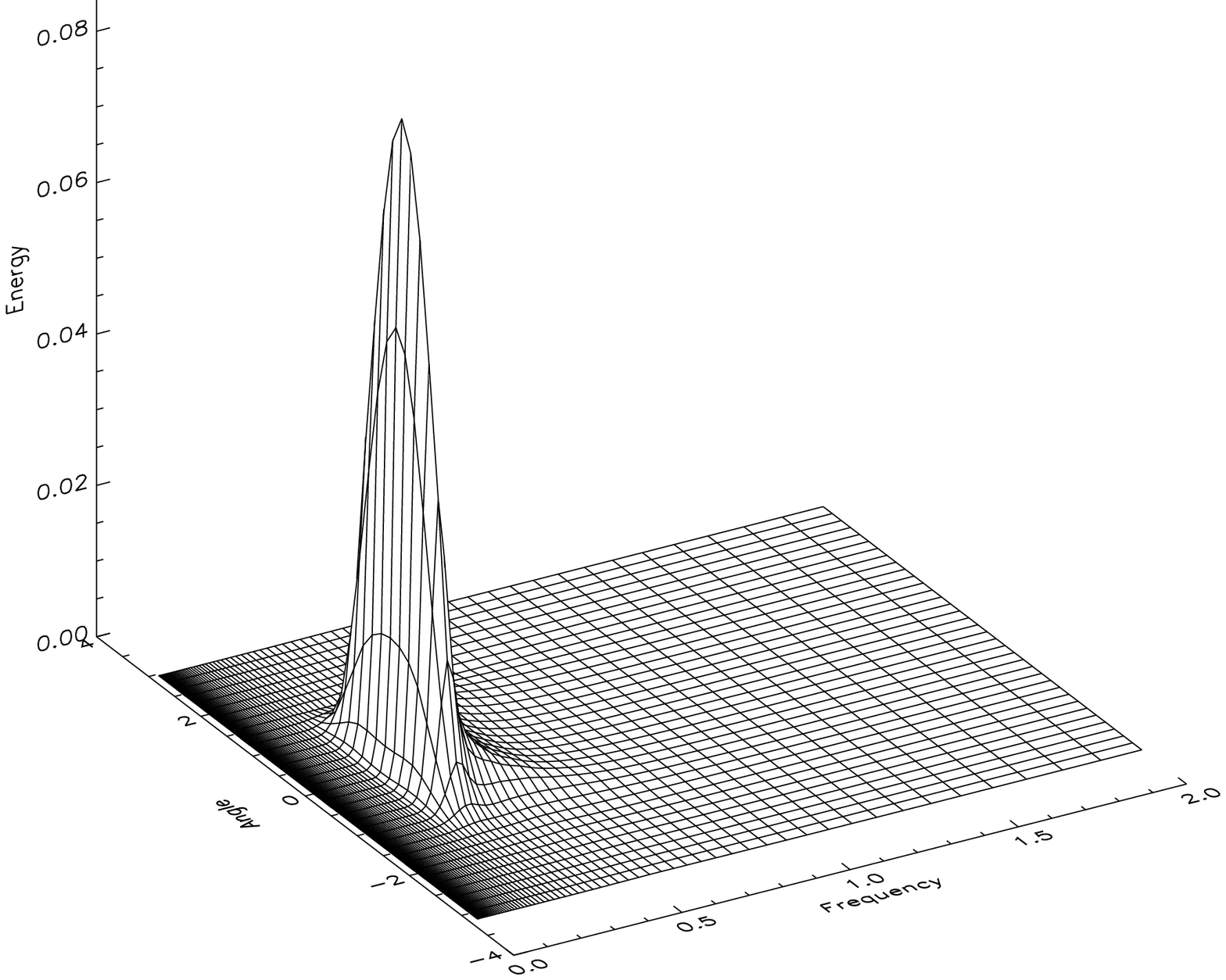} \\
		\end{tabular}
	\captionof{figure}{Spectral energy distribution as the function of the frequency $f$ and angle at the fetch distances 2, 14, 26 and 38 km for time 2 hr} \label{Spectrum3D2h}
	\end{center}
\end{table} 

\begin{table}[htbp]
\centering
	\begin{center} 
		\begin{tabular}{c c}
			\includegraphics[width=0.4\linewidth]{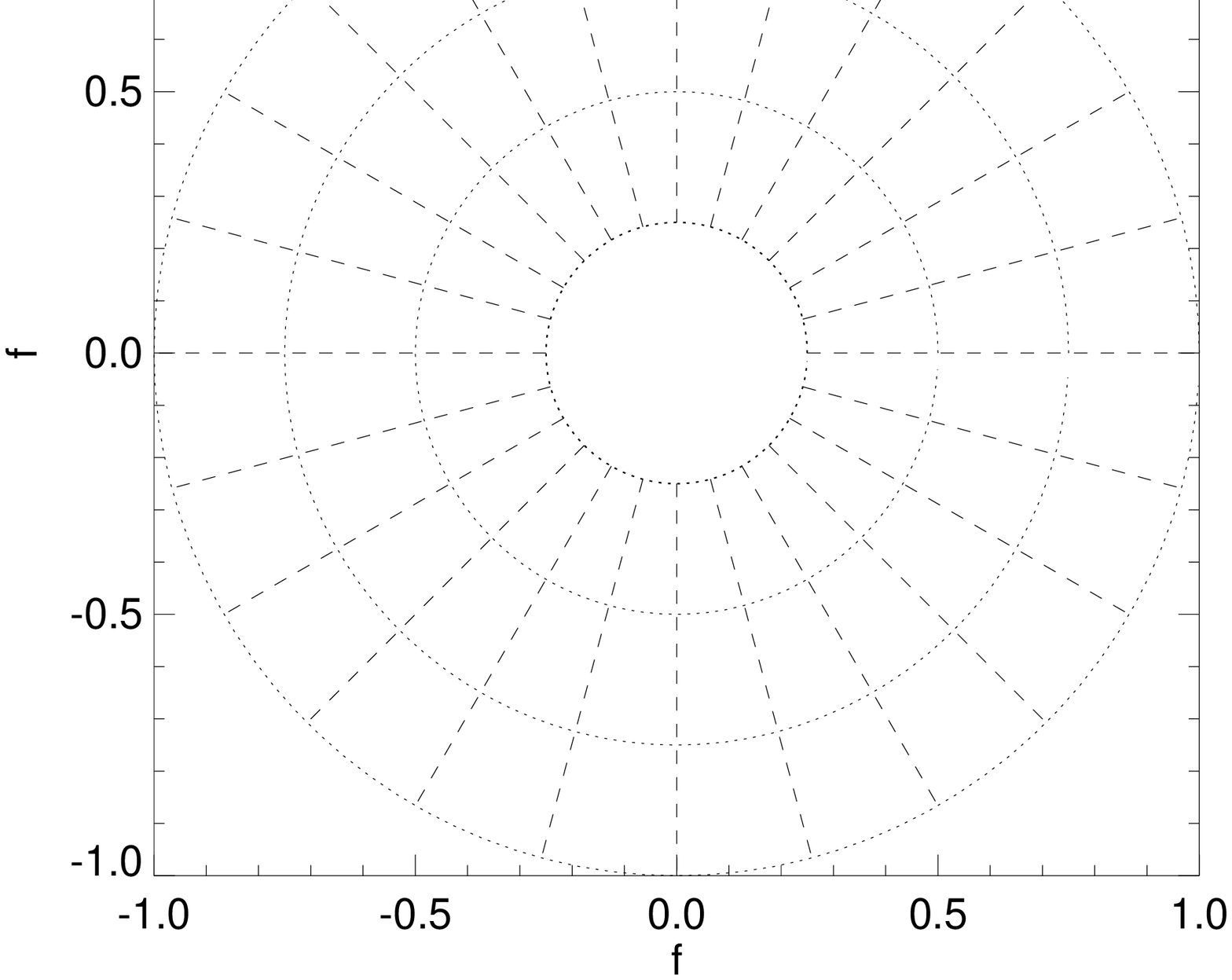} & \includegraphics[width=0.4\linewidth]{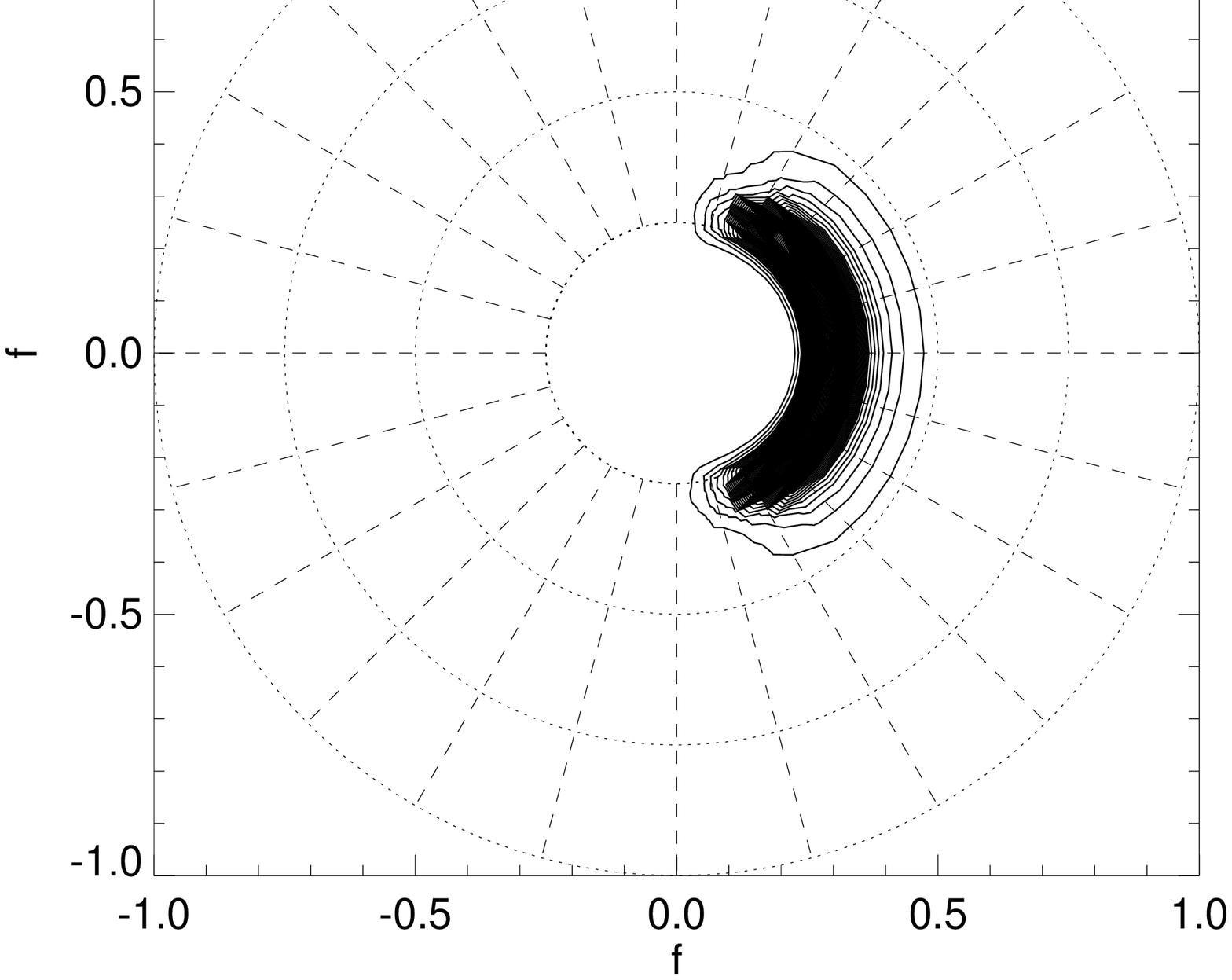}   \\ [1.8 cm]
			\includegraphics[width=0.4\linewidth]{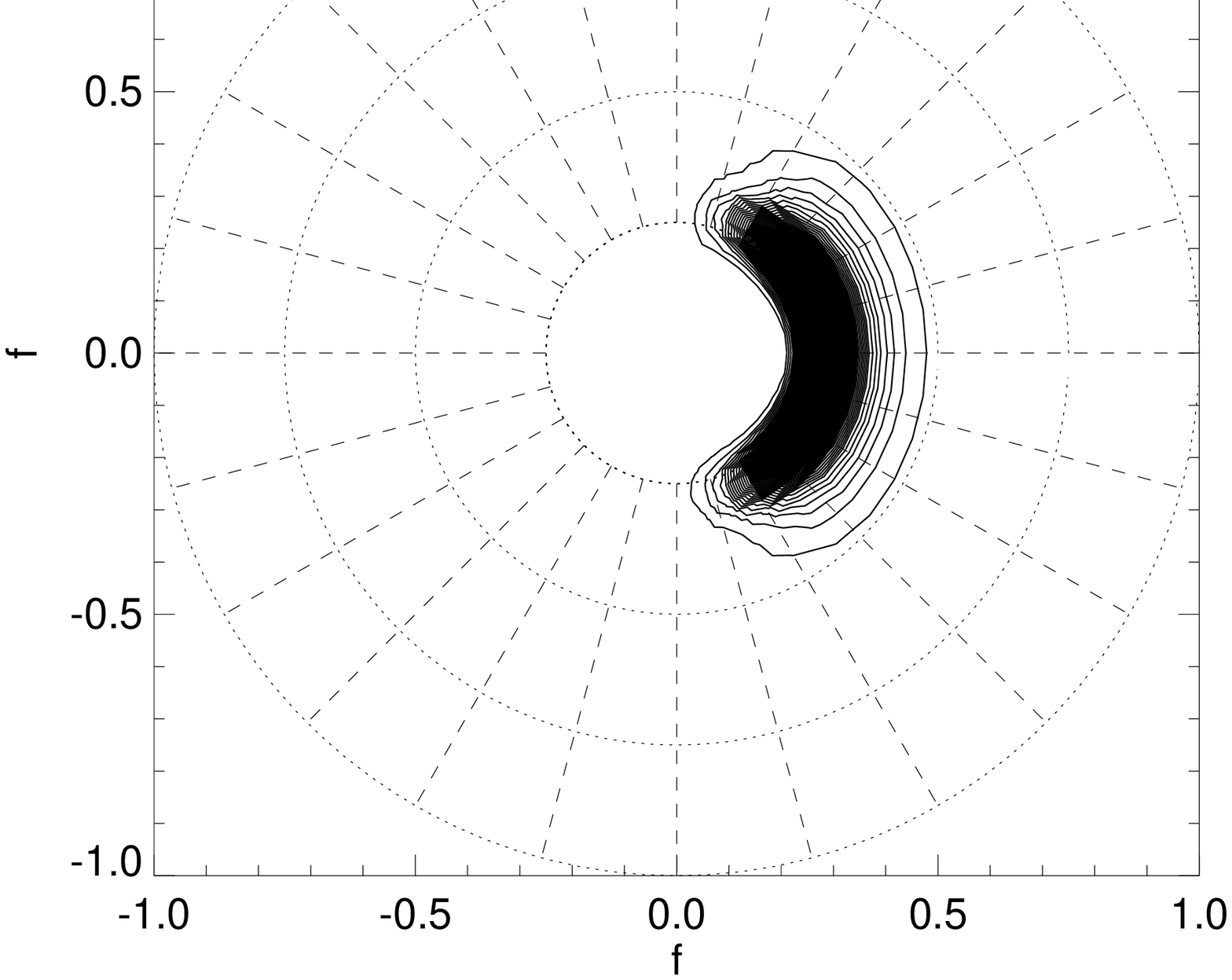} & \includegraphics[width=0.4\linewidth]{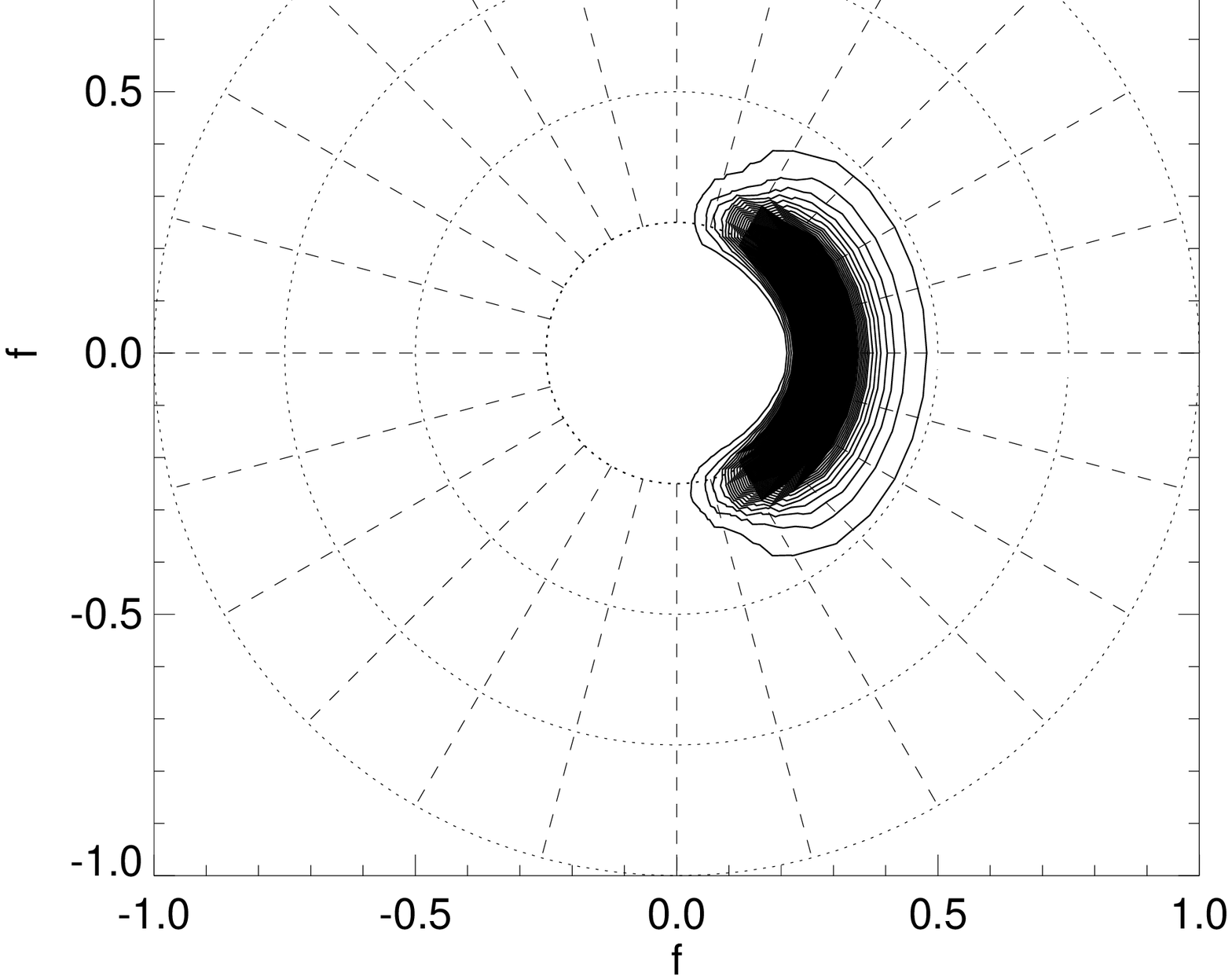} \\
		\end{tabular}
	\captionof{figure}{Spectral energy distribution as the function of the frequency $f$ and angle in polar coordinates at the fetch distances 2, 14, 26 and 38 km for time 2 hr} \label{Polar2h}
	\end{center}
\end{table} 

For time $t=4$ hr, which is close to characteristic time of reaching the east coast due to the advection of the spectral peak, estimated as $L/V_{adv}$, where $L$ is the channel width and $V_{adv}$ is the characteristic advection speed of spectral peak, see Fig.\ref{Spectrum3D4h} and \ref{Polar4h}, the spectral hump starts to develop side lobes. One can interpret this effect as the start of nonlinear interaction development between red and green pipelines at time, corresponding to the maximum of the wave energy concentration in the red pipe, close to the east coast.

\begin{table}[htbp]
\centering
	\begin{center} 
		\begin{tabular}{c c}
			\includegraphics[width=0.4\linewidth]{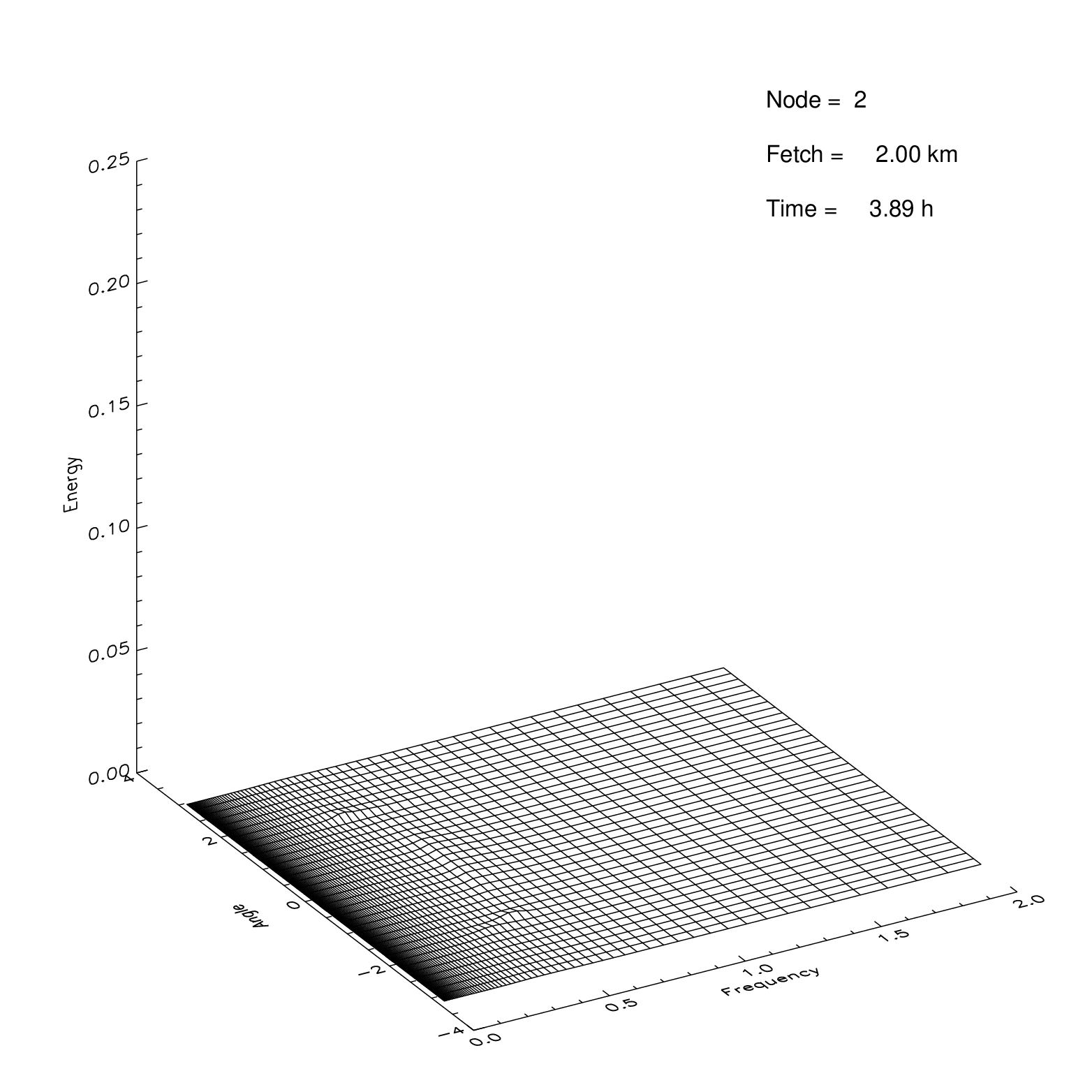} & \includegraphics[width=0.4\linewidth]{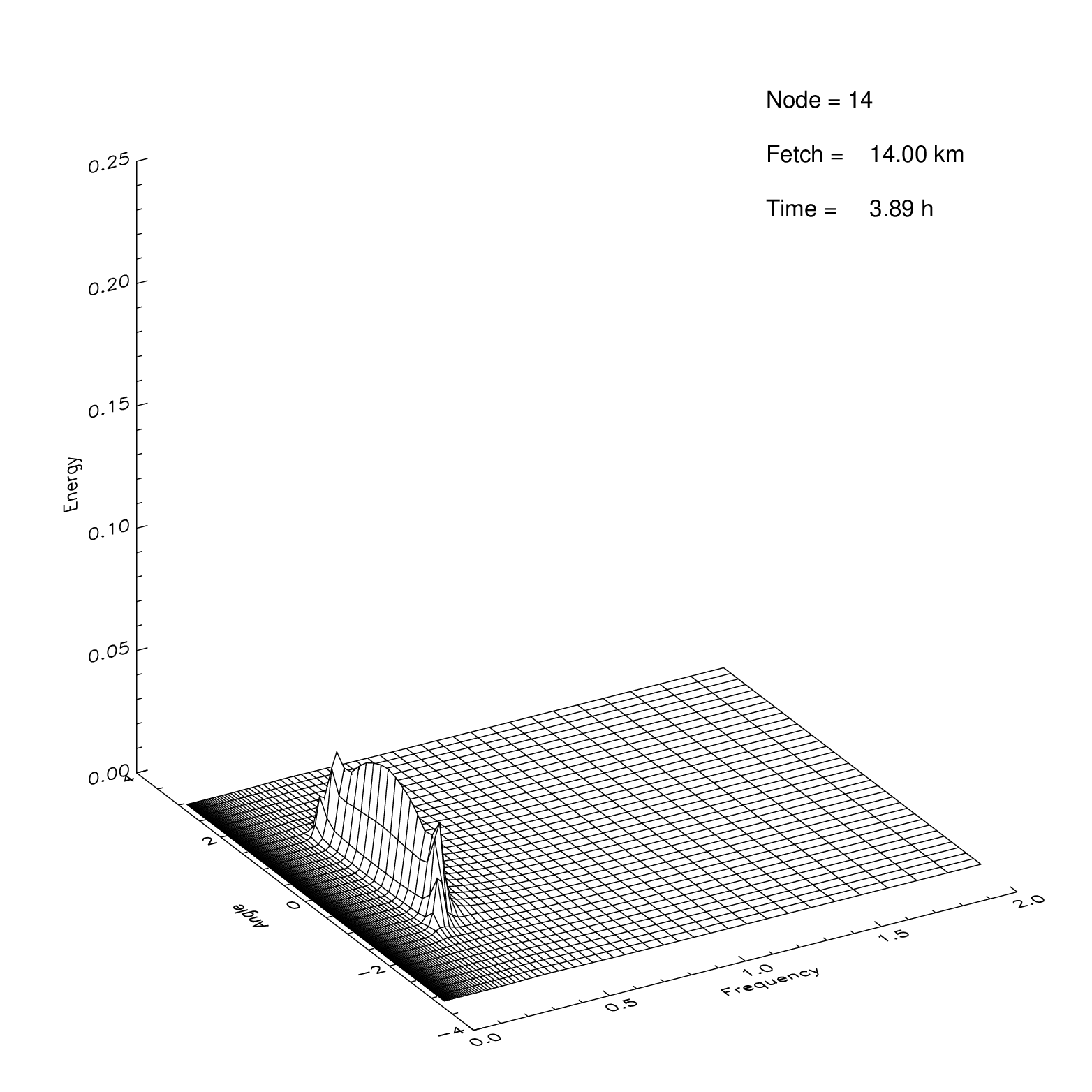}   \\ [1.8 cm]
			\includegraphics[width=0.4\linewidth]{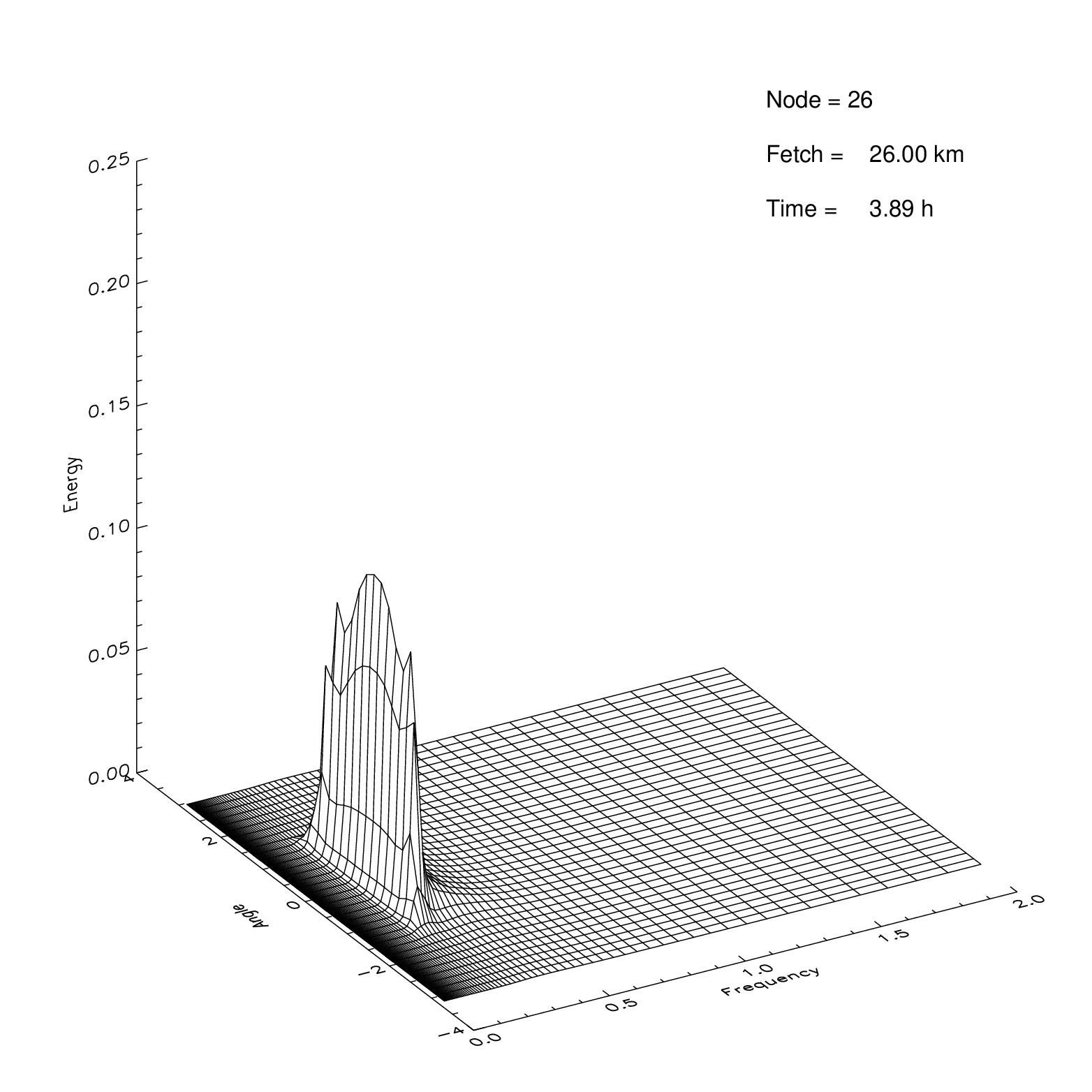} & \includegraphics[width=0.4\linewidth]{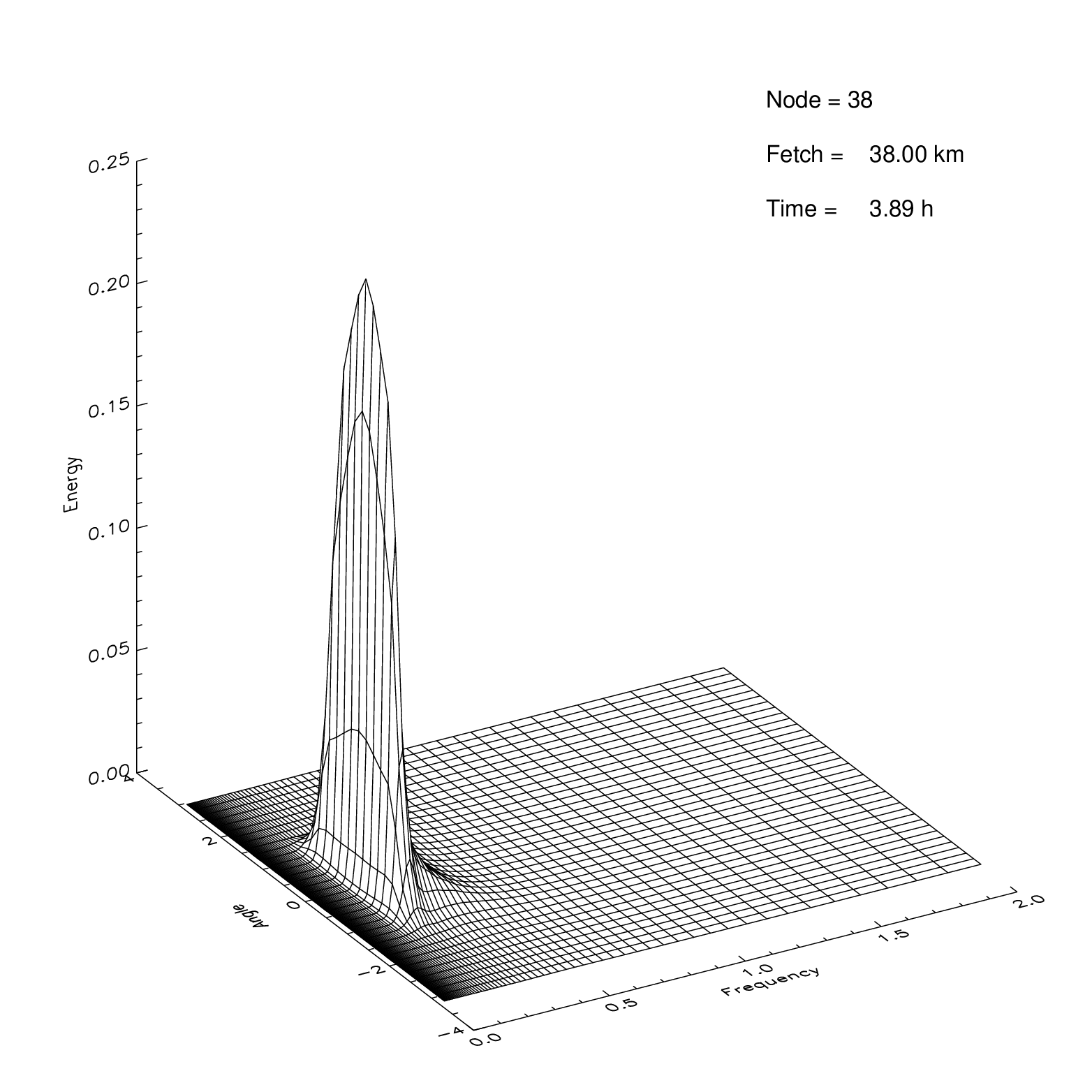} \\
		\end{tabular}
	\captionof{figure}{Spectral energy distribution as the function of the frequency $f$ and angle at the fetch distances 2, 14, 26 and 38 km for time 4 hr} \label{Spectrum3D4h}
	\end{center}
\end{table} 

\begin{table}[htbp]
\centering
	\begin{center} 
		\begin{tabular}{c c}
			\includegraphics[width=0.4\linewidth]{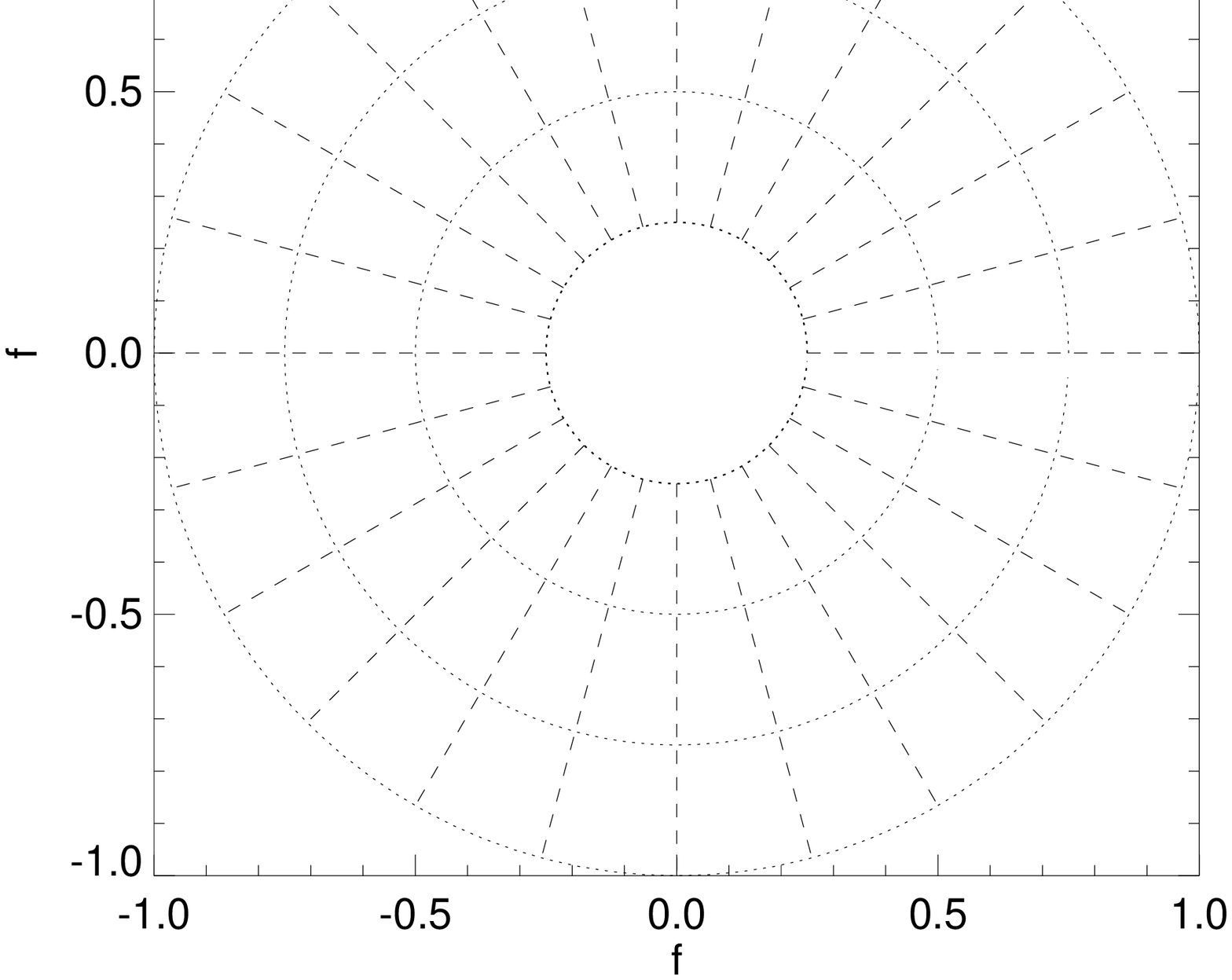} & \includegraphics[width=0.4\linewidth]{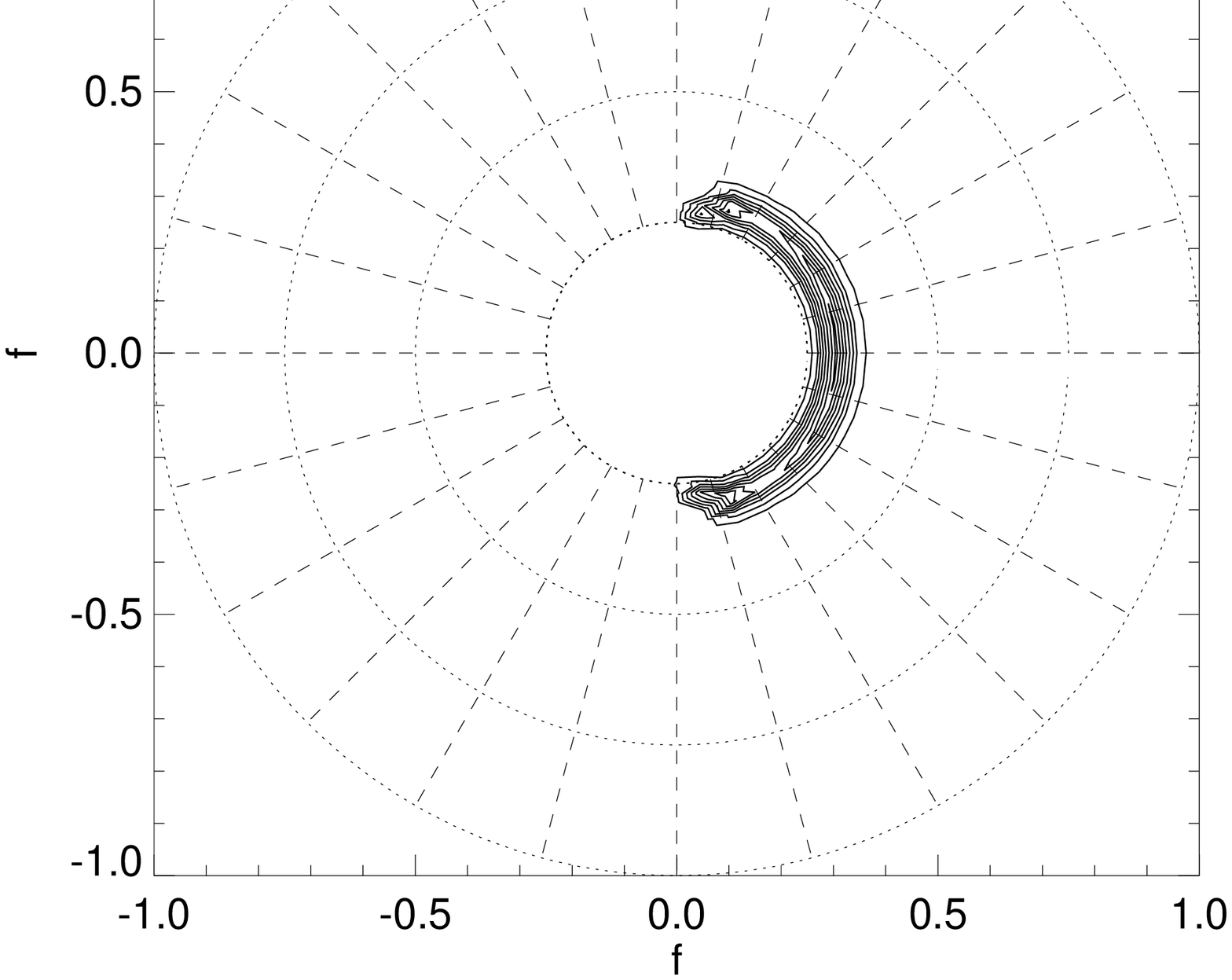}   \\ [1.8 cm]
			\includegraphics[width=0.4\linewidth]{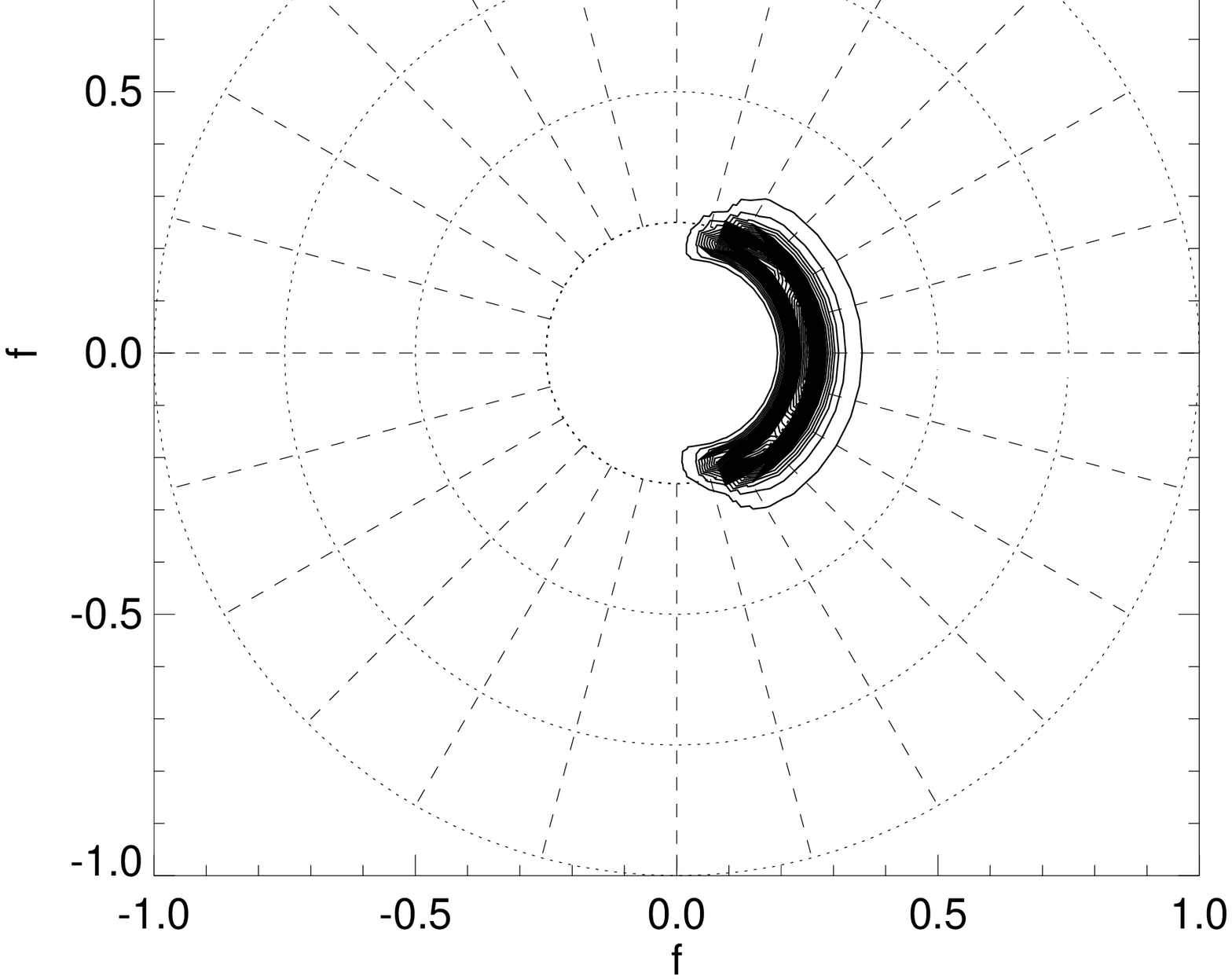} & \includegraphics[width=0.4\linewidth]{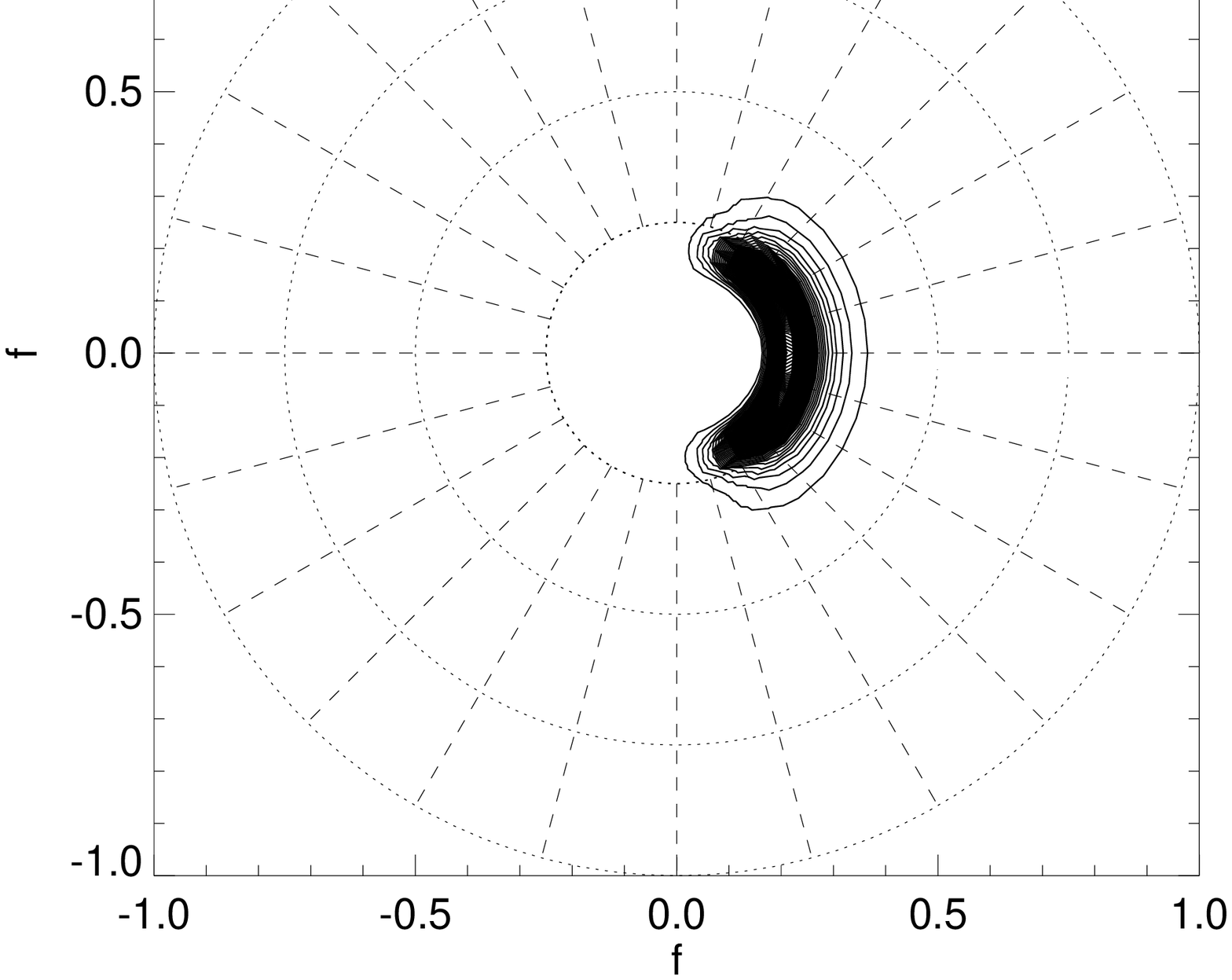} \\
		\end{tabular}
	\captionof{figure}{Spectral energy distribution as the function of the frequency $f$ and angle in polar coordinates at the fetch distances 2, 14, 26 and 38 km for time 4 hr} \label{Polar4h}
	\end{center}
\end{table}

For $t=6$ hr and $t=8$ hr, see Fig.\ref{Spectrum3D6h}, \ref{Polar6h} and Fig.\ref{Spectrum3D8h}, \ref{Polar8h}, the side lobes amplitude exceeds the amplitude of the central hump. The intensity of nonlinear interaction between the fields in the red and green pipelines is apparently relatively high at this stage of the system evolution.

\noindent

\begin{table}[htbp]
\centering
	\begin{center} 
		\begin{tabular}{c c}
			\includegraphics[width=0.4\linewidth]{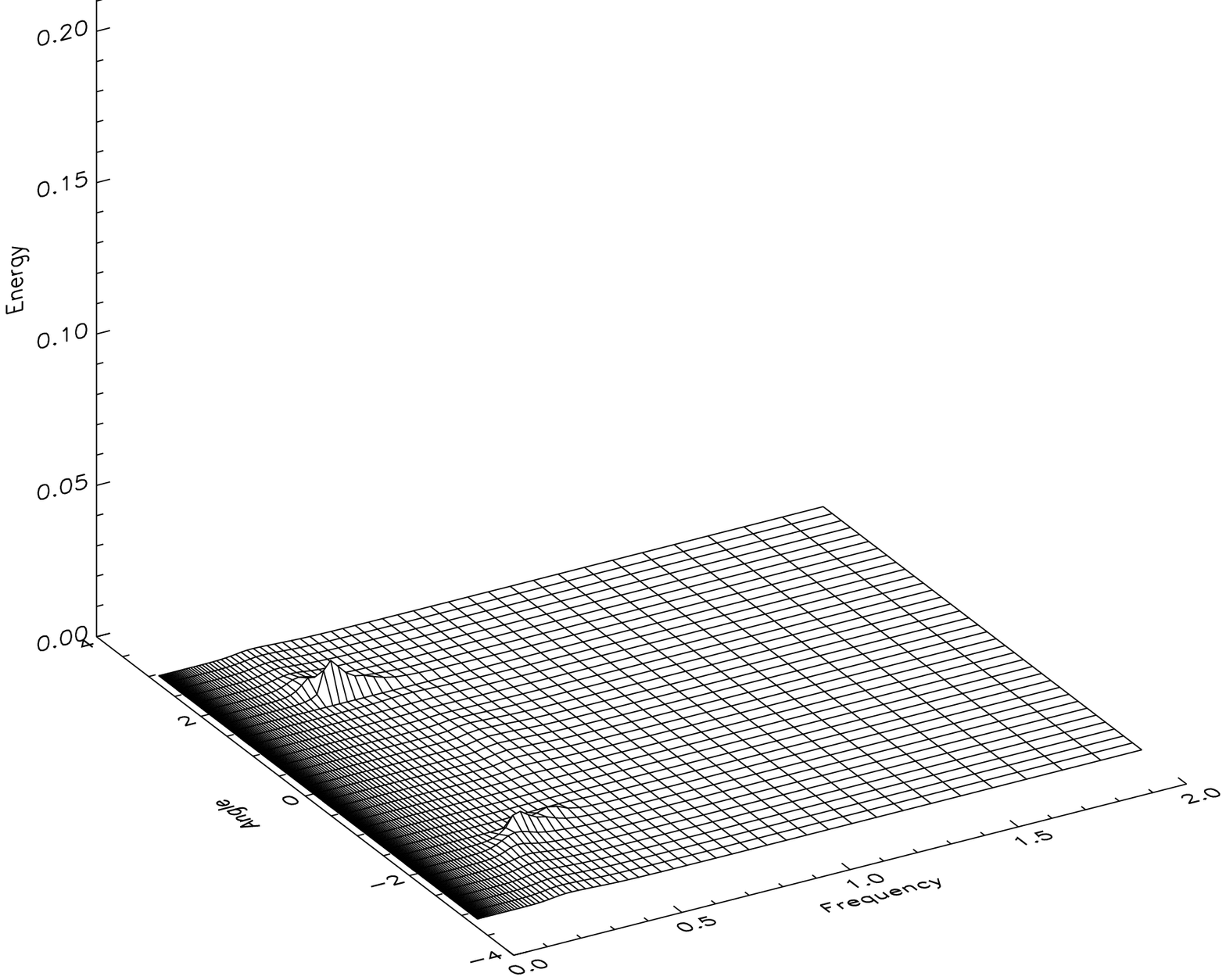} & \includegraphics[width=0.4\linewidth]{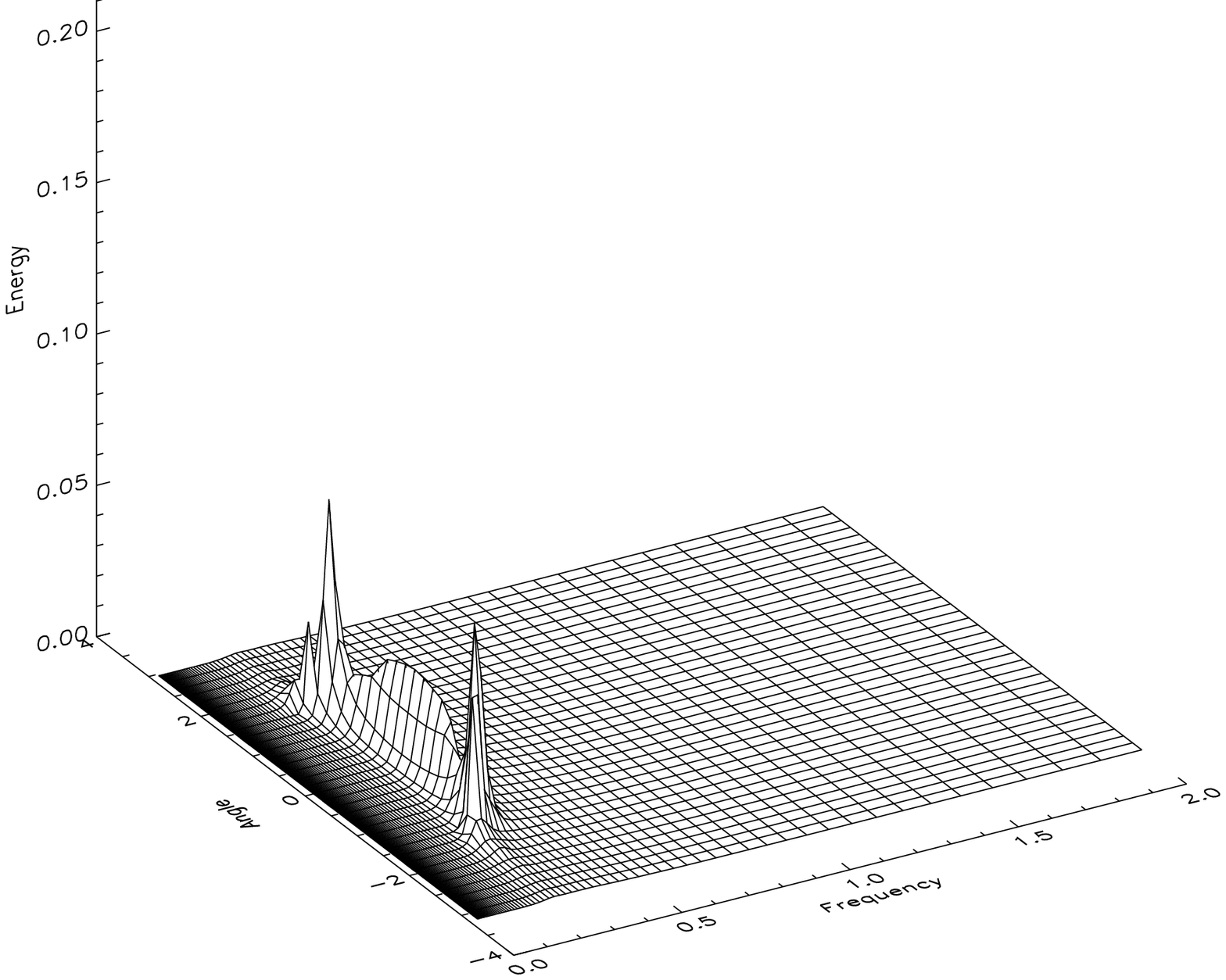}   \\ [1.8 cm]
			\includegraphics[width=0.4\linewidth]{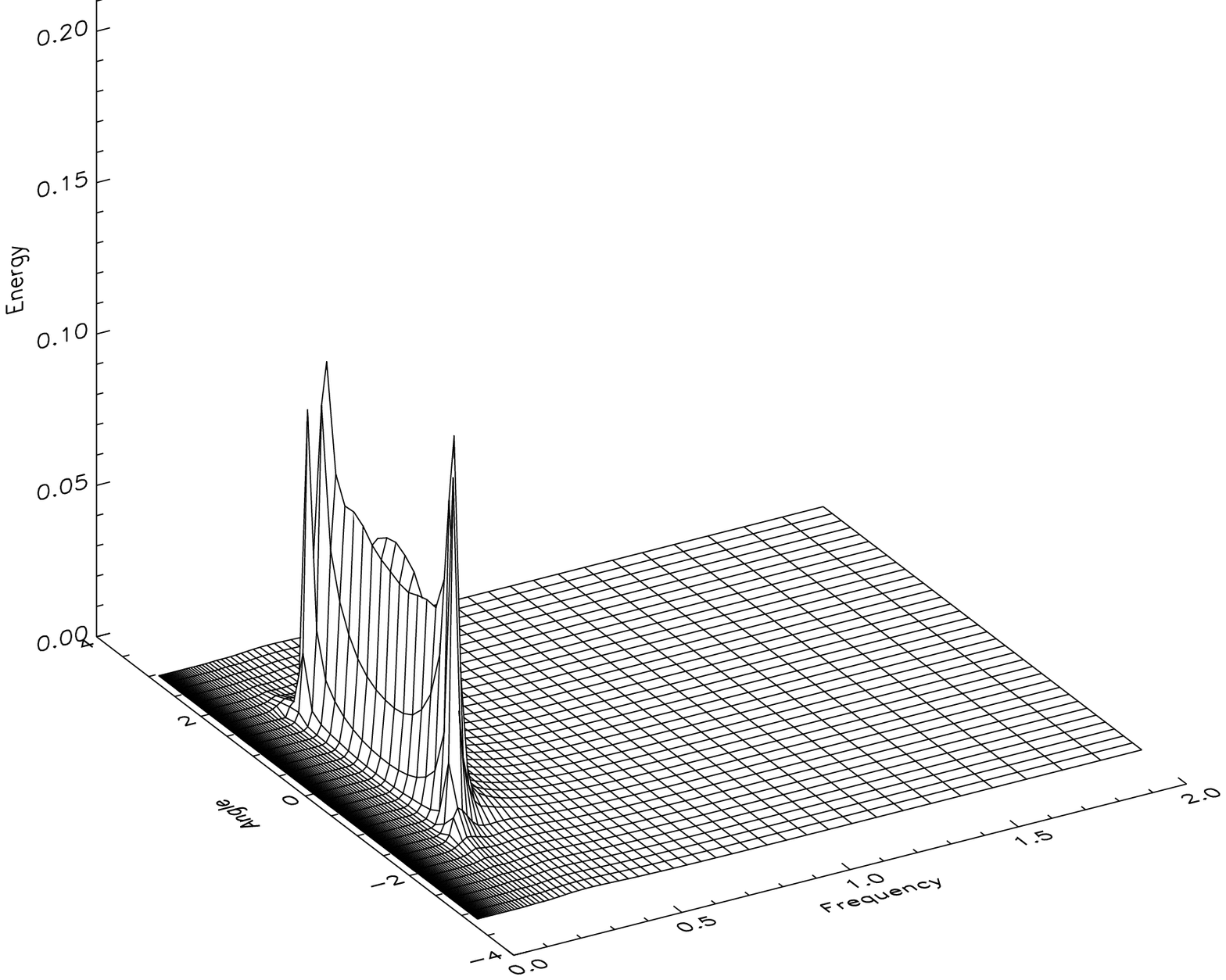} & \includegraphics[width=0.4\linewidth]{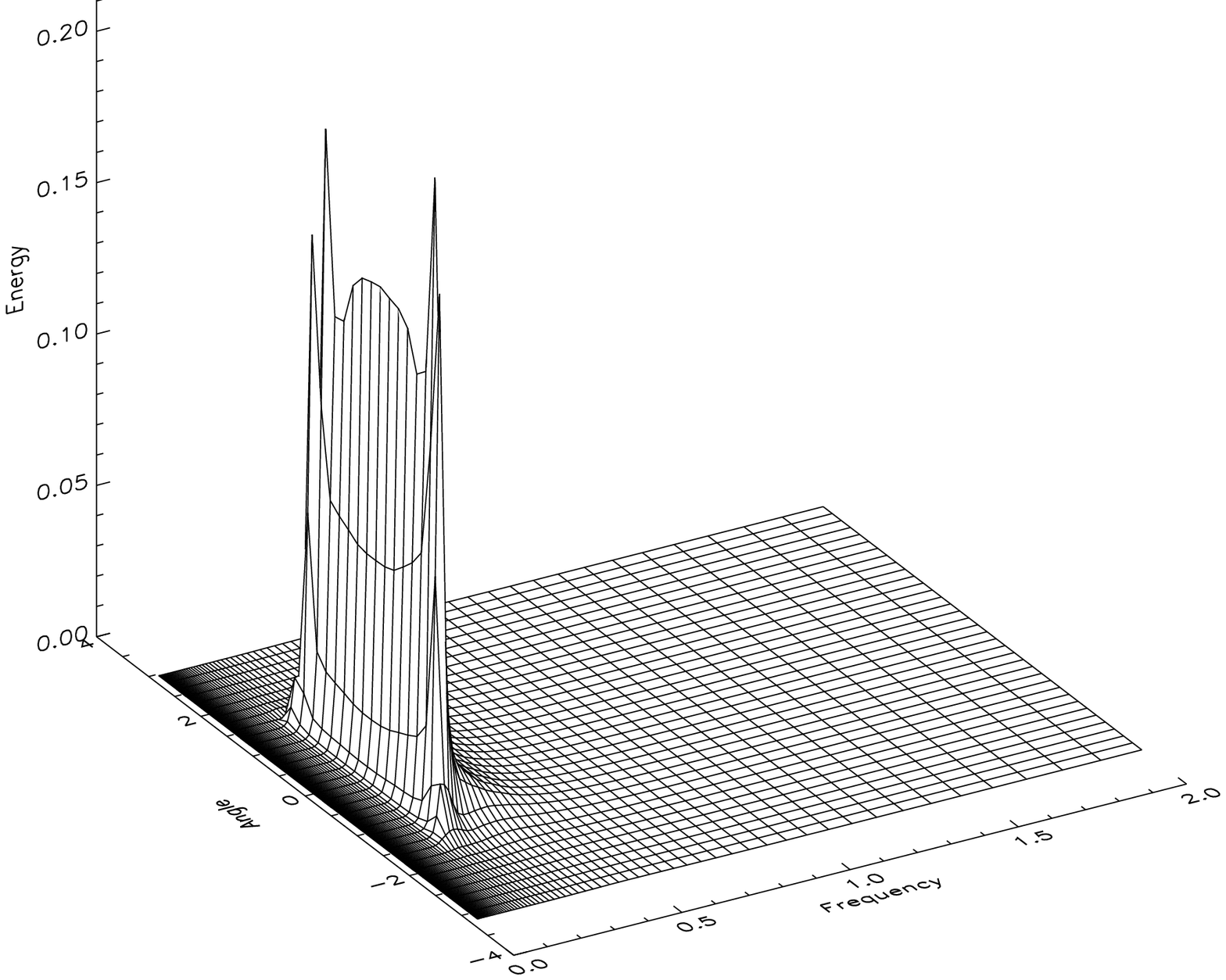} \\
		\end{tabular}
	\captionof{figure}{Spectral energy distribution as the function of the frequency $f$ and angle at the fetch distances 2, 14, 26 and 38 km for time 6 hr} \label{Spectrum3D6h}
	\end{center}
\end{table} 

\begin{table}[htbp]
\centering
	\begin{center} 
		\begin{tabular}{c c}
			\includegraphics[width=0.4\linewidth]{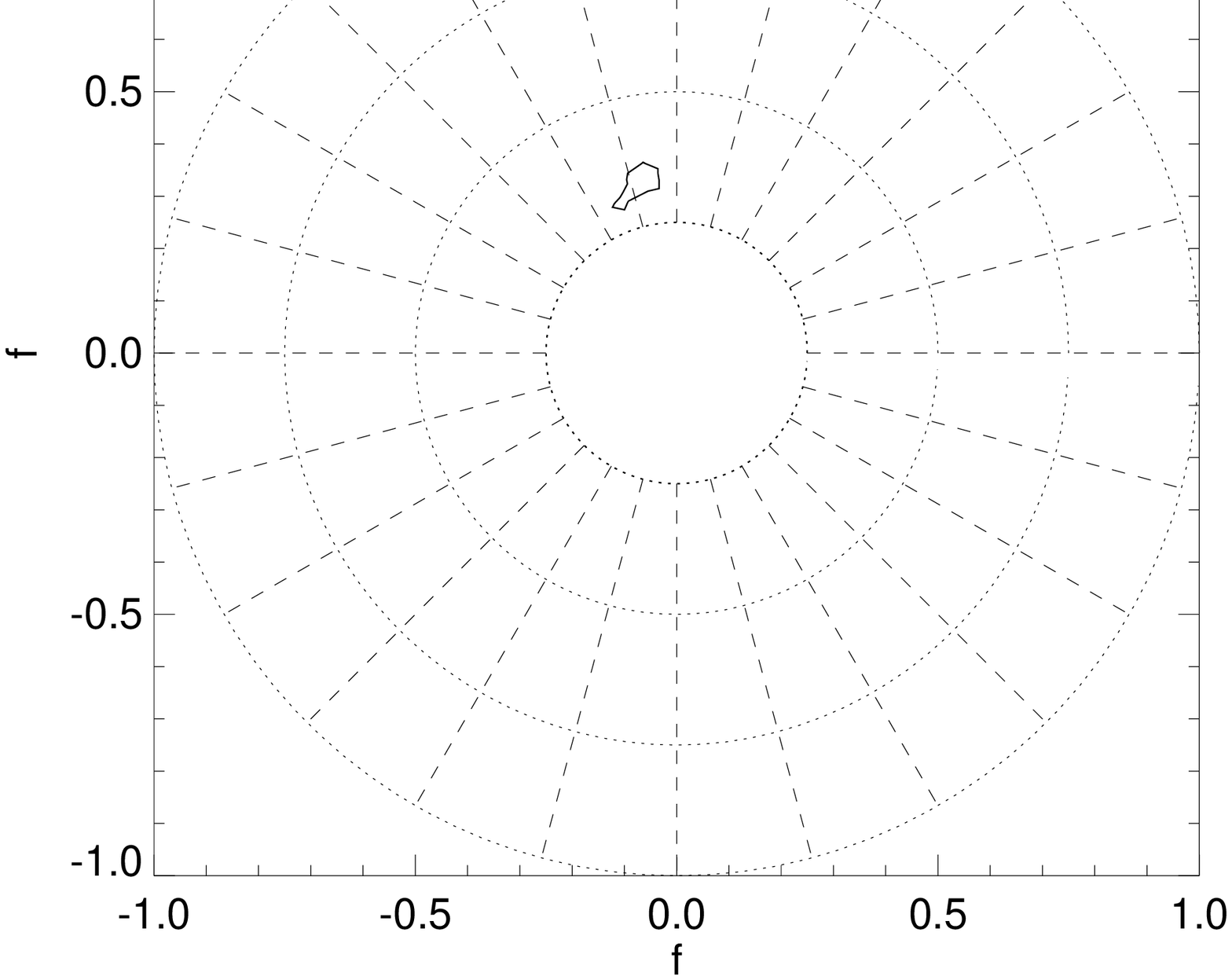} & \includegraphics[width=0.4\linewidth]{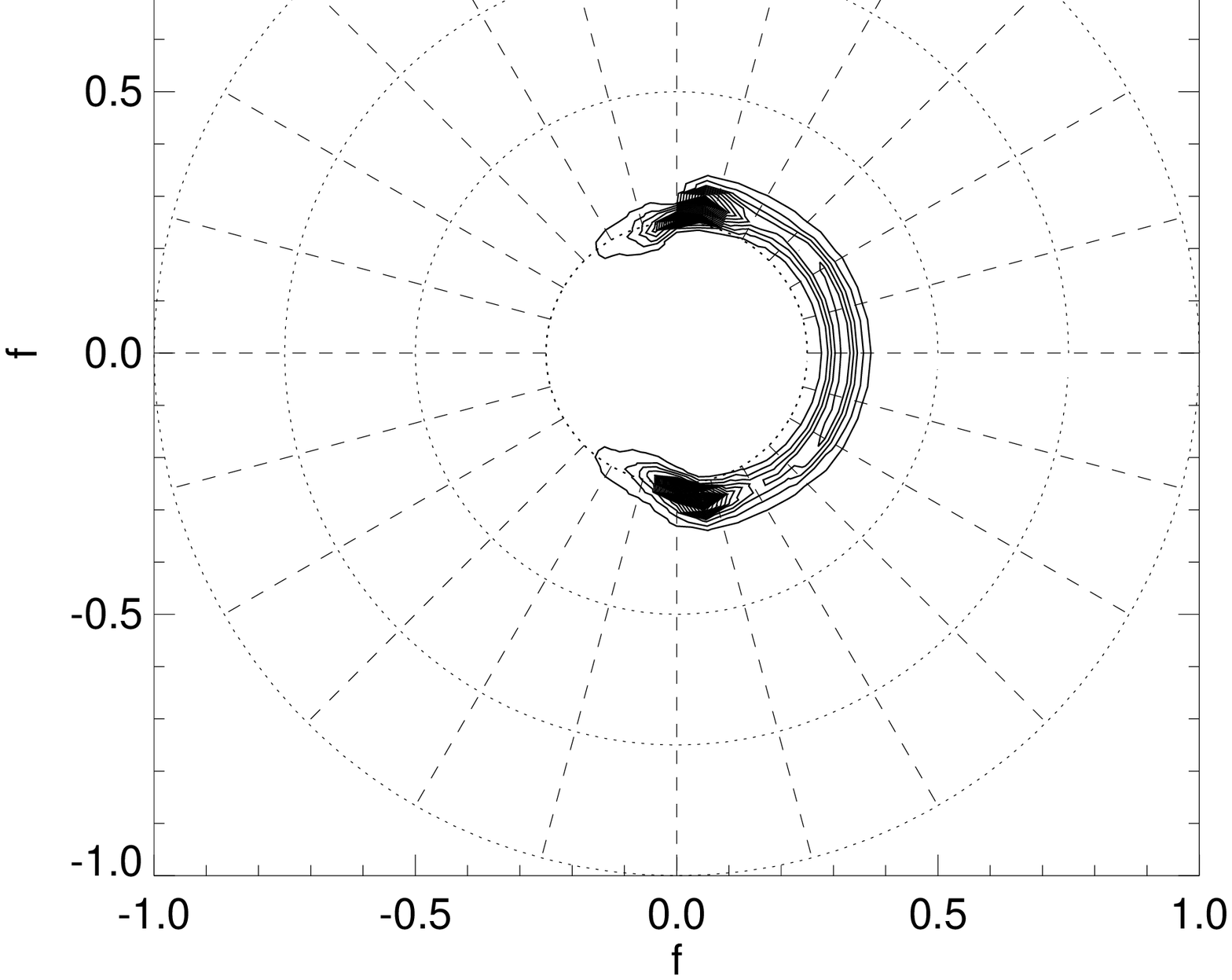}   \\ [1.8 cm]
			\includegraphics[width=0.4\linewidth]{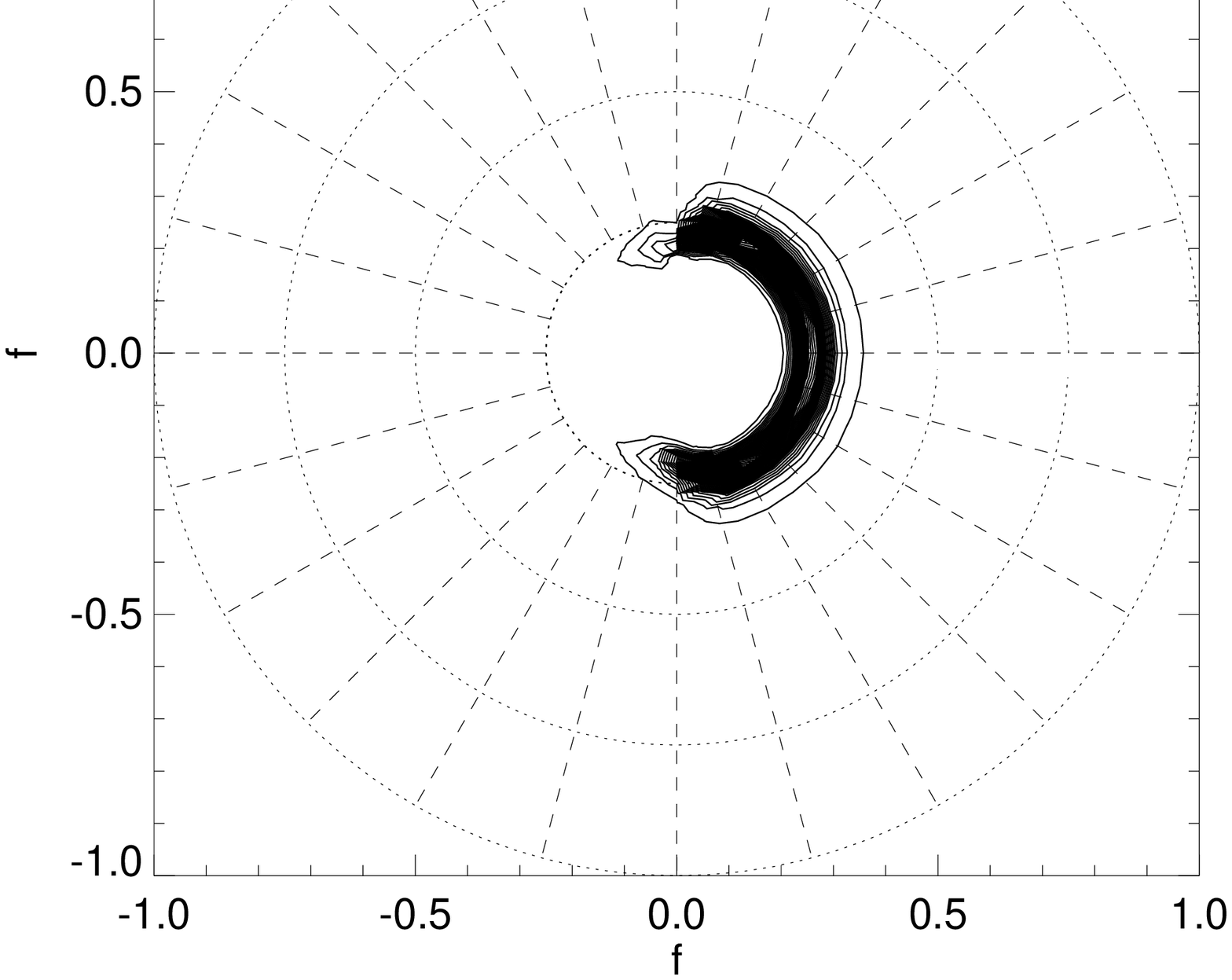} & \includegraphics[width=0.4\linewidth]{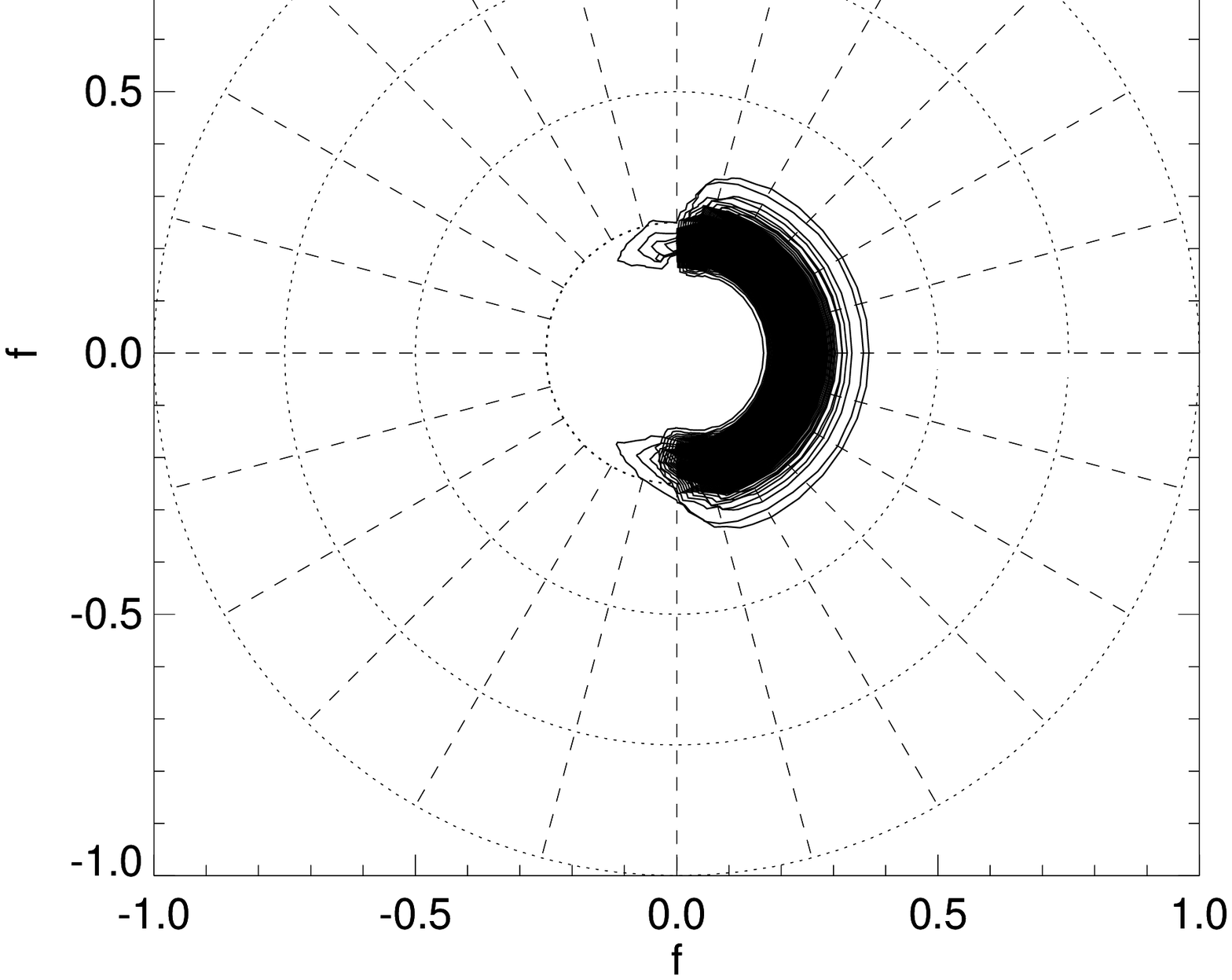} \\
		\end{tabular}
	\captionof{figure}{Spectral energy distribution as the function of the frequency $f$ and angle in polar coordinates at the fetch distances 2, 14, 26 and 38 km for time 6 hr} \label{Polar6h}
	\end{center}
\end{table}

\begin{table}[htbp]
\centering
	\begin{center} 
		\begin{tabular}{c c}
			\includegraphics[width=0.4\linewidth]{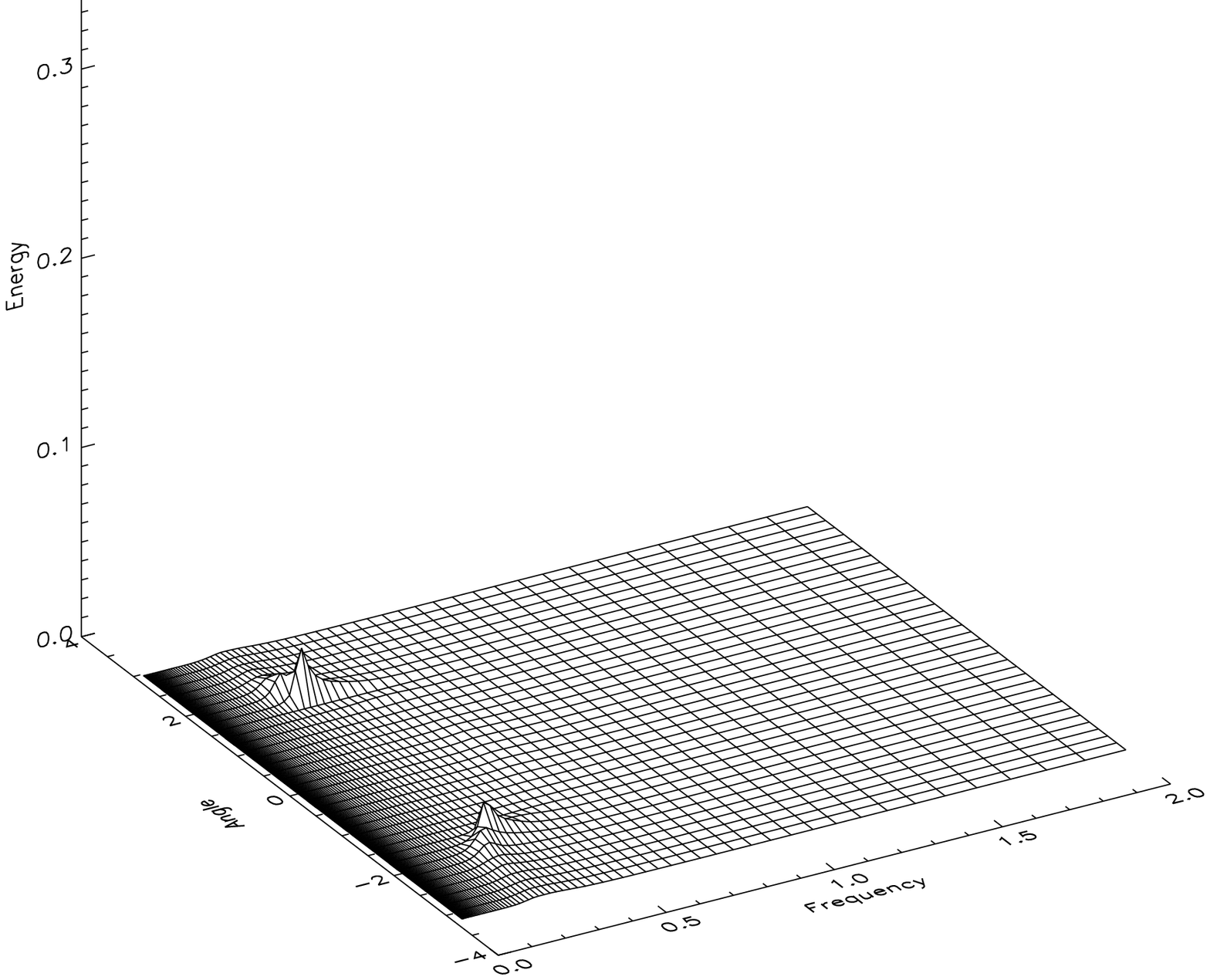} & \includegraphics[width=0.4\linewidth]{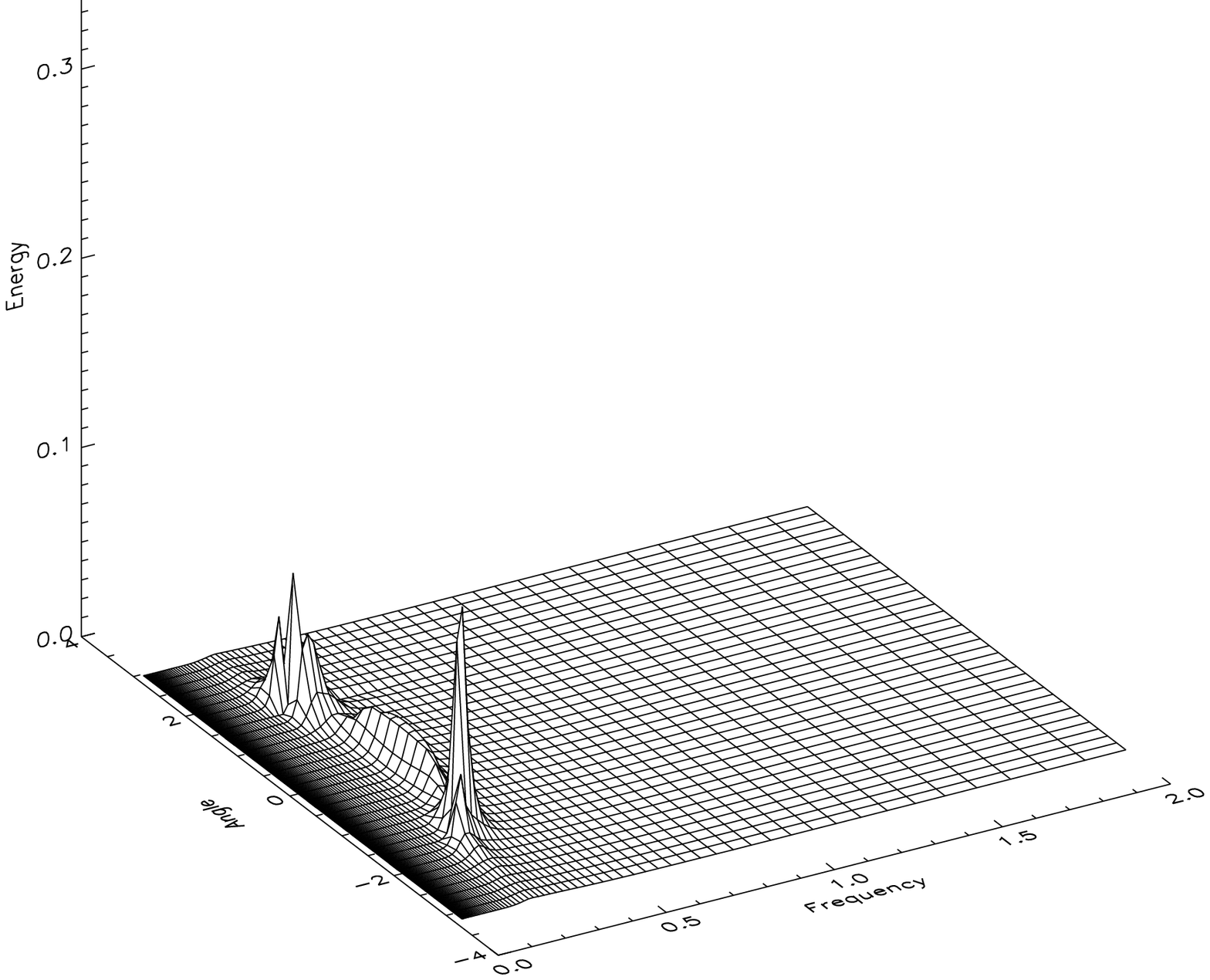}   \\ [1.8 cm]
			\includegraphics[width=0.4\linewidth]{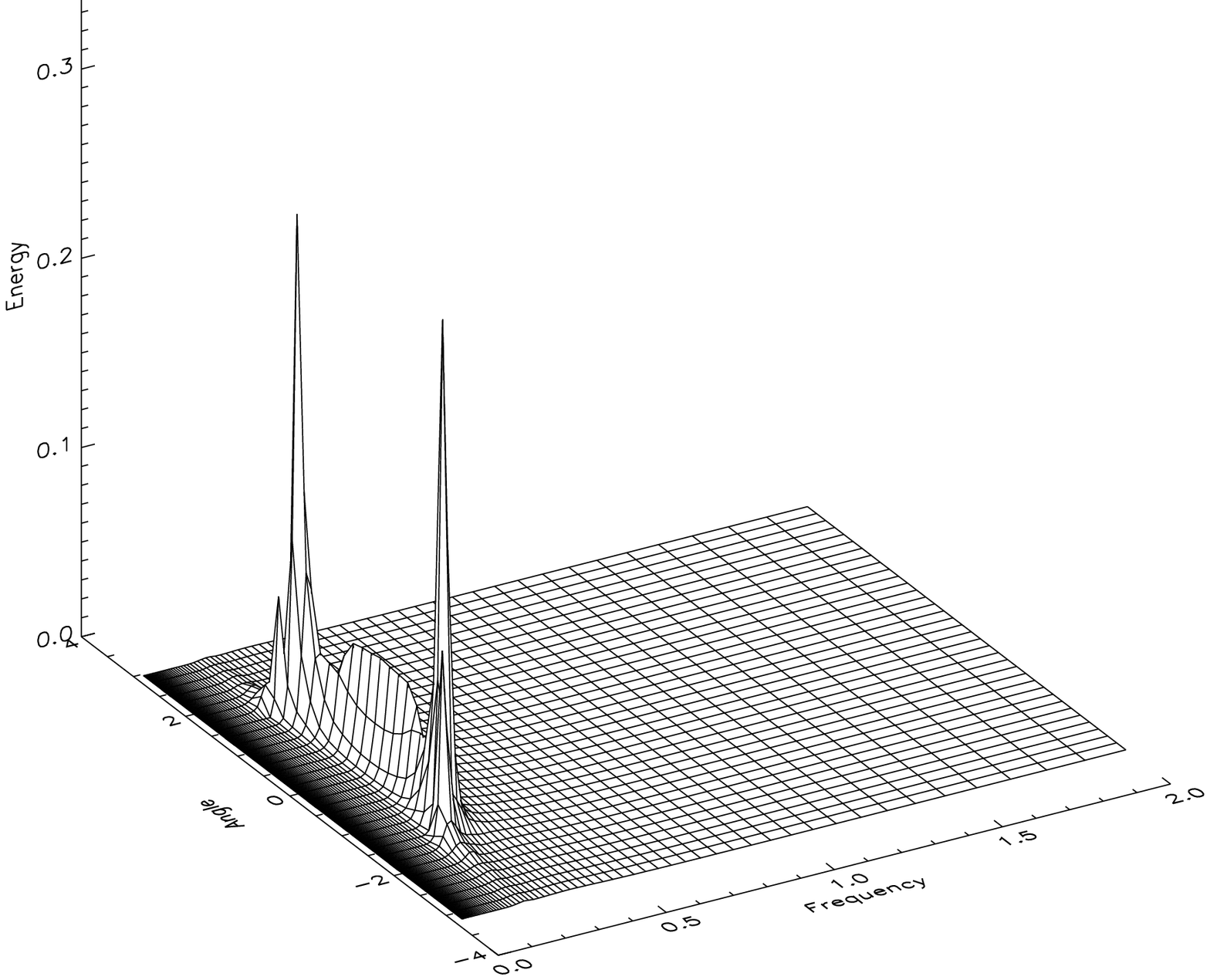} & \includegraphics[width=0.4\linewidth]{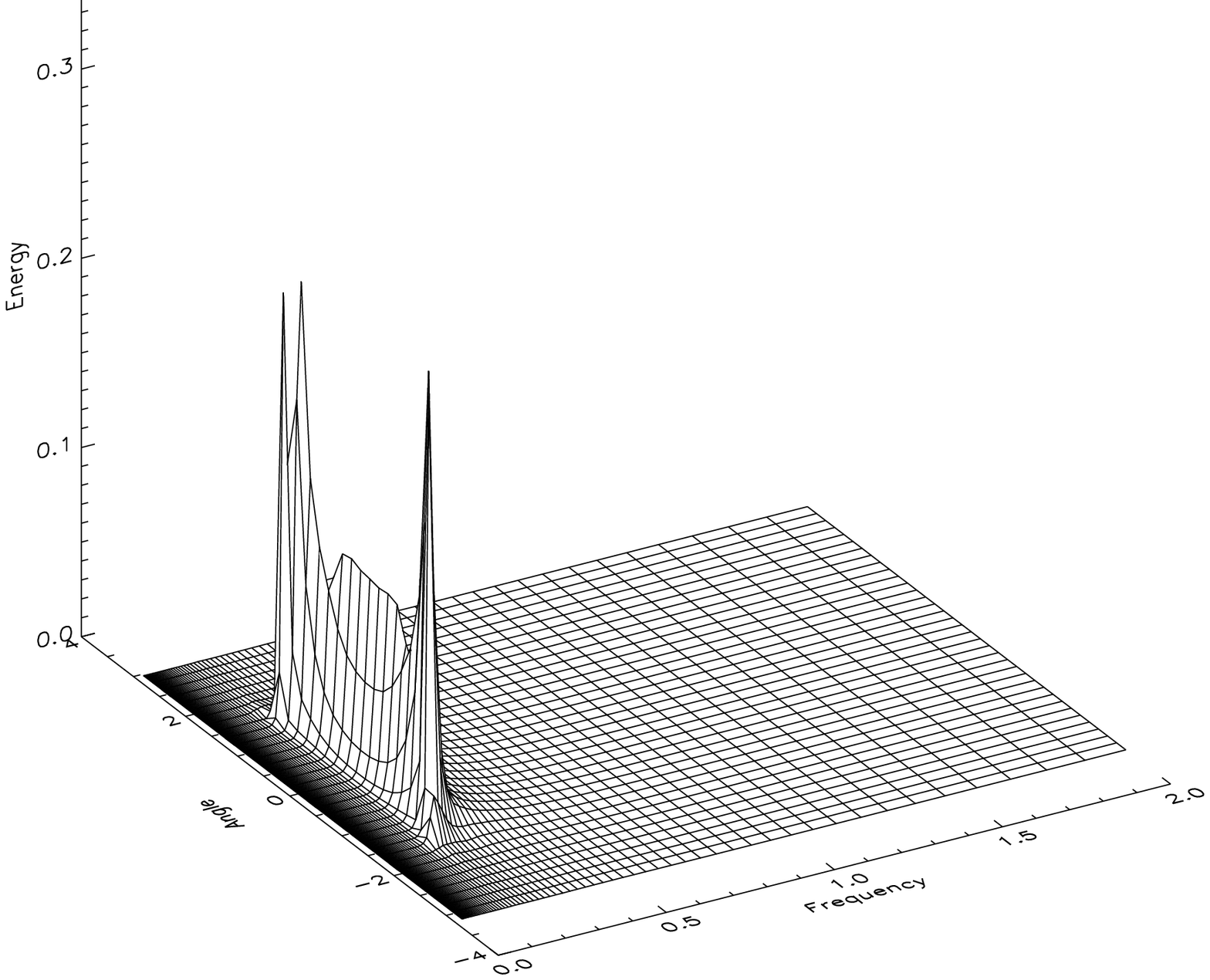} \\
		\end{tabular}
	\captionof{figure}{Spectral energy distribution as the function of the frequency $f$ and angle at the fetch distances 2, 14, 26 and 38 km for time 8 hr} \label{Spectrum3D8h}
	\end{center}
\end{table} 

\begin{table}[htbp]
\centering
	\begin{center} 
		\begin{tabular}{c c}
			\includegraphics[width=0.4\linewidth]{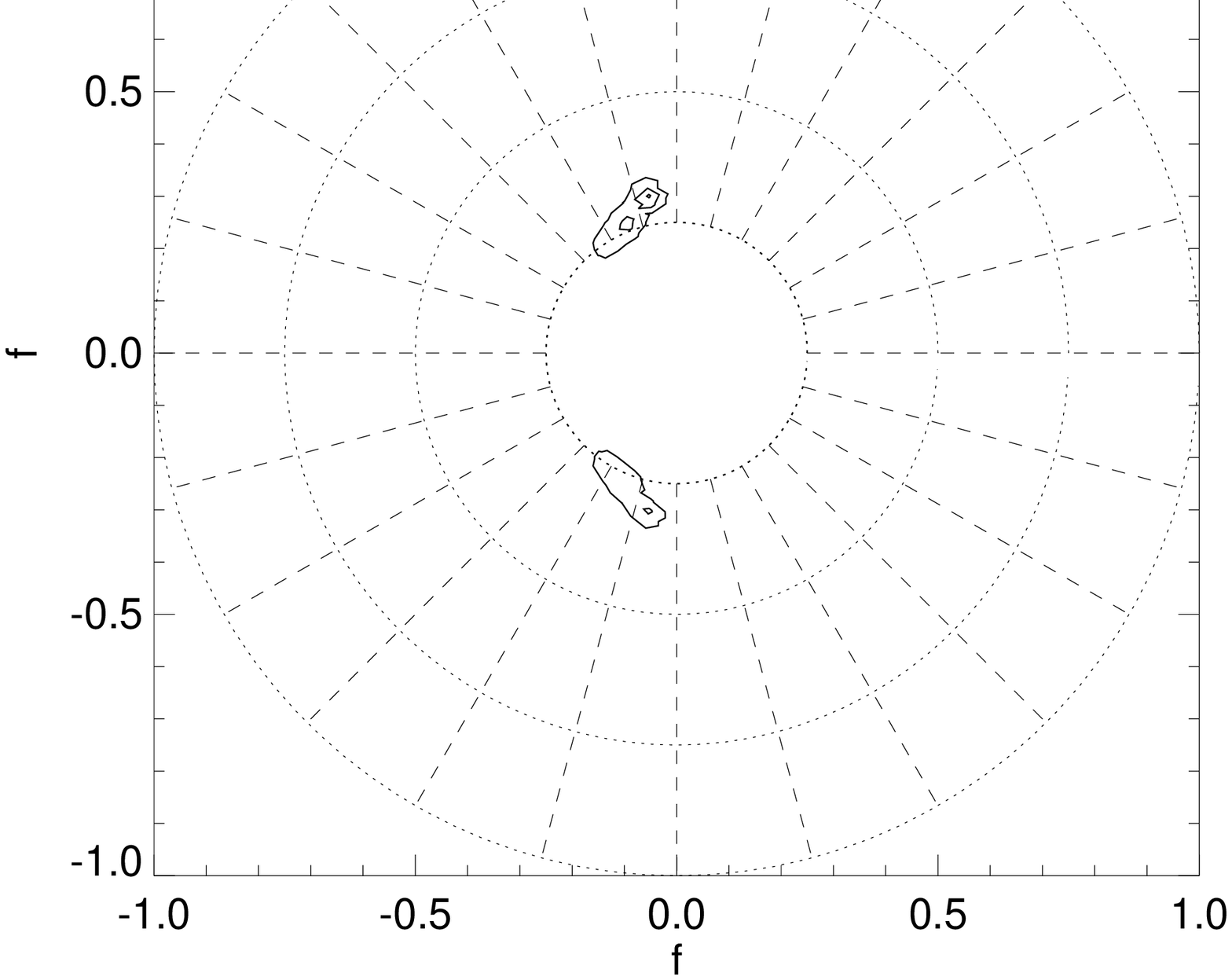} & \includegraphics[width=0.4\linewidth]{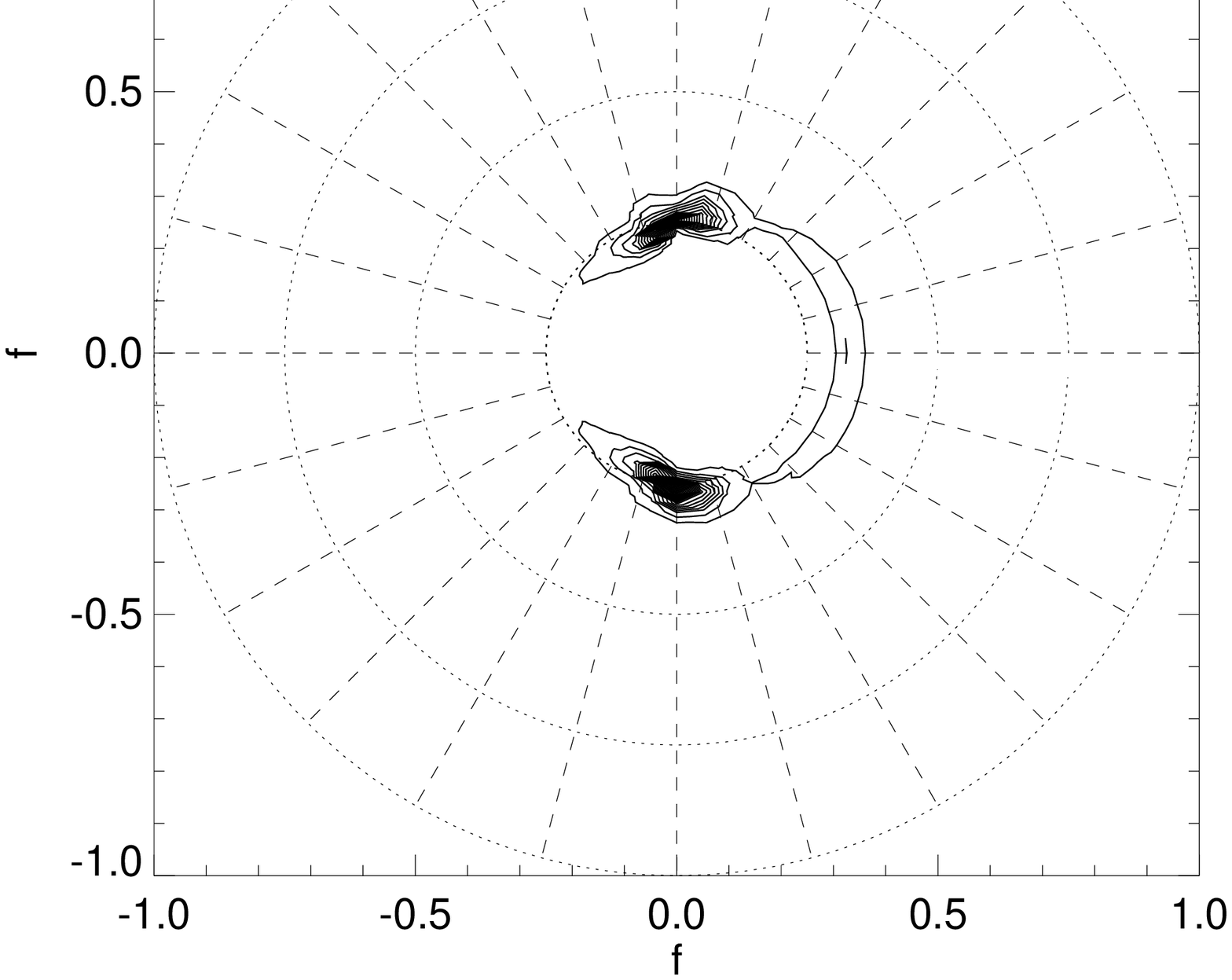}   \\ [1.8 cm]
			\includegraphics[width=0.4\linewidth]{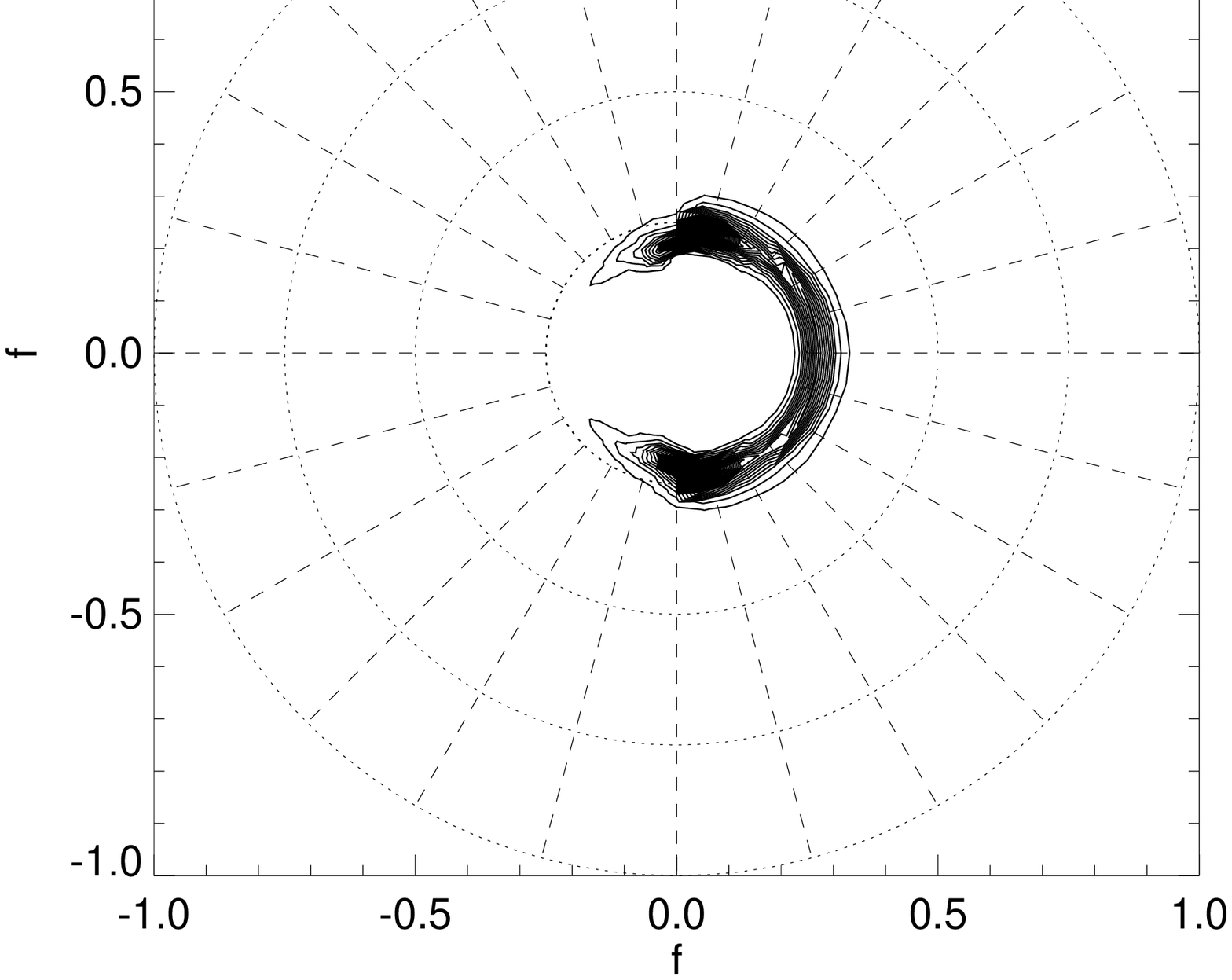} & \includegraphics[width=0.4\linewidth]{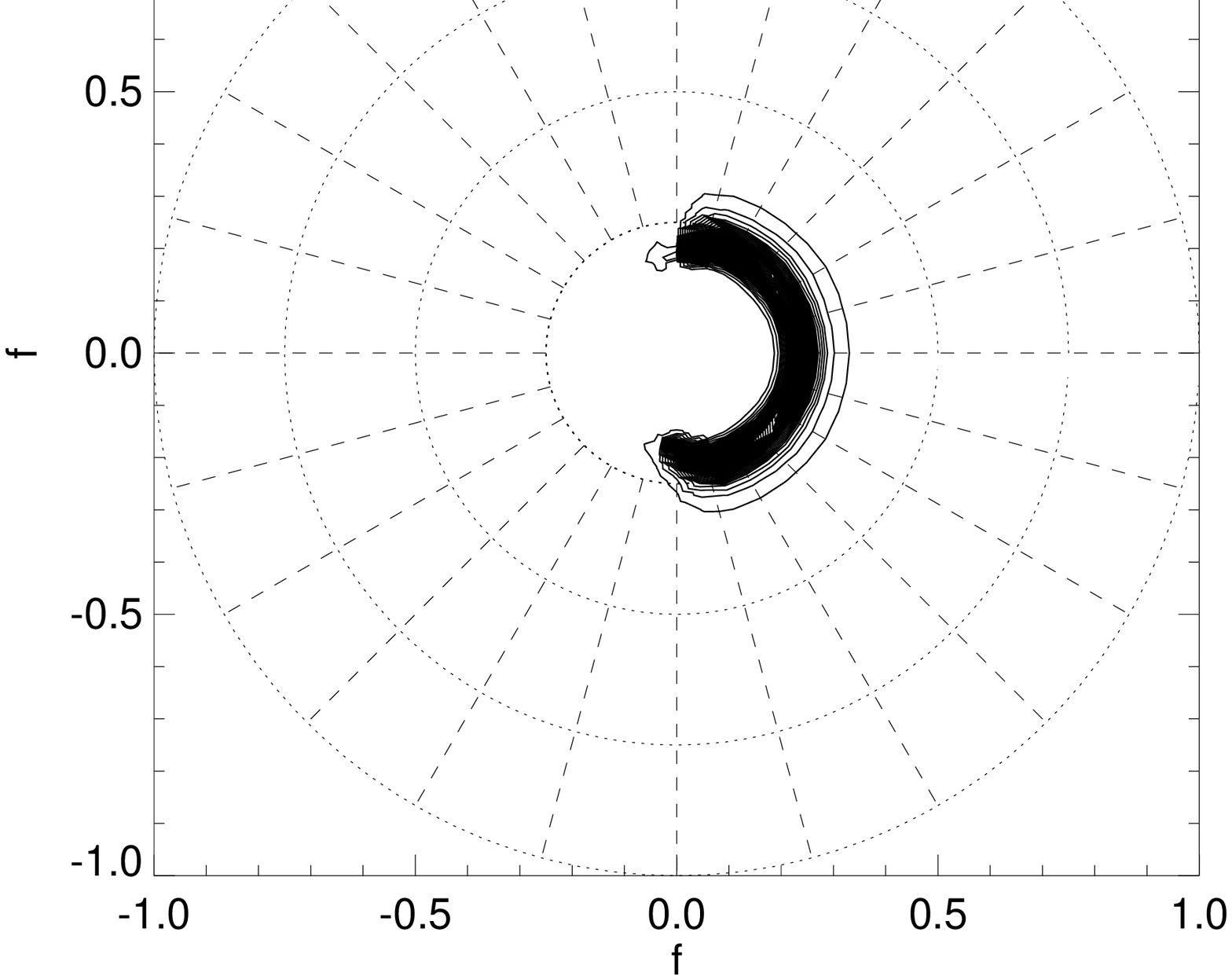} \\
		\end{tabular}
	\captionof{figure}{Spectral energy distribution as the function of the frequency $f$ and angle in polar coordinates at the fetch distances 2, 14, 26 and 38 km for time 8 hr} \label{Polar8h}
	\end{center}
\end{table}

For later time $t=40$ hr, which is close to the stationary state of the wave system, see Fig.\ref{Spectrum3D40h}, \ref{Polar40h}, the shape of the spectrum is quite complex: besides central single-hump energy spectrum, growing away from the west shore, one can see powerful side satellites, corresponding to the waves, propagating predominantly orthogonally to the wind. 

\begin{table}[htbp]
\centering
	\begin{center} 
		\begin{tabular}{c c}
			\includegraphics[width=0.4\linewidth]{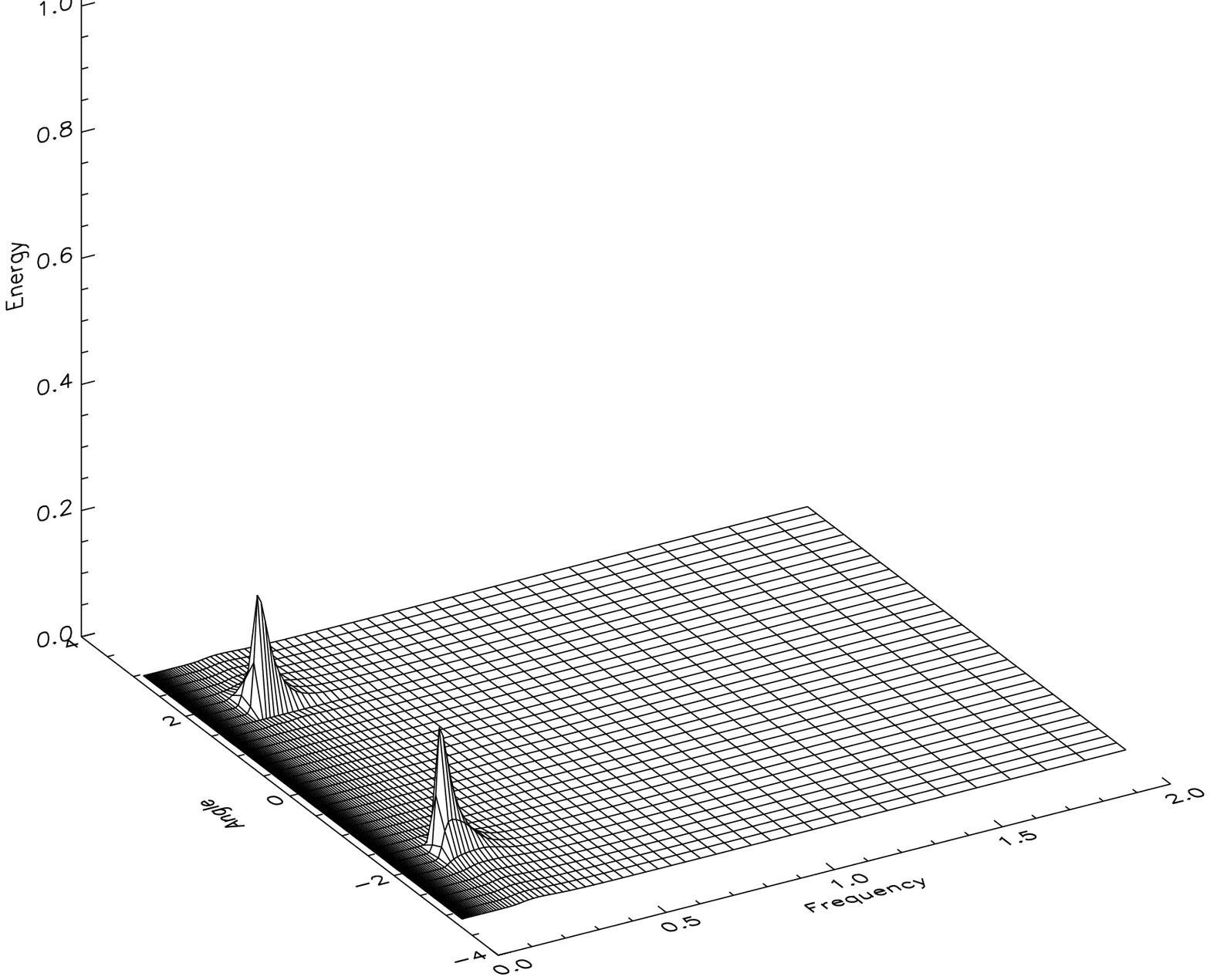} & \includegraphics[width=0.4\linewidth]{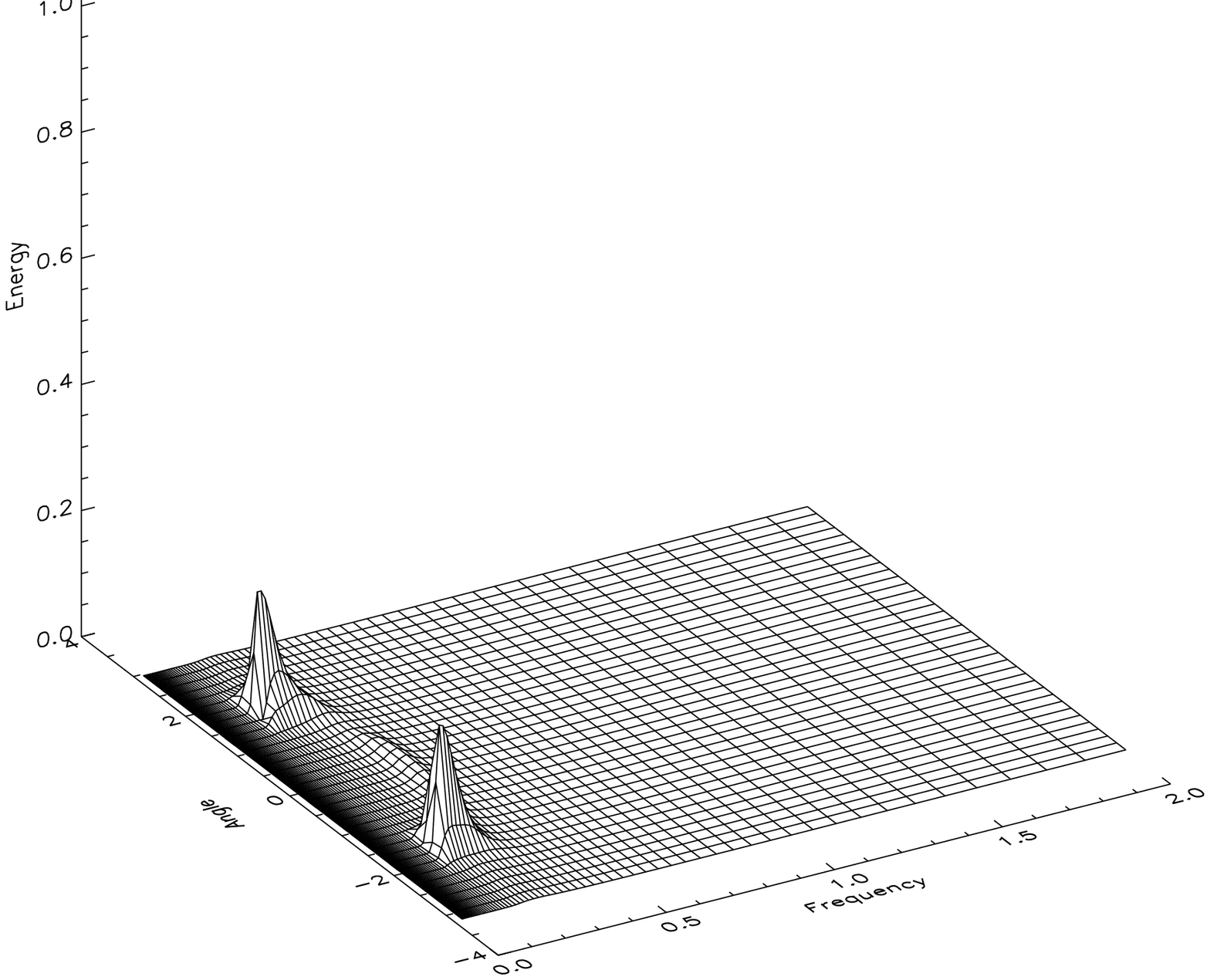}   \\ [1.8 cm]
			\includegraphics[width=0.4\linewidth]{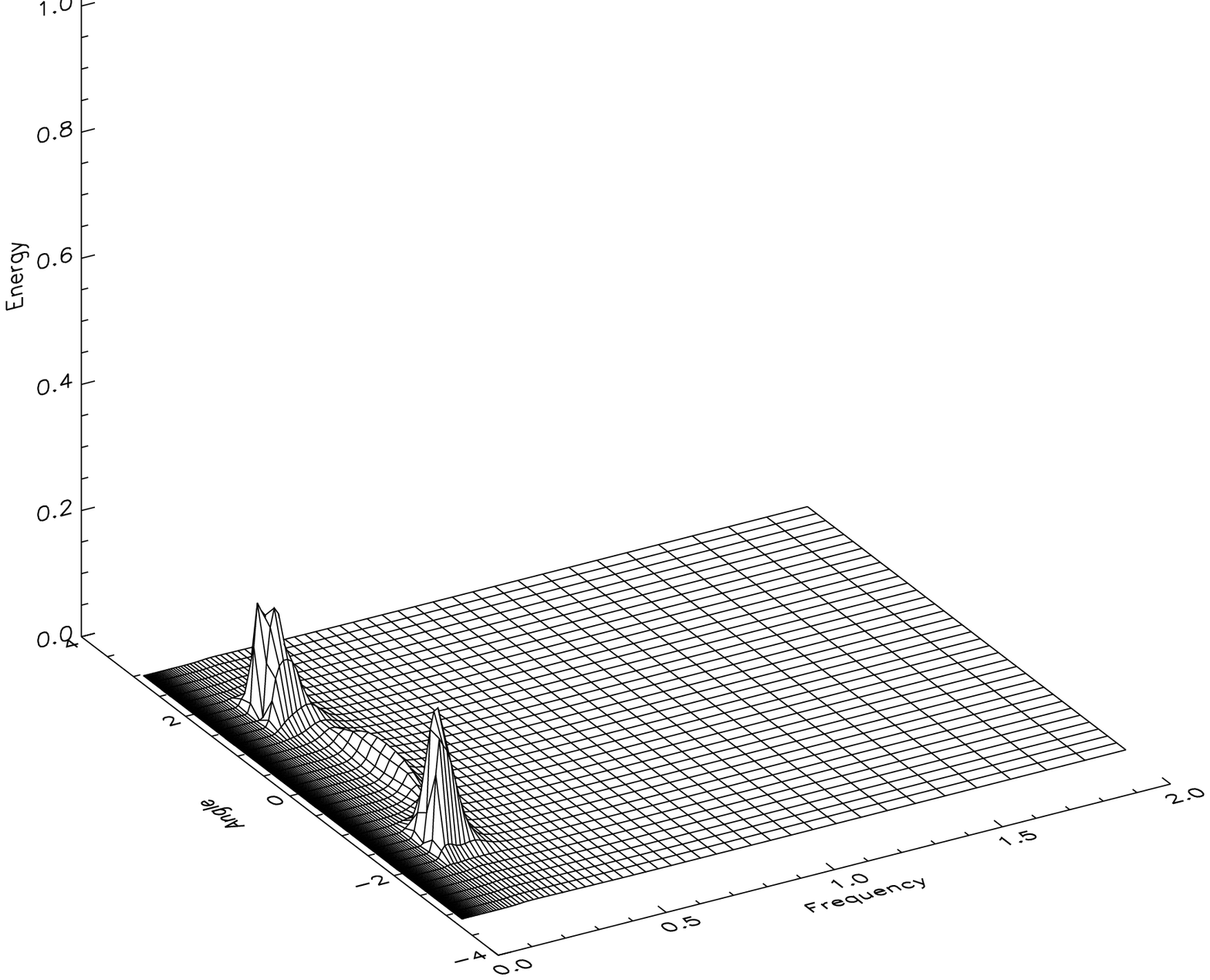} & \includegraphics[width=0.4\linewidth]{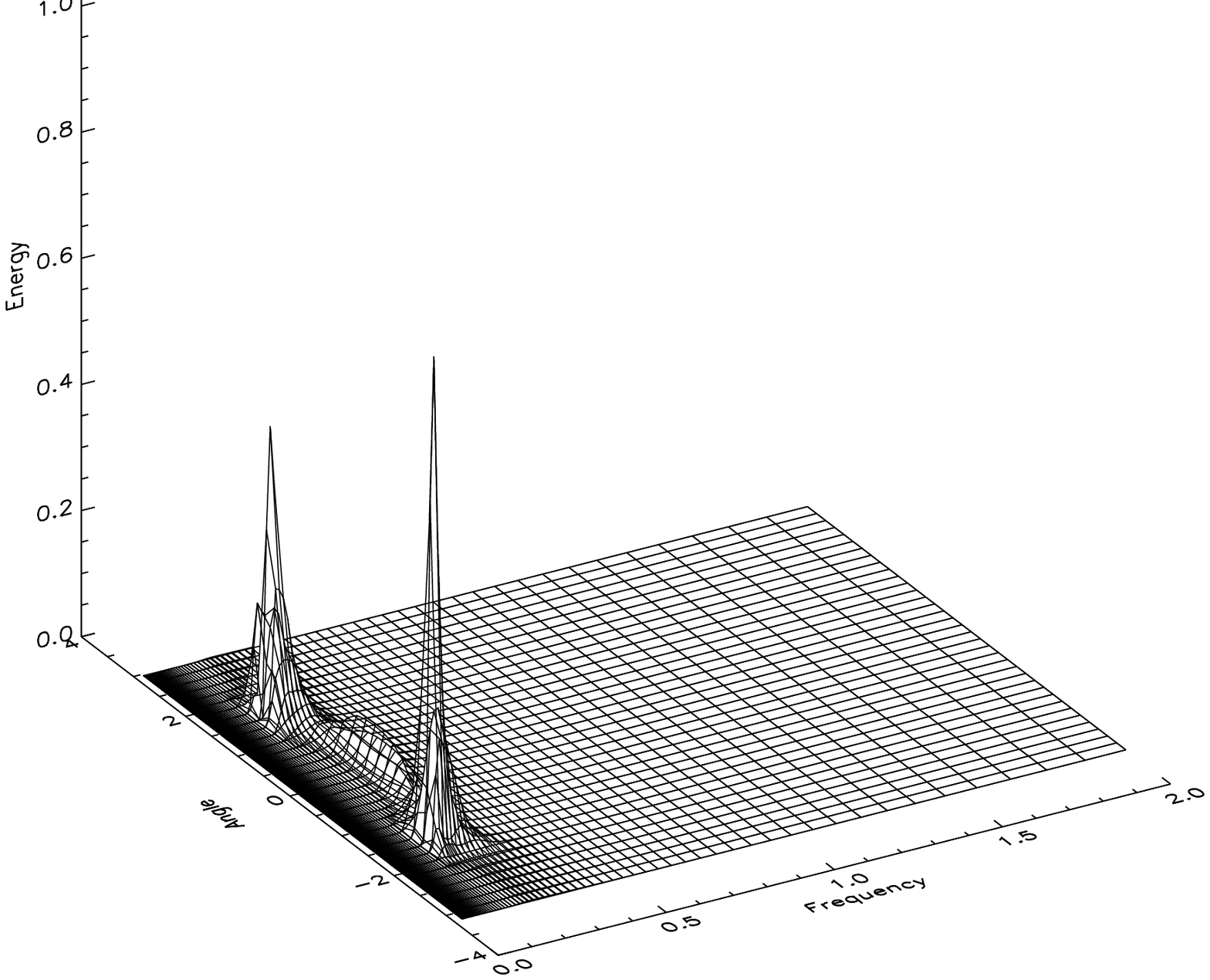} \\
		\end{tabular}
	\captionof{figure}{Spectral energy distribution as the function of the frequency $f$ and angle at the fetch distances 2, 14, 26 and 38 km for time 40 hr} \label{Spectrum3D40h}
	\end{center}
\end{table} 

\begin{table}[htbp]
\centering
	\begin{center} 
		\begin{tabular}{c c}
			\includegraphics[width=0.4\linewidth]{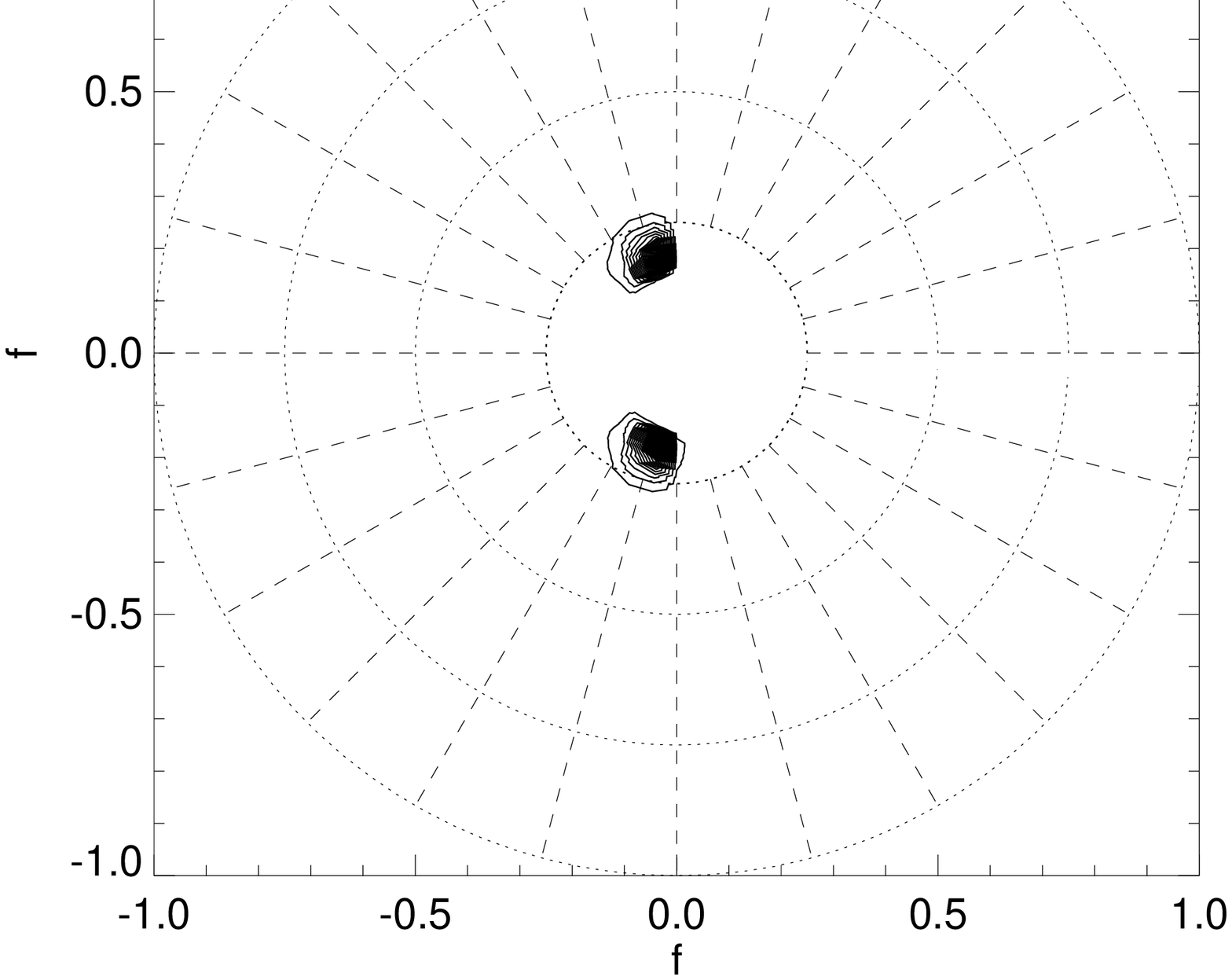} & \includegraphics[width=0.4\linewidth]{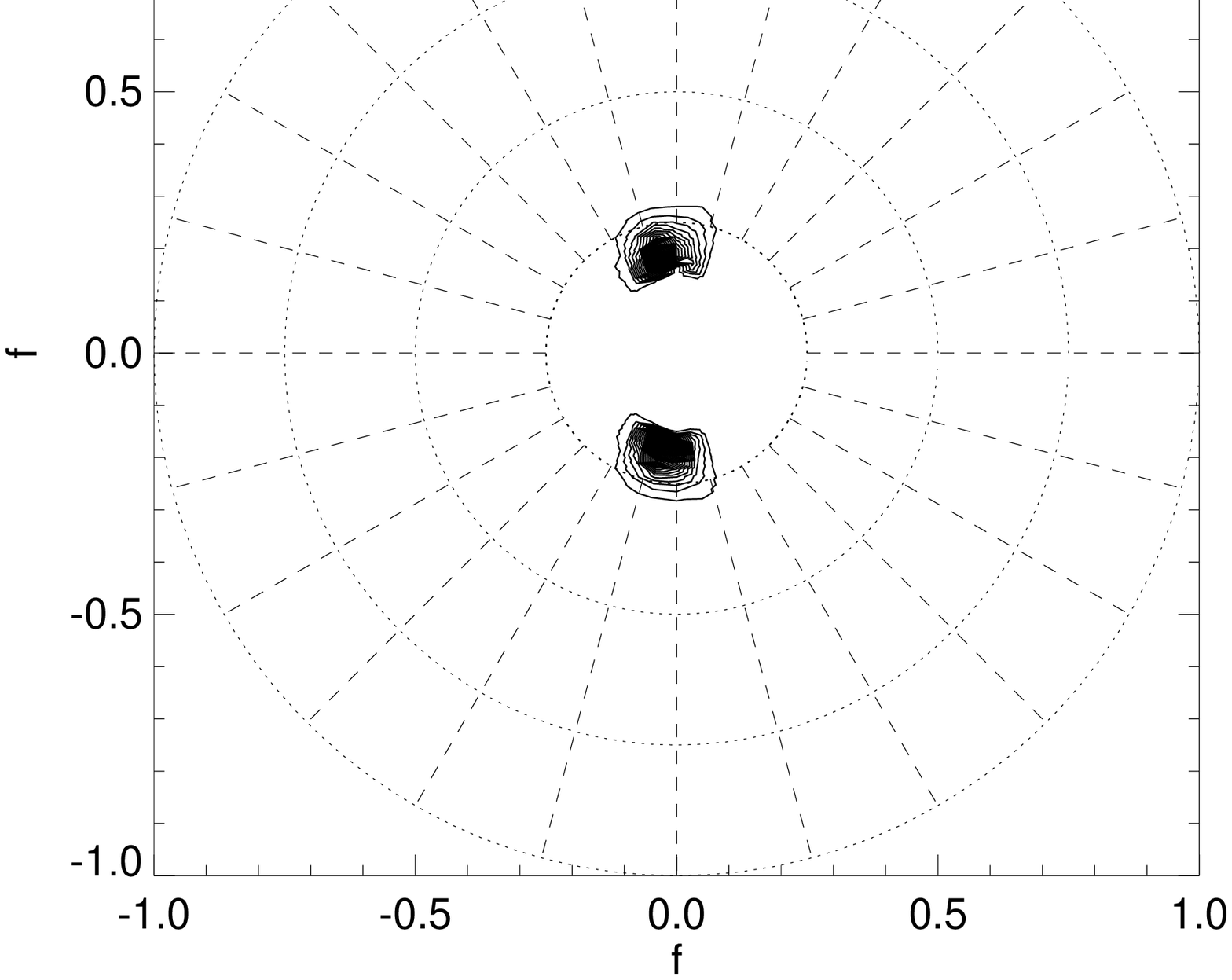}   \\ [1.8 cm]
			\includegraphics[width=0.4\linewidth]{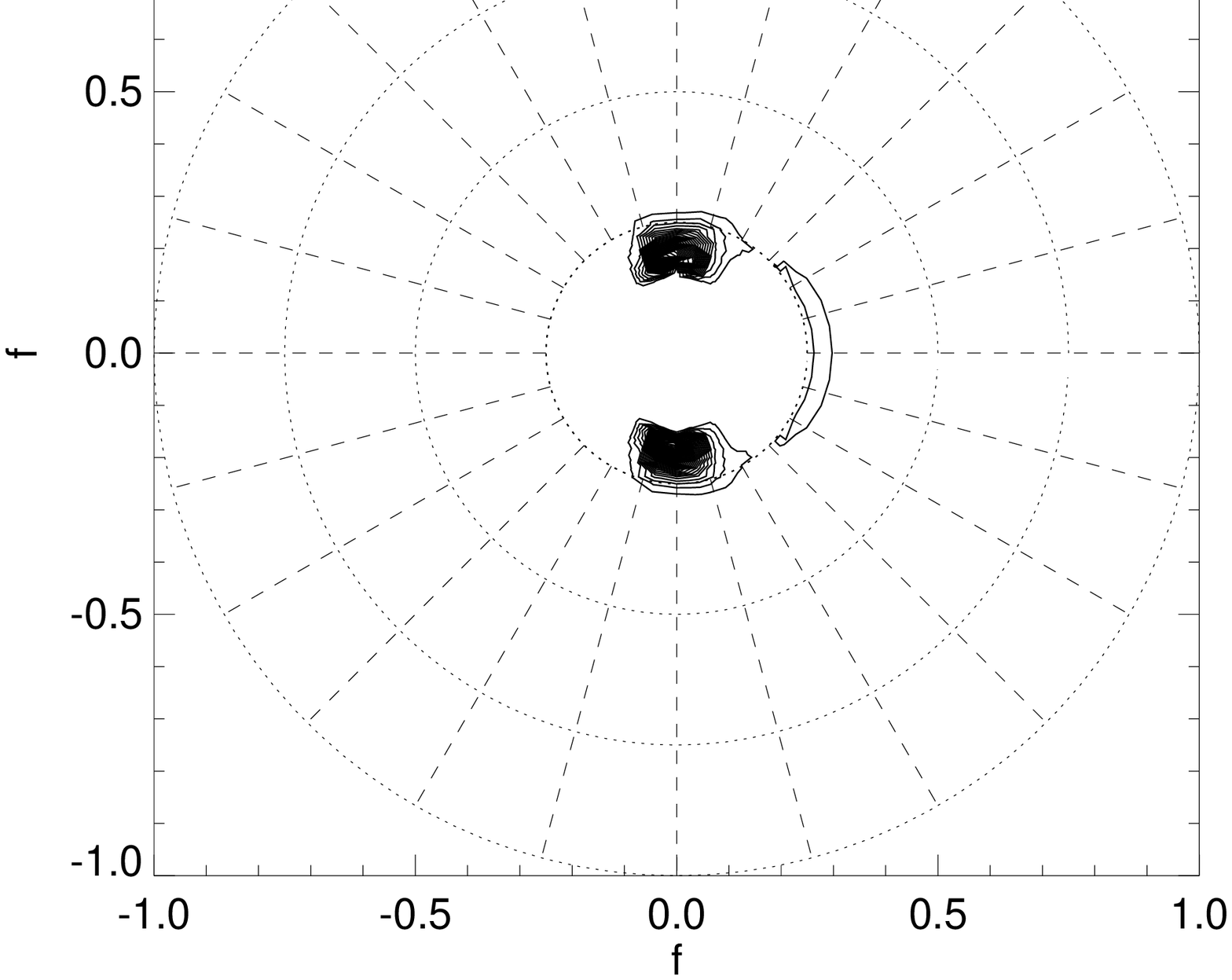} & \includegraphics[width=0.4\linewidth]{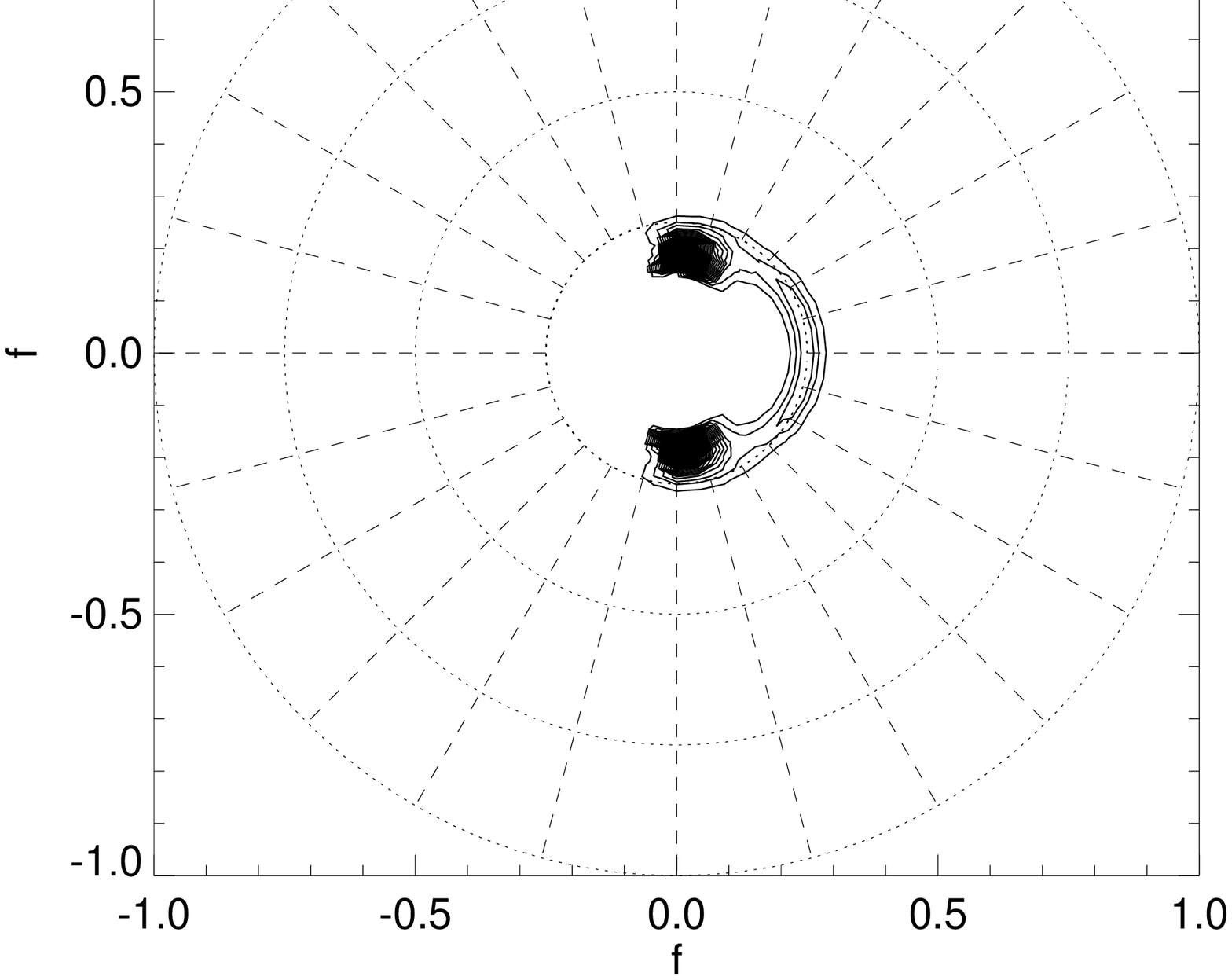} \\
		\end{tabular}
	\captionof{figure}{Spectral energy distribution as the function of the frequency $f$ and angle in polar coordinates at the fetch distances 2, 14, 26 and 38 km for time 40 hr} \label{Polar40h}
	\end{center}
\end{table}

This observation is quite remarkable: it means that long enough in time excitation of the waves by the wind, blowing perpendicular to the straits shorelines, asymptotically excites the stationary wave turbulence, consisting of two components: waves, propagating in the wind direction, similar to the classical self-similar regime with the single spectral maximum and growing along the fetch from the west to the east coast, and another component, consisting of quazi-monochromatic waves, propagating almost perpendicular to the direction of the wind. There is the slant of the second wave system with respect to the shore lines forection, which increases in the direction toward the west shore, and reaching the maximum anglular slant of $15^{0}$ with respect to the shore line in the vicinity of the west coast.  The formation of this slant could mean that the system needs to get adjusted in this way to find an extra dissipation to bring the wave energy system into the balanced stationary state.

The splitting of the system into subsystems in space in time is also confirmed by Fig.\ref{FreqAvSpectra}, which plots frequency-averaged wave energy spectra as the function of the angle at different fetch locations for $t=$ 2, 4 ,8 and 110 hours. One can see that the structure of the wave system is significantly different for differnt time spans: it has the single maximum shape for $t<4$ hours and has multi-maximum shape for $t>8$ hours, consisting asymptotically of two quazi-monochromatic waves orthogonal to the wind and the hump in the direction of the wind.

\begin{table}[htbp]
\centering
	\begin{center} 
		\begin{tabular}{c c}
			\includegraphics[width=0.4\linewidth]{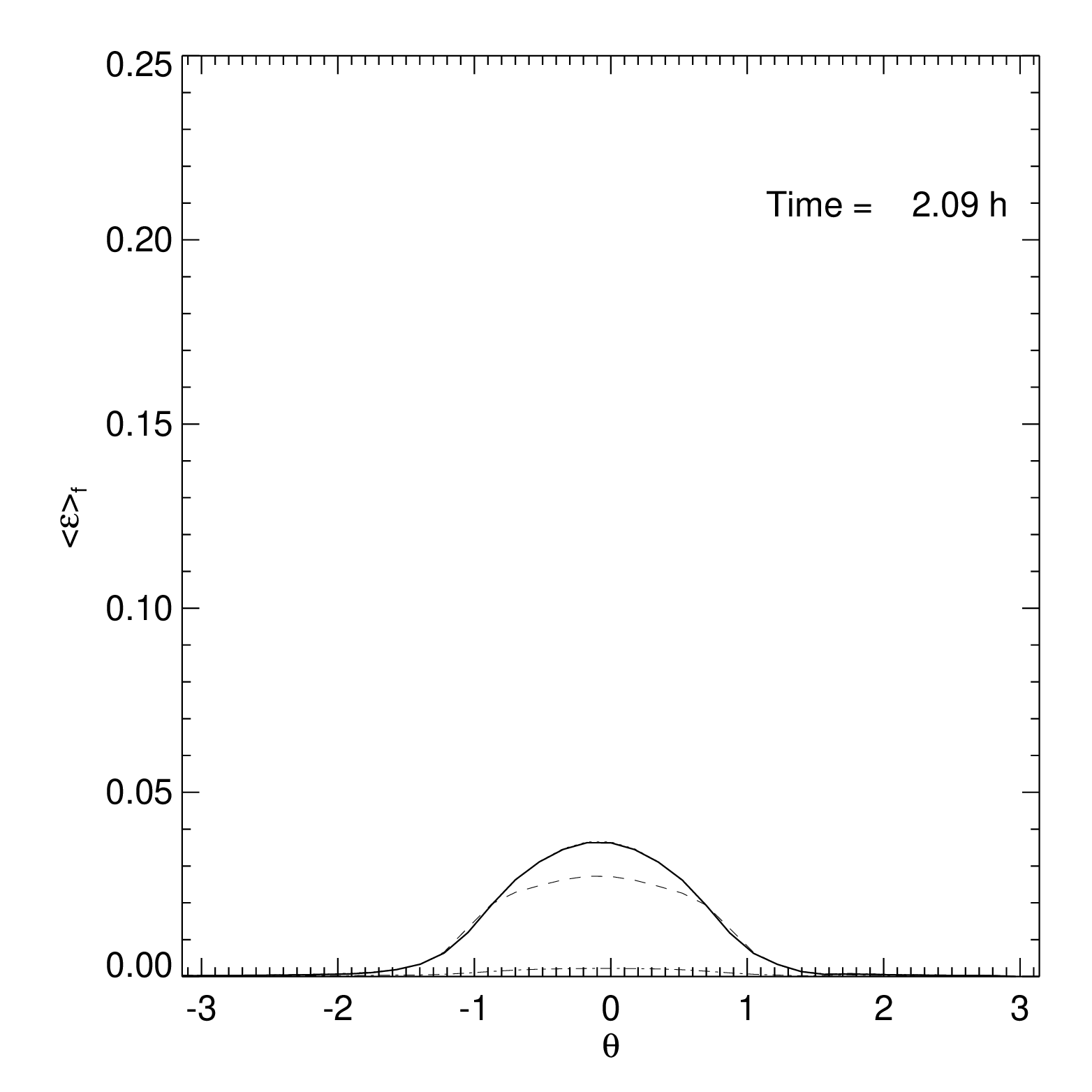} & \includegraphics[width=0.4\linewidth]{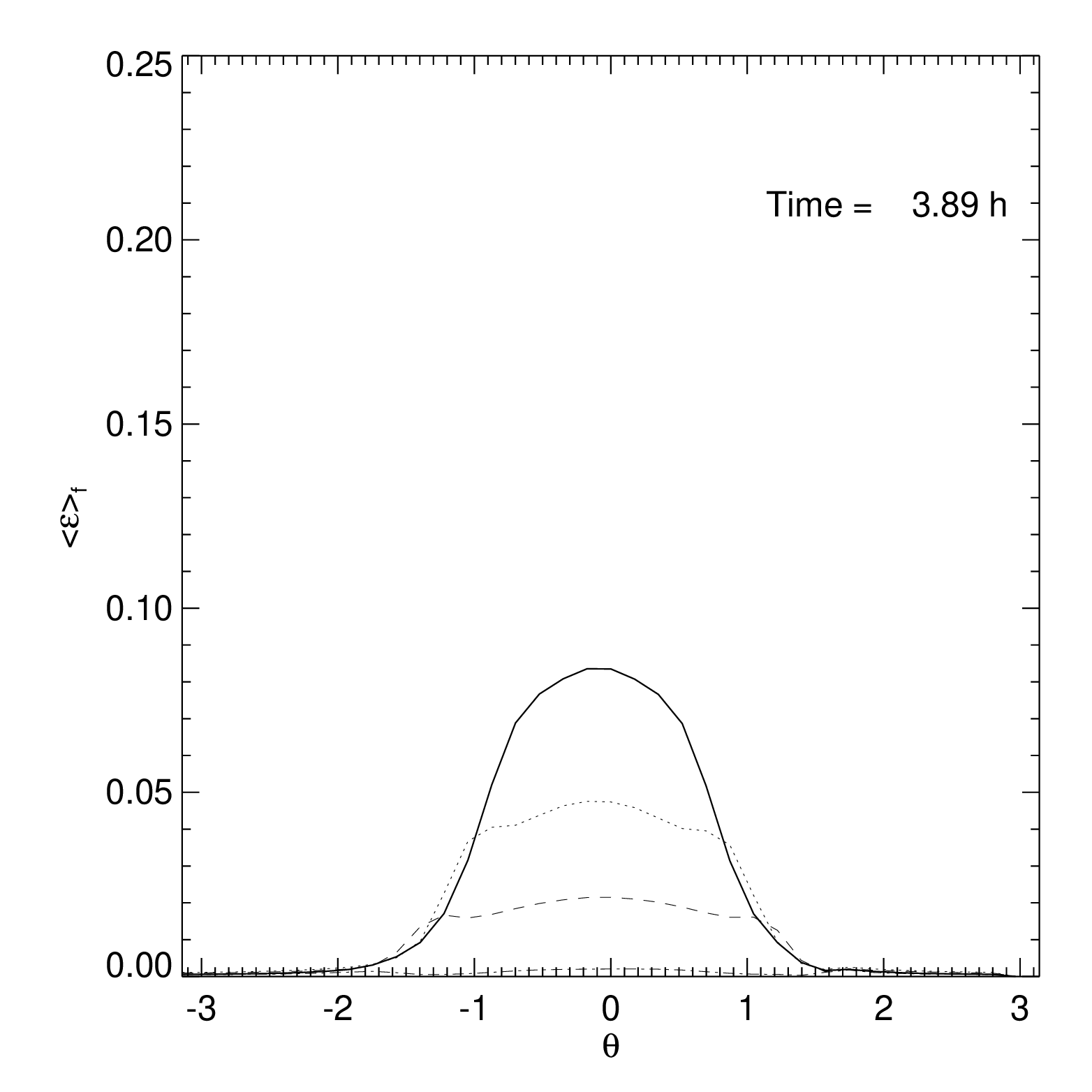}   \\ [1.8 cm]
			\includegraphics[width=0.4\linewidth]{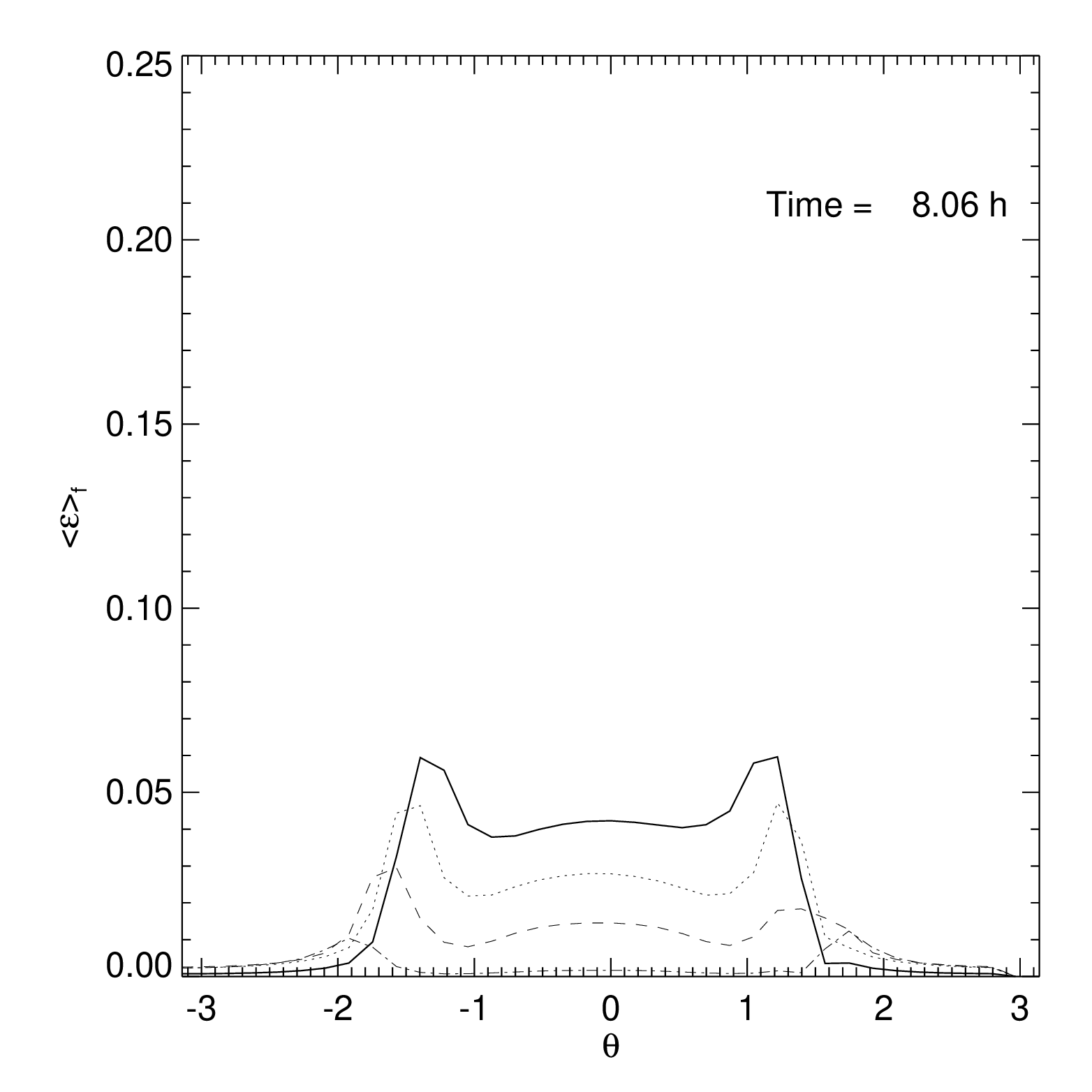} & \includegraphics[width=0.4\linewidth]{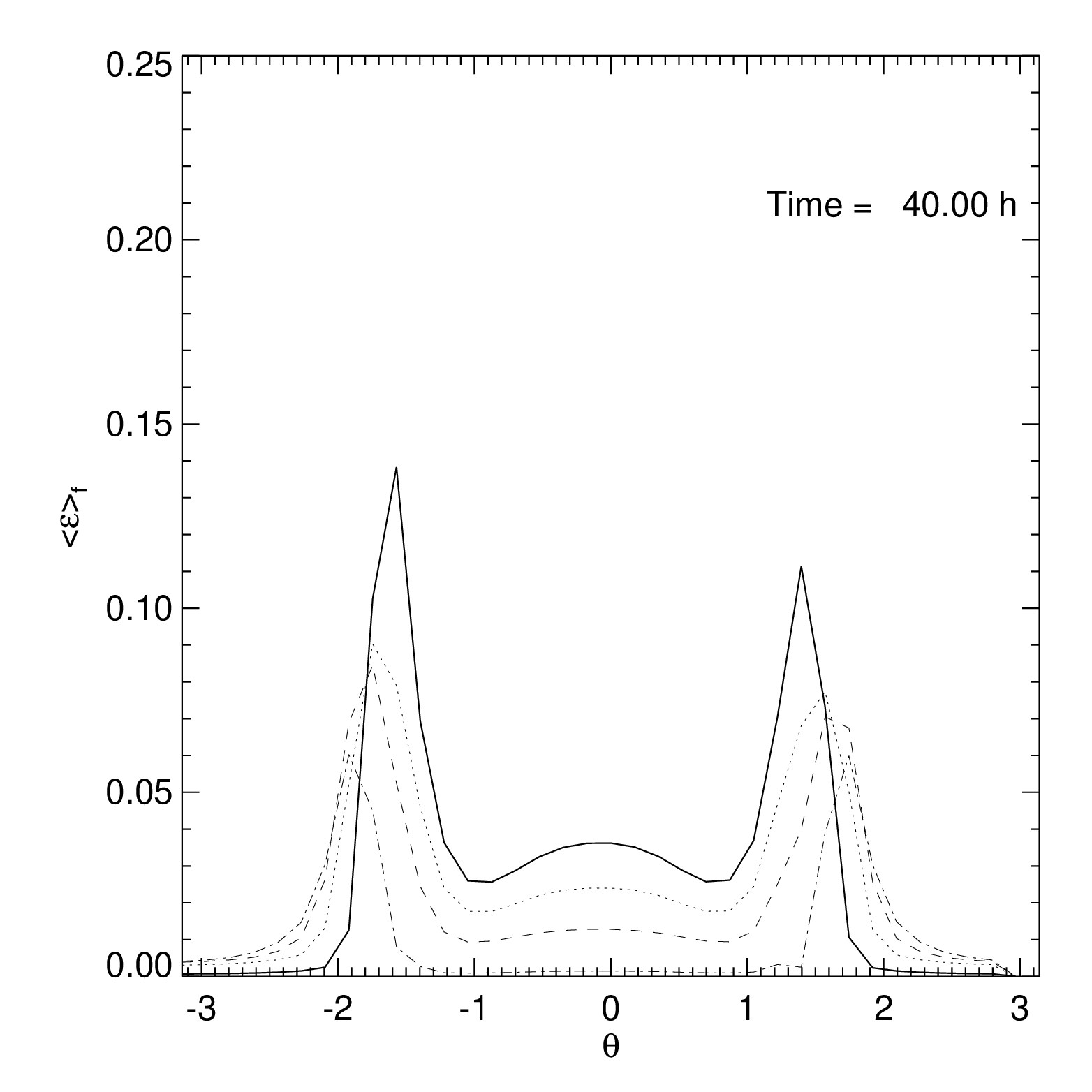} \\
		\end{tabular}

			\captionof{figure}{Frequency averaged spectra $<\varepsilon>_f = \int_{-\pi}^{\pi} \varepsilon (f,\theta) d\theta$ as the function of the angle $\theta$ at different fetch coordinates: $x=2$ km - dash-dotted line;  $x=14$ km - dashed line; $x=26$ km - dotted line; $x=38$ km - solid line at times $t=$ 2, 4, 8 and 110 hours.} \label{FreqAvSpectra}
	\end{center}
\end{table}

We are led to infer that the wave system in the asymptotic stationary state consist of two subsystems: first qualitatively similar to the classical limited fetch situation without the secondary boundary, which we call "wind sea", and the second, consisting of the standing quasi-monochromatic wave, directed almost perpendicular to the wind direction.

The described wave system works as the laser-like {\bf N}onliner {\bf O}cean {\bf W}ave {\bf A}mplifier of quasi-monochromatic waves, pumped by the orthogonally blowing wind, which we reduce to the acronym {\bf NOWA}. One should notice that these quasi-monochromatic waves are "condensated" predominantely near the separatix of the advection velocity profile, i.e. in the locations where the advection velocity is equal zero. Such waves have no chance to be advected out of theconsidered domain to get absorbed at the shorelines.

\begin{figure}[ht]
\noindent\center\includegraphics[scale=0.7, angle=90]{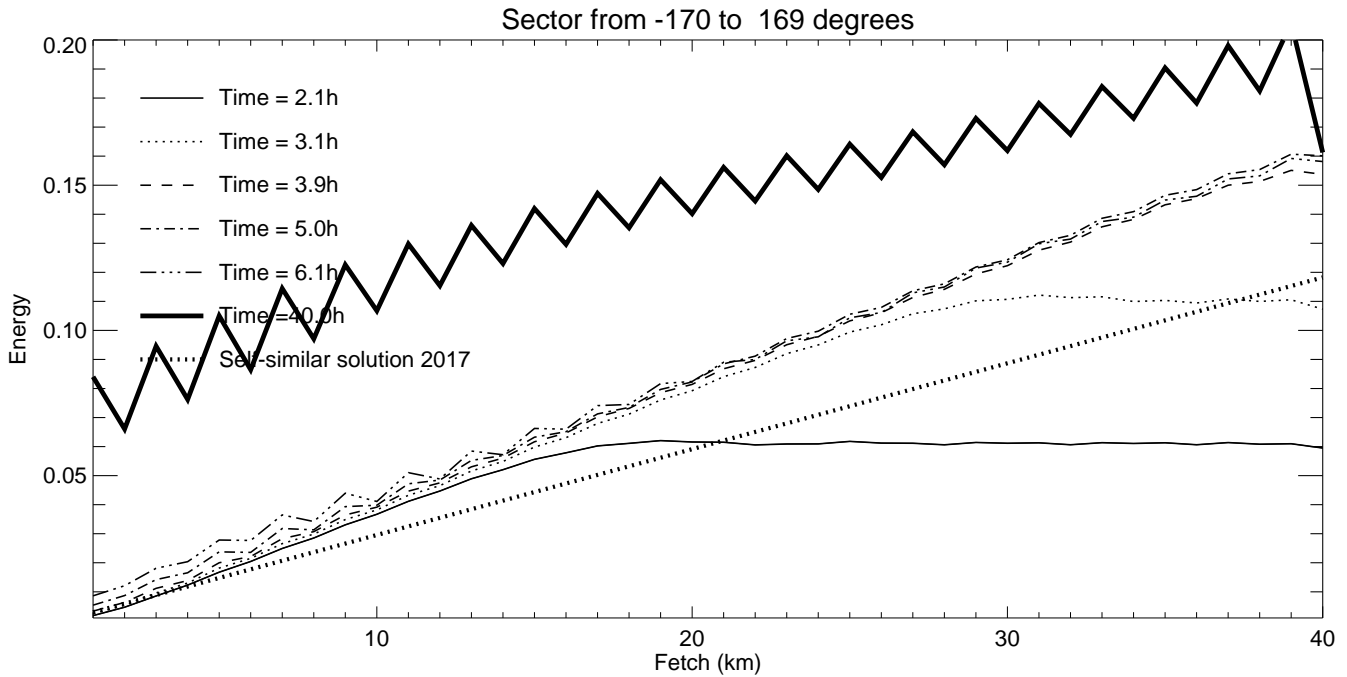} 
\caption{Wave energy distribution along the fetch for different moments of time. }
\label{TotalEnergyOnFetch}
\end{figure}

Fig.\ref{TotalEnergyOnFetch} presents distribution of the total wave energy along the fetch in logarithmic scale for different times. For $t < 5$ hr the energy evolution is described by threshold-like function, propagating along the fetch, in qualitative correspondence with Fig.\ref{FetchTest2}. For $t \simeq 5$ hr, the total wave energy distribution along the fetch comes into close to  self-similar, i.e. linear shape in accordance with Eqs.(\ref{Indices}), (\ref{EnFetchSelfSim}).

For $ t > 8$ hr we observe relatively slow evolution of the total wave energy to the asymptotic state, still having the tendency of total wave energy growth from west to east coast, though with much slower than linear growth rate, observed for self-similar regime.

\begin{figure}[ht]
\noindent\center\includegraphics[scale=0.7, angle=90]{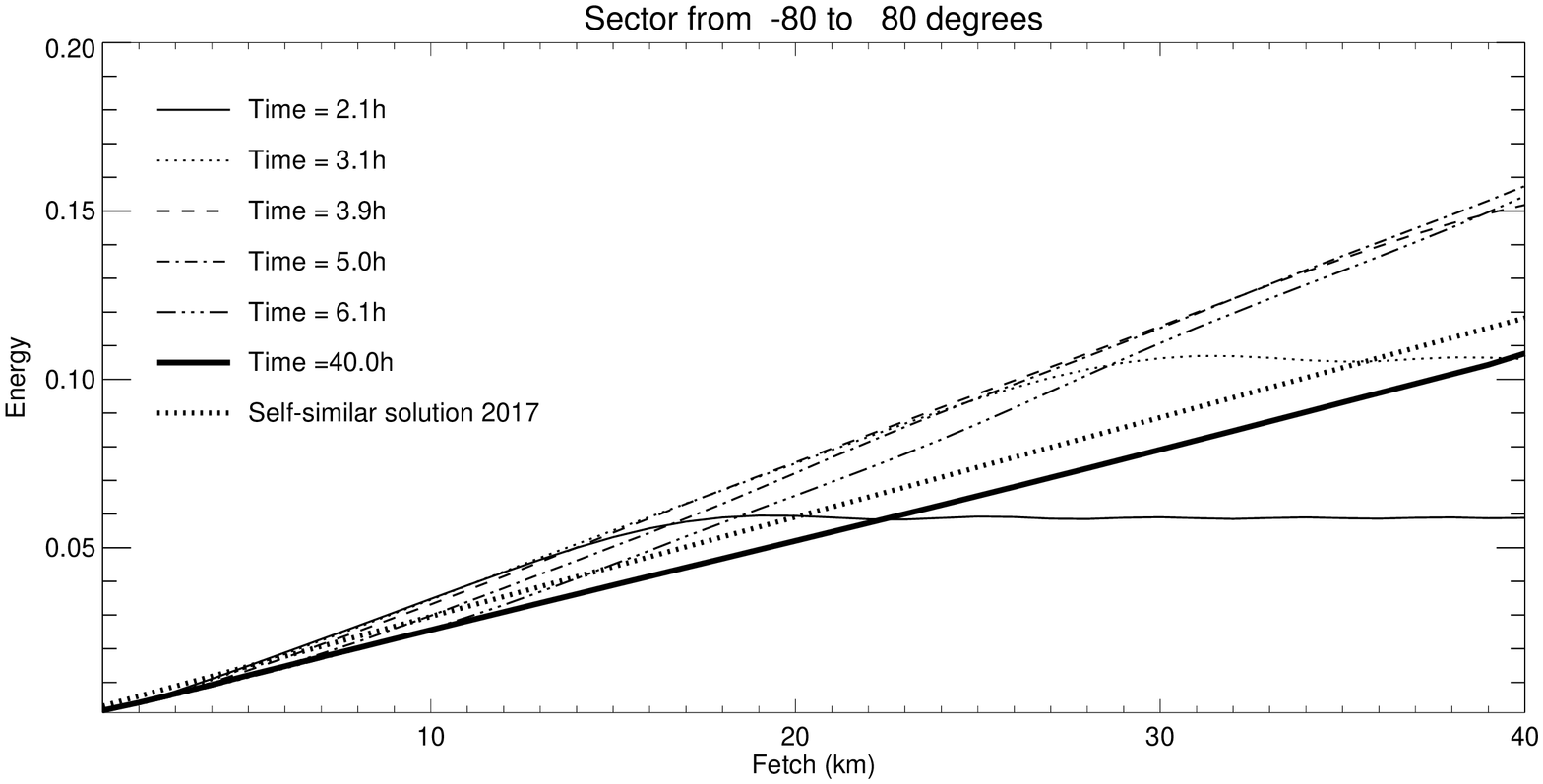} 
\caption{Decimal logarithm of wave energy distribution along the fetch for different moments of time, calculated in the angle spread $-90^o < < 90^o$.}
\label{TotalEnergyCentralOnFetch}
\end{figure}

While the Fig.\ref{TotalEnergyOnFetch} presents the mixed state  of the wind sea and monochromatic waves, Fig.\ref{TotalEnergyCentralOnFetch} shows decimal logarithm of wave energy distribution along the fetch for different moments of time, calculated in the angle spread $-\pi/8 < \theta < \pi/8$, which takes into account only the wave sea effects, exluding the quazi-monochromatic waves. One can see that this wind sea distribution is asimptotically close in time to the classical self-similar form, than the distribution on Fig.\ref{TotalEnergyOnFetch}, except closely adjacent to the west coast region of 3 km width. This effect is quite natural to expect, since the intensity of quazi-monochromatic waves grows approaching to the west coast, along with degree of their interaction, leading to the deformation of self-similar behavior.

%

\begin{figure}[ht] 
\noindent\center\includegraphics[scale=0.7, angle=90]{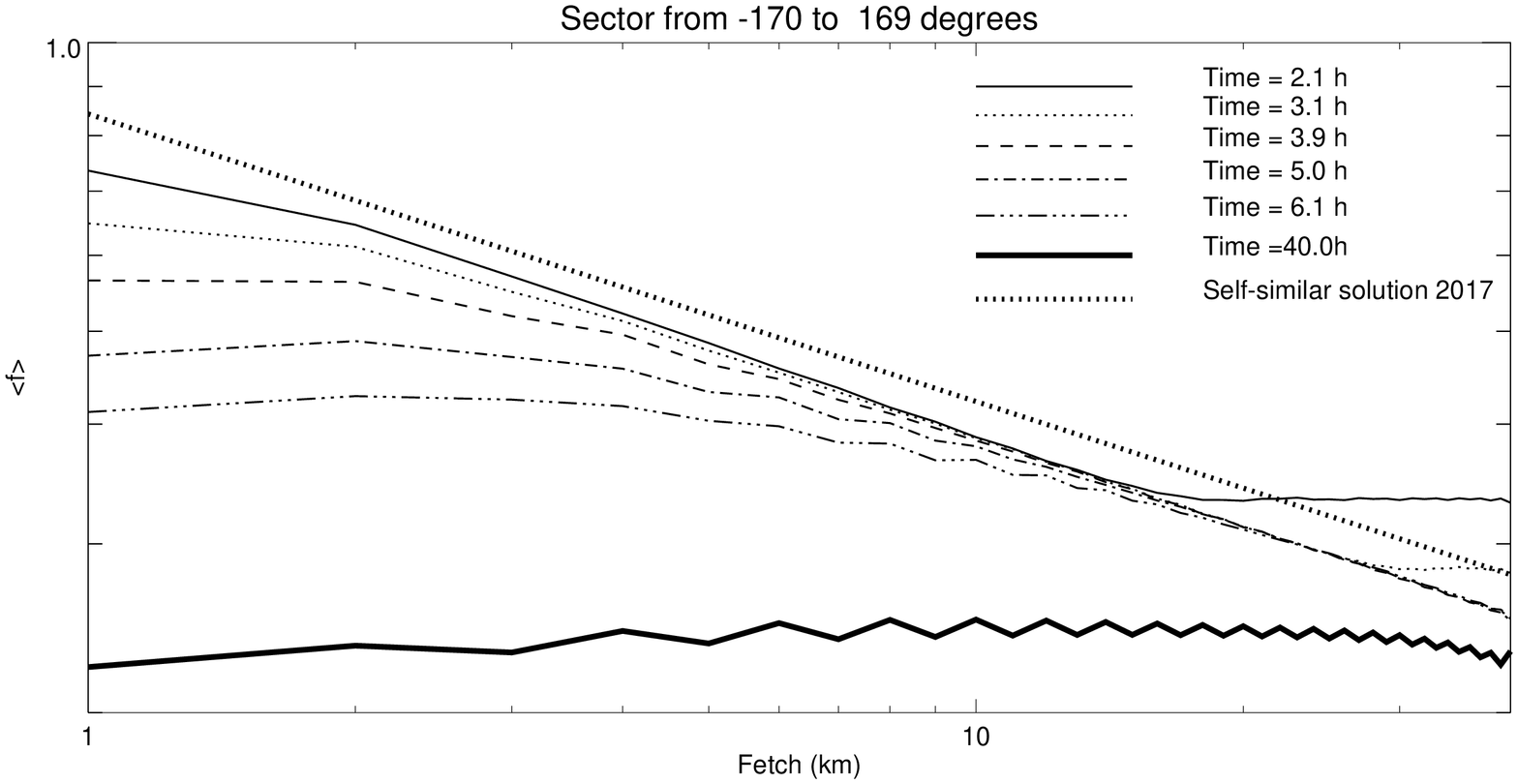} 
\caption{ Decimal logarithm of mean frequency distribution as the function of the decimal logarithm of the fetch for different moments of time.}
\label{MeanFreqOnFetch}
\end{figure}

Fig.\ref{MeanFreqOnFetch} shows the logarithm of the mean frequency distribution as the function of the logarithm of the fetch for different moments of time. One can see that the linear portion of the "threshold-like" function, propagationg from the east to west coast is closely described by self-similar solution Eqs.(\ref{Indices}), (\ref{FrFetchSelfSim}), though the complete correspondence with the self-similar solution along the whole fetch is reached  for earlier  time 3 hr, which is different from time 5 hr of reaching the same state for the  total wave energy energy on Fig.\ref{TotalEnergyOnFetch}. Asymptotically in time, the mean frequency distribution along the fetch comes to almost constant value.

\begin{figure}[ht] 
\noindent\center\includegraphics[scale=0.7, angle=90]{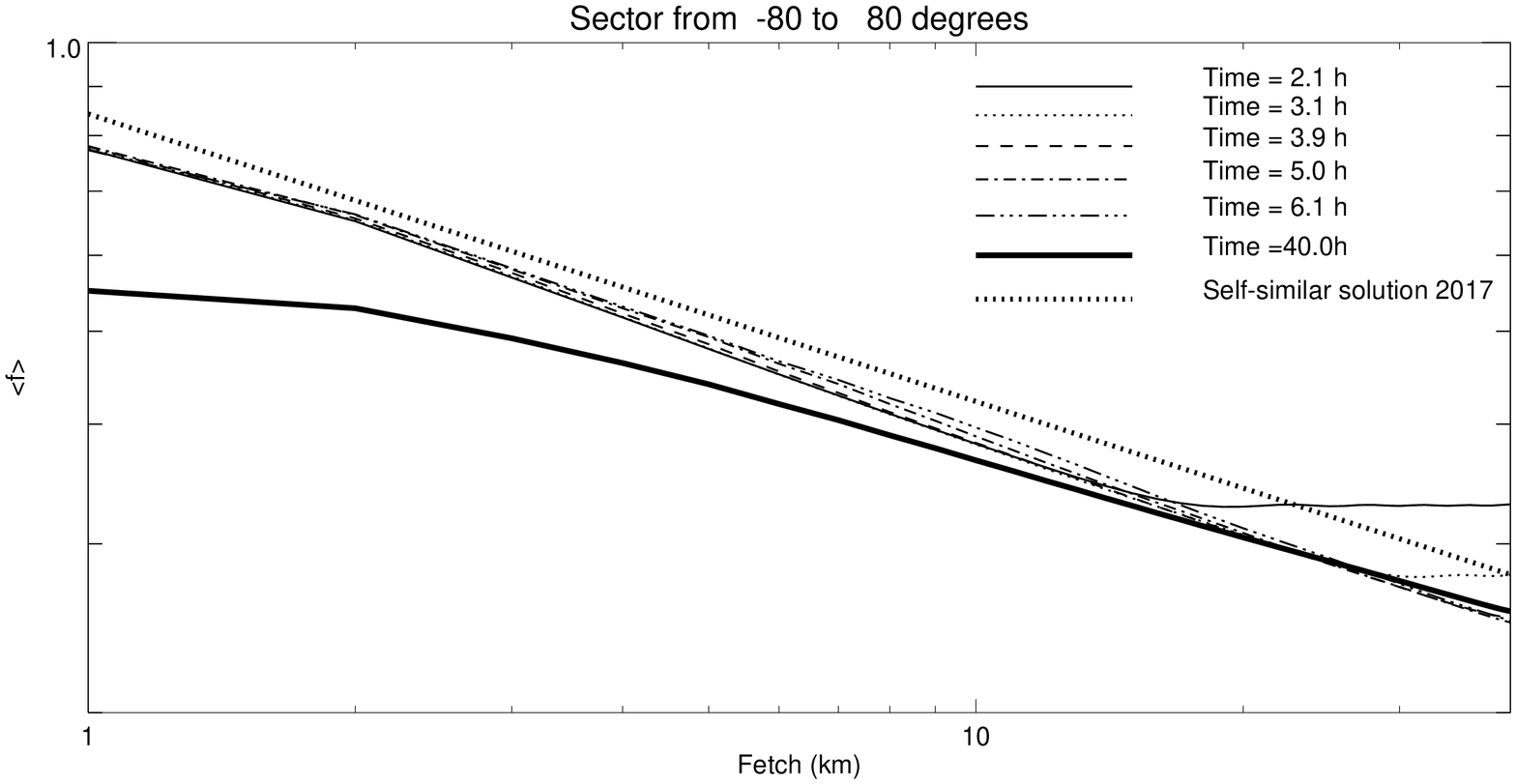} 
\caption{ Decimal logarithm of mean frequency distribution as the function of the decimal logarithm of the fetch for different moments of time.}
\label{MeanFreqCentralOnFetch}
\end{figure}

Similar to the separate analysis of the total wave energy,  Fig.\ref{MeanFreqCentralOnFetch} shows decimal logarithm of wave sea mean frequency distribution along the fetch for different moments of time, calculated in the angle spread $-\pi/8 < \theta < \pi/8$, which takes into account only the wave sea effects, exluding the quazi-monochromatic waves. One can see that this wind sea distribution is asimptotically closer in time to the classical self-similar form, than the distribution on Fig.\ref{TotalEnergyOnFetch}, except closely adjacent to the west coast region of 3 km width. This effect is quite natural to expect, since the intensity of quazi-monochromatic waves grows approaching to the west coast, along with degree of their interaction with the wave sea (i.e. interaction of the "red" and "green" pipes), leading to the deformation of self-similar behavior.

\begin{figure}[ht] 
\noindent\center\includegraphics[scale=0.7, angle=90]{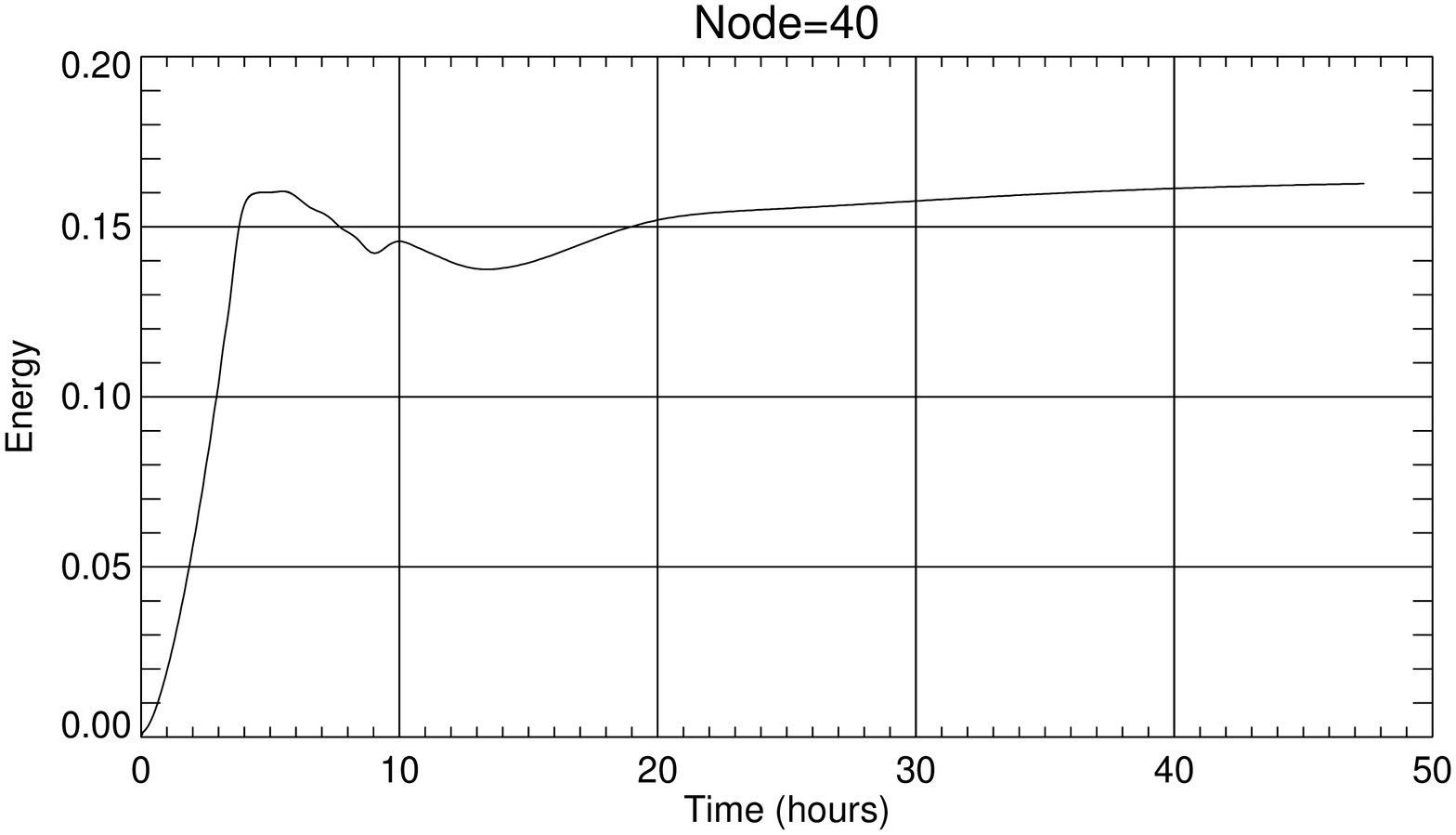} 
\caption{Energy dependence on time at the fetch distance $x=40$ km.}
\label{EnergyVsTime40}
\end{figure}

\begin{figure}[ht] 
\noindent\center\includegraphics[scale=0.7, angle=90]{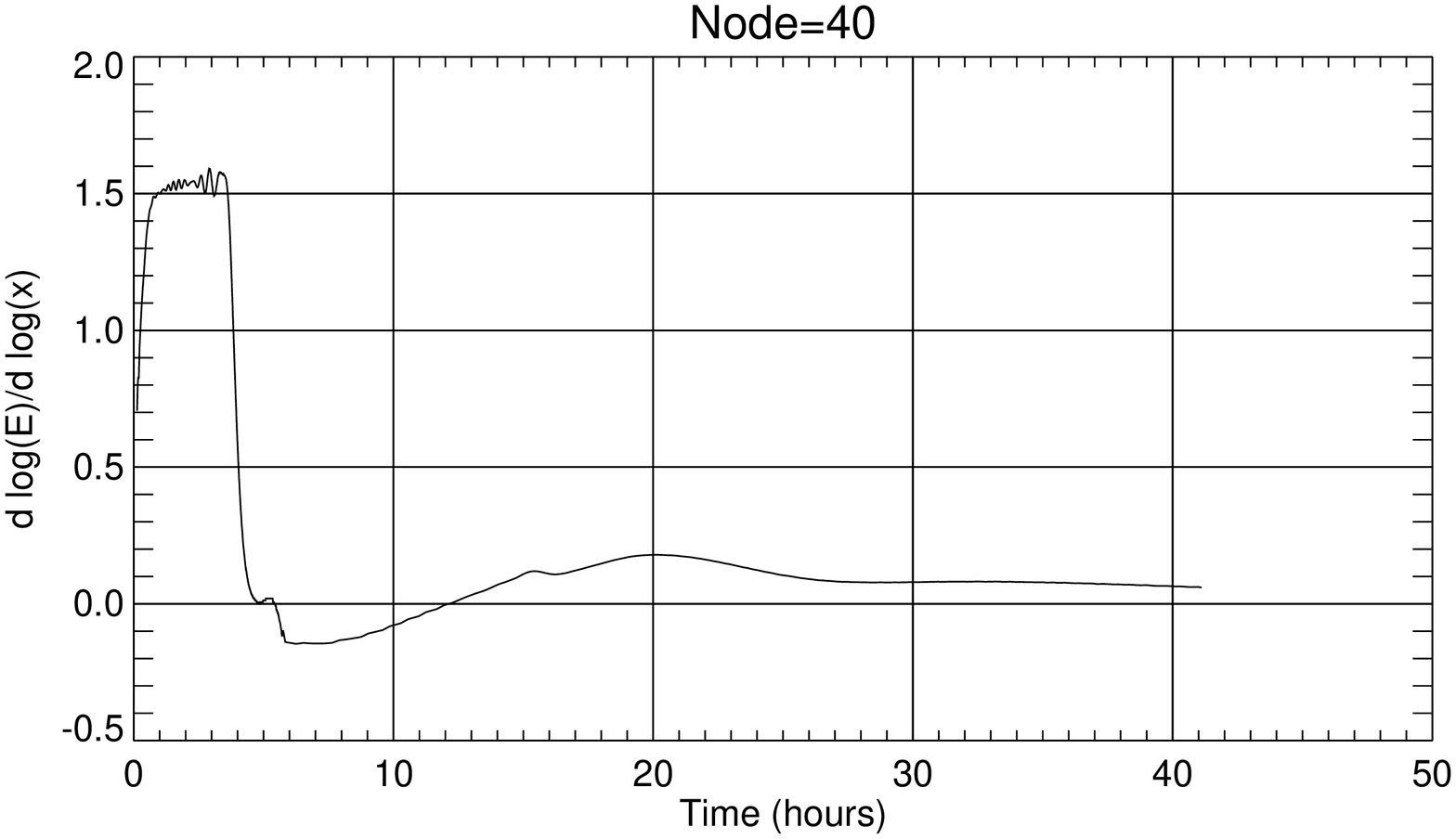} 
\caption{Energy dependence on time at the fetch distance $x=40$ km.}
\label{EnergyVsTime40Index}
\end{figure}

Fig.\ref{EnergyVsTime40} shows energy dependence on time at the fetch distance $x=40$ km, the east coast. Fig.\ref{EnergyVsTime40Index} presents local power index of energy dependence on time at the fetch distance $x=40$ km, the east coast, corresponding to the dependence Fig.\ref{EnergyVsTime40}. One can see that for time span between 1 and 4 hours the power index is close to $10/7$, which value is within $10\%$ deviation from the self-similar solution index for the duration limited case.

\conclusions \label{Concl}

We presented numerical simulation of ocean surface wave turbulence in the channel with the constant wind, blowing prependicular to the shore lines.  It shows that limited fetch growth in the straits can be splitted in space and time into different processes.

Initial process consists of threshold-like  wave energy front propagation in the form of spectral energy single hump from the west to east coast for characteristic times, defined by the ratio of the channel width to the characteristic spectral peak advection velocity. This regime is localized in the positive advection velocity part of the Real-Fourier space, corresponding to the waves, moving in the wind direction, and has similarities with already knows self-similar limited fetch regimes for unlimited domains. 

The second regime occurs later in time, after the wave energy threshold-like front reaches the east coast, and its start is caused by initial nonlinear interference of the seeding white noise wave energy, contained in the negative velocity advection region (green pipeline) with already formed spectrum in the positive velocity advection region (red pipeline). The exhibition of the second regime consists in the amplification of quasi-monochromatic waves, propagating orthogonally to the wind. The described wave system works as the laser-like \textit{ Nonliner Ocean Wave Amplifier} of the quazi-monochromatic waves in the wind-orthogonal direction, which we reduce to the acronym \textit{NOWA}, and is apparently predominantly connected with the wave energy condensation on the separatrix, dividing the regions of zero advection velocity.

It is quite surprising that the wave system eventually reaches asymptotic equilibrium, or 'mature sea state', due to balancing of the energy coming through the wind input channel by another two channels of energy dissipation: wave-breaking dissipation and wave energy absorption at shorelines. The described mechanism of 'mature sea' formation yield the good physically based alternative to the dubious 'mature sea' concept, circulating in oceanographic literature. The part of the stationary wave energy spectrum closer to the beginning of the fetch tends to slant against the wind at the angle of $15^{\circ}$ with respect to the shore line. Those waves, having velocity component opposite to the wind, demonstrate, in particular, the fact of nonlinear waves generation against the wind. 

We also show that surface wave turbulence in the channels is separated to already known wind-sea self-similar regime and low-frequency quazi-monochromatic waves, propagating almost ortogonally to the wind.

Asymtotically in time, the part of the wave energy, propagating in the wind direction, is approximately equal to the part of the energy, propagating perpendicular and against the wind. It is the most important result of the presented numerical experiments, demonstrating the importance of correct taking into account of nonlinear interaction in HE, which is also responsible for backscattering of the wind-driven waves as well as quazi-monochromatic waves radiation generation orthogonally to the wind.

The obtained results have multiple concequences. The first one hold the promise for explanation of the seiches, presenting significant problem for moored ships in ports as well as prediction  of the amplitude and localization structure in confined basins.

Another lesson of the presented research epmphasizes the importance of proper understanding of the boundary conditions effects, inevitably existing in wave forecasting models, which should not be interpreted as numerical artifacts. Our next plan is changing of boundary conditions and including of some reflection from the coastal line. As the result, we expect strong amplification of this laser-like effect.

\begin{acknowledgements} 

The research presented in the section \ref{Int} has been accomplished due to the support of the grant “Wave turbulence: the theory, mathematical modeling andexperiment” of the Russian Scientific Foundation No 14-22-00174. The research  set forth in the section \ref{Int}, was funded by the program of the presidium of RAS "Nonlinear dynamics in mathematical and physical sciences". The research presented in other chapters was supported by ONR grant N00014-10-1-0991.

The authors gratefully acknowledge the support of these foundations.

\end{acknowledgements}

\bibliographystyle{copernicus}
\bibliography{references}

\end{document}